%% file: arxiv.tex
\documentclass{article}
\def\ARXIV{}
\usepackage[letterpaper,top=2cm,bottom=2cm,left=3cm,right=3cm,marginparwidth=1.75cm]{geometry}
\usepackage{natbib}
\usepackage{booktabs,caption}
\usepackage[flushleft]{threeparttable}
\usepackage[toc,page]{appendix}
\usepackage{hhline}
\usepackage[utf8]{inputenc}
\usepackage[T1]{fontenc}
\usepackage{amsmath}
\usepackage{rotating}
\usepackage{multirow}
\usepackage{algorithm}
\usepackage{algpseudocode}
\usepackage{fbox}
\usepackage{physics}
\usepackage{mathtools}
\usepackage{graphicx}
\usepackage{float}
\usepackage{amssymb}
\usepackage{listings} 
\usepackage{amsfonts}
\usepackage{bbm}
\usepackage{listings}
\usepackage{longtable}
\usepackage{hyperref}
\usepackage{multirow}
\usepackage{amsthm}
\usepackage{caption}
\usepackage{subcaption}
\usepackage{tabularx}
\usepackage{makecell}
\usepackage{booktabs}
\usepackage{pdflscape}
\usepackage{afterpage}
\usepackage{longtable}
\usepackage{capt-of}% or use the larger `caption` package
\usepackage{tabu}
\usepackage{comment}
\usepackage{bibunits}
\defaultbibliographystyle{abbrvnat}

\usepackage{xcolor}
\theoremstyle{definition}

\definecolor{codegreen}{rgb}{0,0.6,0}
\definecolor{codegray}{rgb}{0.5,0.5,0.5}
\definecolor{codepurple}{rgb}{0.58,0,0.82}
\definecolor{backcolour}{rgb}{0.95,0.95,0.92}

\lstdefinestyle{mystyle}{
    backgroundcolor=\color{backcolour},   
    commentstyle=\color{codegreen},
    keywordstyle=\color{magenta},
    numberstyle=\tiny\color{codegray},
    stringstyle=\color{codepurple},
    basicstyle=\ttfamily\footnotesize,
    breakatwhitespace=false,         
    breaklines=true,                 
    captionpos=b,                    
    keepspaces=true,                 
    numbers=left,                    
    numbersep=5pt,                  
    showspaces=false,                
    showstringspaces=false,
    showtabs=false,                  
    tabsize=2
}

\title{Scalable non-separable spatio-temporal Gaussian process models for large-scale short-term weather prediction}

\author{
  Tim Gyger\thanks{Institute of Financial Services Zug, Lucerne University of Applied Sciences and Arts}
  \thanks{Department of Mathematical Modeling and Machine Learning, University of Zurich}\and Reinhard Furrer\footnotemark[2]\and Fabio Sigrist
  \thanks{Seminar for Statistics, ETH Zurich} \footnotemark[1]
}

\date{\today}
\begin{document}

\maketitle

\input{abstract.tex}

\input{main.tex}

\bibliographystyle{abbrvnat}
\bibliography{bib_IterartiveFSA}

\clearpage

\include{appendix.tex}

\end{document}

%% file: abstract.tex
\begin{abstract}

Monitoring daily weather fields is critical for climate science, agriculture, and environmental planning, yet fully probabilistic spatio-temporal models become computationally prohibitive at continental scale. We present a case study on short-term forecasting of daily maximum temperature and precipitation across the conterminous United States using novel scalable spatio-temporal Gaussian process methodology. Building on three approximation families — inducing-point methods (FITC), Vecchia approximations, and a hybrid Vecchia-inducing-point full-scale (VIF) approach — we introduce three extensions that address key bottlenecks in large space-time settings: (i) a scalable correlation-based neighbor selection strategy for Vecchia approximations with point-referenced data, enabling accurate conditioning under complex dependence structures, (ii) a space-time kMeans++ inducing-point selection algorithm, and (iii) GPU-accelerated implementations of computationally expensive operations, including matrix operations and neighbor searches. Using both synthetic experiments and a large NOAA station dataset containing more than one million space-time observations, we analyze the models with respect to predictive performance, parameter estimation, and computational efficiency. Our results demonstrate that scalable Gaussian process models can yield accurate continental-scale forecasts while remaining computationally feasible, offering practical tools for weather applications. %All methods are available in an open-source C++ library with R and Python interfaces (\url{https://github.com/fabsig/GPBoost}).

\end{abstract}

%% file: main.tex
\section{Introduction}\label{sect_Intro}

Reliable short-term weather forecasts are essential for applications such as crop management, water allocation, hydrological modeling, risk assessment, and utility planning. Advances in meteorological observing systems and remote sensing have created new opportunities for statistical modeling of atmosphere-land processes. Daily temperature and precipitation fields exhibit substantial variability across spatial and temporal scales, are strongly influenced by topography and large-scale circulation, and evolve seasonally. Modeling and forecasting such fields therefore requires approaches that account for joint space-time dependence rather than treating locations or time points independently.

Operational weather forecasting is dominated by numerical weather prediction (NWP) models \citep{bauer2015quiet}, which solve nonlinear physical equations governing atmospheric dynamics and quantify uncertainty through ensemble simulation. These systems have achieved remarkable forecasting skill but require specialized high-performance computing infrastructure and substantial operational resources. Recently, purely data-driven artificial intelligence (AI)-based forecasting models have emerged as competitive alternatives to traditional NWP systems \citep{ben2024rise}. Prominent examples include FourCastNet \citep{pathak2022fourcastnet}, Pangu-Weather \citep{bi2022pangu}, and GraphCast \citep{lam2023learning}. Unlike physics-based NWP approaches, these methods learn relationships directly from massive historical weather archives, often consisting of global gridded reanalysis and simulation data. %Once trained, they can provide forecasts at substantially lower computational cost and with faster inference times.

Despite their success, both modern NWP and AI-based forecasting systems are primarily designed for gridded forecasting, often at spatial resolutions of approximately $0.25^\circ$ (roughly 20--30\,km across the continental United States), and rely on extensive computational resources, high-dimensional atmospheric inputs, and very large training datasets. In many applications, however, the primary interest lies in accurately modeling local fine-scale variability, flexibly incorporating covariates, and obtaining interpretable probabilistic forecasts with principled uncertainty quantification. Moreover, operational forecasting products may be unavailable, too coarse in spatial resolution, difficult to recalibrate, computationally inaccessible, or lack publicly available historical forecast data needed for retrospective evaluation. In such settings, statistical models trained directly on observational station data provide an attractive complementary approach.

Gaussian process (GP) models offer a flexible, fully probabilistic framework for representing complex spatio-temporal dependence. In particular, non-separable covariance functions such as those of \citet{gneiting2002nonseparable} allow spatial correlation to vary with temporal lag and temporal dependence to vary with spatial separation, thereby capturing key interactions underlying realistic weather dynamics. These properties have led to widespread use of GPs in environmental and climate statistics, including short-term weather and spatio-temporal modeling \citep{cressie2011statistics, porcu202130}. However, applying exact GP inference to continental-scale monitoring networks observed over many time points is computationally infeasible, since likelihood evaluation and prediction require factorizing an $n \times n$ covariance matrix with $\mathcal{O}(n^3)$ computational cost and $\mathcal{O}(n^2)$ memory requirements.

To overcome these limitations, numerous scalable GP approximations have been proposed \citep[see, e.g.,][]{heaton2019case}. Low-rank inducing-point methods such as the modified predictive process approximation, also known as the fully independent training conditional (FITC) approximation, \citep{quinonero2005unifying, banerjee2008gaussian, finley2009improving} capture large-scale dependence efficiently. Sparse local approximations exploit conditional independence through covariance tapering \citep{furrer2006covariance} or the Vecchia framework \citep{vecchia1988estimation,datta2016hierarchical,katzfuss2017general}, which induces sparse precision matrices and often yields highly accurate approximations in spatial applications \citep{guinness2019gaussian,rambelli2025accuracy}. Hybrid approaches combine low-rank and sparse ideas, including the full-scale approximation \citep{sang2012full} and the Vecchia-inducing-point full-scale (VIF) approximation \citep{gyger2025vecchia}. Additional scalable GP methodologies include multi-resolution approximations \citep{MRA} and stochastic partial differential equation (SPDE)-based methods \citep{SPDE}.
%, where residual dependence is represented using Vecchia conditioning with correlation-based neighbor selection \citep{kang2023correlation}

Despite substantial progress, two practical challenges remain particularly acute for large non-separable space-time applications involving Vecchia and inducing-point approximations: (a) constructing informative conditioning sets when dependence is not well characterized by purely local space-time proximity, and (b) selecting inducing points that adequately capture both spatial and temporal variability without extensive manual tuning. In this paper, we address these challenges in a continental-scale weather forecasting study using observational station data across the conterminous United States.

Our main methodological contributions are as follows. First, we propose a scalable correlation-based neighbor-search algorithm for standalone Vecchia approximations under non-separable space-time dependence. This contribution builds on the correlation-based neighbor selection used by \citet{gyger2025vecchia} for the residual covariance structure within the VIF framework. Second, we introduce a space-time separated kMeans++ inducing-point selection algorithm (sts-kMeans++) for predictive process and VIF approximations. Unlike standard kMeans++ selection, the proposed procedure explicitly accounts for the distinct spatial and temporal structure of the data and is designed to represent both dimensions efficiently. Third, as part of this work, we develop and integrate new GPU implementations into the \texttt{GPBoost} library (\url{https://github.com/fabsig/GPBoost}). These implementations combine custom \texttt{CUDA} routines developed for this work with optimized NVIDIA \texttt{CUDA} libraries for the main computational bottlenecks, enabling efficient estimation and prediction for spatio-temporal datasets containing millions of observations. Using these methods, we investigate how far scalable GP approximations can be pushed for short-term forecasting of temperature and precipitation using only observational station data and evaluate their predictive performance relative to the operational High-Resolution Rapid Refresh (HRRR) model \citep{dowell2022high}, a convection-allowing NWP system operated by the National Oceanic and Atmospheric Administration (NOAA) that provides hourly forecasts on a 3 km grid across North America. %We therefore consider three families of approximations, namely Vecchia, FITC, and hybrid VIF approximations, within the non-separable covariance framework of \citet{gneiting2002nonseparable}. We introduce two methodological enhancements motivated by large-scale weather forecasting: (i) correlation-based neighbor selection for Vecchia conditional sets, and (ii) a space-time separated kMeans++ (sts-kMeans++) algorithm for inducing-point selection within FITC and VIF. Furthermore, we implement and analyze the effect of GPU acceleration for these methods. We evaluate these strategies using both synthetic experiments and a large real-world application.
%, extending existing work on ordering and conditioning-set construction for Vecchia methods \citep{kang2023correlation, datta2016nonseparable} beyond purely distance-based approaches
 % in large-scale settings and complementing existing inducing-point selection approaches \citep{quinonero2005unifying,banerjee2008gaussian,terenin2022numerically}

\section{Scientific question and dataset}\label{sect_data_sq}

Daily weather conditions at specific locations arise from evolving seasonal cycles, physiographic features, and complex interactions among regional circulation. Consequently, statistical models for weather forecasting must capture dependence not only across large spatial scales but also through time, where local dynamics can change rapidly. This joint space-time structure poses a substantial modeling challenge. While GP models are well-suited for representing such multiscale dependencies, direct GP inference becomes computationally intractable for continental-scale networks with millions of space-time observations. Our focus is therefore on determining whether scalable GP approximations can retain the essential characteristics of space-time dependence, including non-separable interactions between space and time, realistic spatial and temporal correlation ranges, and reliable uncertainty quantification, while making real-time forecasting feasible at continental resolution. Concretely, we treat feasibility as the ability to (i) estimate covariance and mean parameters reliably on large expanding space-time training sets, and (ii) produce well-calibrated predictive distributions with runtimes compatible with routine refitting and forecasting.

\begin{figure}[ht!]
\centering
\ifdefined\ARXIV
  \includegraphics[width = \linewidth]{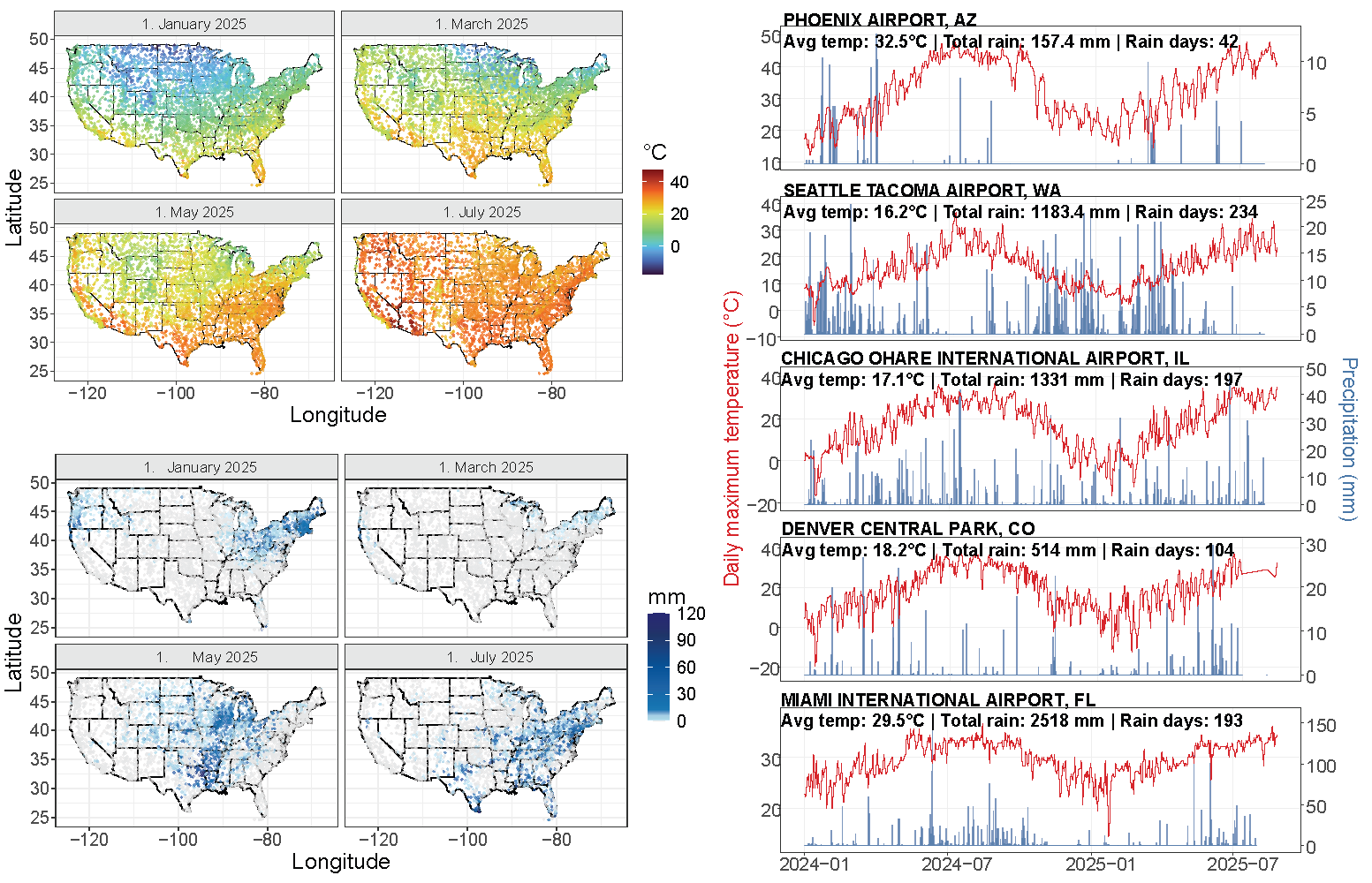}
\else
  \includegraphics[width = \linewidth]{figures/Real_World_all.pdf}
\fi
\caption{Daily maximum temperature (top-left; $^\circ$C, resolution $0.1\,^\circ$C) and precipitation (bottom-left; mm) across the conterminous United States are shown for four selected dates in 2025. The right panels display the corresponding time series at five representative GHCNd stations over the full study period, together with the average temperature, cumulative precipitation (mm), and number of rainy days (precipitation $> 0$ mm). Note that the panel scales differ across stations.}\label{fig:RWall}
\end{figure}

To investigate this question, we use an open-source dataset from NOAA containing daily maximum temperature and precipitation from approximately 3,000 monitoring stations across the conterminous United States. The network exhibits highly heterogeneous spatial coverage, with dense sampling in the eastern states and much sparser across the West. Data were retrieved using the \texttt{R} package \texttt{rnoaa} \citep{chamberlain2023package} from the Global Historical Climatology Network daily (GHCNd) \citep{menne2012overview}. We analyze 608 consecutive days from 2024-01-01 to 2025-08-31, yielding approximately 1.7 million space-time observations. 
Figure \ref{fig:RWall} presents representative snapshots of daily maximum temperature and precipitation across the station network, highlighting the pronounced spatial variability in the dataset. The figure further illustrates the temporal variability and heterogeneity of the observations through daily temperature and precipitation time series at five representative stations covering distinct U.S.\ climate regimes: humid subtropical Miami, arid Phoenix, temperate coastal Seattle, continental Chicago, and high-elevation Denver. Daily maximum temperatures exhibit strong annual seasonality at all locations, with comparatively smooth temporal evolution in the western United States and more pronounced winter-summer contrasts in continental regions such as Chicago and Denver. Precipitation patterns are considerably more intermittent and heterogeneous, differing substantially in both frequency and intensity across stations. Miami and Seattle experience frequent precipitation events, whereas Phoenix and Denver are characterized by extended dry periods interrupted by occasional intense storms. 
%The episodic nature and right-skewed distribution of precipitation amounts, together with the smoother yet spatially varying behavior of temperature, highlight the need for models that simultaneously handle zero inflation, skewness, and non-separable space-time dependence.
This dataset provides a demanding benchmark: it contains non-Gaussian precipitation responses and strong spatio-temporal residual variation. It therefore offers a realistic testing ground for evaluating the scalability and predictive performance of Vecchia, FITC, and hybrid VIF GP models under a non-separable space-time structure. Importantly, the application spans both a continuous response modeled using a Gaussian likelihood (temperature) and an intermittent, right-skewed response with many zeros (precipitation), allowing us to assess approximation behavior under markedly different distributional regimes. %The complete dataset and fully reproducible code are publicly available at \url{https://github.com/TimGyger/SpaceTimeGPApprox}. 
%(\texttt{rnoaa::meteo\_pull\_monitors})

As covariates, we include spatial coordinates projected into the USGS Contiguous US Albers Equal Area system (\textit{EPSG:5070}, in meters), together with physiographic covariates capturing large-scale and local terrain effects. These include elevation, distance to the coastline, slope, and terrain orientation, represented through northness and eastness. Nonlinear topographic influences are accommodated by quadratic terms in spatial coordinates and elevation, as well as interactions between elevation, slope, and coastline distance. Seasonal variability is modeled using periodic sine-cosine basis functions at annual and semiannual frequencies. To allow seasonal patterns to vary across space, these basis functions interact with elevation, distance to the coastline, and spatial coordinates, enabling regionally heterogeneous seasonal amplitudes and phases. Large-scale climatological regimes are further represented using Köppen-Geiger climate classifications \citep{koppen2011thermal, beck2018present}, obtained via the \texttt{R} package \texttt{kgc} \citep{bryant2017kgc}. For temperature forecasting, we additionally include 30-year NOAA temperature normals to anchor the mean structure to long-term climatology. Additional dataset visualizations and fully reproducible code are provided in Appendix \ref{App:DATA}.

\section{Spatio-temporal Gaussian process model}\label{sect2}

We model observations at space-time locations $\{(\boldsymbol{s}_i,t_i)\}_{i=1}^{n}$, with $\boldsymbol{s}_i \in \mathcal{S}\subset\mathbb{R}^2$ and $t_i\in\mathcal{T}\subset\mathbb{R}$, as the sum of a structured mean component driven by known geographical, temporal, and climatological covariates, and a random effect that captures residual spatio-temporal variability. The model for the temperature data is
\begin{align}
    \boldsymbol{Y} = \boldsymbol{X}\boldsymbol{\beta} + \boldsymbol{b} + \boldsymbol{\epsilon},
    \qquad
    \boldsymbol{b} \sim \mathcal{N}(\boldsymbol{0}, \mathbf{\Sigma}),
    \qquad
    \boldsymbol{\epsilon} \sim \mathcal{N}\!\left(\boldsymbol{0}, \sigma^2 \boldsymbol{I}_n\right),
    \label{Eq:Model}
\end{align}
where $\boldsymbol{Y} = \big(Y((\boldsymbol{s}_1, t_1)), \ldots, Y((\boldsymbol{s}_n, t_n))\big)^{\mathrm{T}}\in \mathbb{R}^n$, $\boldsymbol{X} \in \mathbb{R}^{n \times p}$ is the design matrix constructed from $p$ predictors such as elevation, climate zone, and seasonality, $\boldsymbol{\beta}\in \mathbb{R}^{p}$ is the corresponding vector of regression coefficients, and $\boldsymbol{\epsilon}$ represents independent microscale variation (a nugget). The latent process $\boldsymbol{b}$ captures structured space-time dependence with covariance elements $\Sigma_{ij}
=
c\big((\boldsymbol{s}_i,t_i), (\boldsymbol{s}_j,t_j)\big)$, for a parametric covariance function $c(\cdot,\cdot)$ defined on $\mathcal{S}\times\mathcal{T}$.
%\subsection{Space-time covariance structure}
Constructing valid space-time covariance functions is challenging because they must remain positive definite over both spatial and temporal domains simultaneously. A common simplifying assumption is separability, where $c\big((\boldsymbol{s}_i,t_i), (\boldsymbol{s}_j,t_j)\big)
= c_S(\boldsymbol{s}_i,\boldsymbol{s}_j)\cdot c_T(t_i,t_j)$, which decouples spatial and temporal dependence \citep{gneiting2006geostatistical, chen2021space}. While separability facilitates scalable computation, it prohibits interaction between space and time, which is often unrealistic for weather processes such as moving fronts. 
% Another simplification is space-time isotropy, where dependence is assumed to depend only on a combined distance, e.g., $\sqrt{\|\boldsymbol{s}_i-\boldsymbol{s}_j\|_2^2 + C\cdot|t_i-t_j|^2}$, where $C>0$ is a constant and $\|\cdot\|_2$ the Euclidean norm. Relaxing isotropy allows directions or evolving spatial correlation structures, which is particularly beneficial for environmental processes.
To allow for such interactions, \citet{gneiting2002nonseparable} propose flexible non-separable covariance functions. A common form \citep{datta2016nonseparable} is
\begin{align}
\begin{split}
    c\big((\boldsymbol{s}_i, t_i), (\boldsymbol{s}_j, t_j)\big) = 
    &\frac{\sigma^2_1 2^{1 - \nu}}{\Gamma(\nu)(a |t_i - t_j|^{2\alpha} + 1)^{\delta + \gamma d/2}}
    \left( \frac{c \|\boldsymbol{s}_i - \boldsymbol{s}_j\|_2}{(a |t_i - t_j|^{2\alpha} + 1)^{\gamma/2}} \right)^\nu
    \\ &K_\nu \!\left( \frac{c \|\boldsymbol{s}_i - \boldsymbol{s}_j\|_2}{(a |t_i - t_j|^{2\alpha} + 1)^{\gamma/2}} \right),
\end{split}
    \label{Gneiting}
\end{align}
which reduces to an isotropic Matérn model with smoothness parameter \( \nu > 0 \) and spatial decay parameter $c > 0$ at zero temporal lag $|t_i - t_j| = 0$. The parameters \(\alpha \in (0,1]\) and \(a > 0\) control the smoothness and scaling of the temporal component, respectively, while \(\gamma \in [0,1]\) governs the degree of non-separability between space and time, yielding a separable covariance function when \(\gamma = 0\). The parameter \(\delta \geq 0\) further affects the temporal scaling of the overall covariance, independently of the non-separability effect.

\subsection{Estimation and prediction}

We adopt an empirical Bayes approach \citep{robbins1964empirical} and estimate the parameters $\boldsymbol{\theta}=\{\sigma^2,\sigma_1^2,a,c,\alpha,\nu,\gamma,\delta\}$ and regression coefficients $\boldsymbol{\beta}$ by maximizing the marginal likelihood. For Gaussian responses, this corresponds to minimizing the negative log-marginal likelihood
\begin{align*}
\mathcal{L}(\boldsymbol{\beta},\boldsymbol{\theta};\boldsymbol{Y},\boldsymbol{X})
=
\frac{n}{2} \log (2 \pi)
+\frac{1}{2}\log \det\!\big(\widetilde{\mathbf{\Sigma}}\big)
+\frac{1}{2}(\boldsymbol{Y}-\boldsymbol{X}{\boldsymbol{\beta}})^{\mathrm{T}}
\widetilde{\mathbf{\Sigma}}^{-1}(\boldsymbol{Y}-\boldsymbol{X}{\boldsymbol{\beta}}),
\end{align*}
where $\widetilde{\mathbf{\Sigma}}={\mathbf{\Sigma}} + \sigma^2 \boldsymbol{I}_n$. Gradients used in gradient-based optimization involve terms such as
\begin{align*}
\frac{\partial}{\partial \boldsymbol{\theta}}\mathcal{L}(\boldsymbol{\beta},\boldsymbol{\theta};\boldsymbol{Y},\boldsymbol{X})
=
\frac{1}{2}\mathrm{Tr}\!\left(\widetilde{\mathbf{\Sigma}}^{-1}
\frac{\partial \widetilde{\mathbf{\Sigma}}}{\partial \boldsymbol{\theta}}\right)
-\frac{1}{2}(\boldsymbol{Y}-\boldsymbol{X}{\boldsymbol{\beta}})^{\mathrm{T}}
\widetilde{\mathbf{\Sigma}}^{-1}  
\frac{\partial \widetilde{\mathbf{\Sigma}}}{\partial \boldsymbol{\theta}} 
\widetilde{\mathbf{\Sigma}}^{-1}(\boldsymbol{Y}-\boldsymbol{X}{\boldsymbol{\beta}}),
\end{align*}
whose evaluation requires $\mathcal{O}(n^3)$ operations. This motivates the scalable approximations studied in Section~\ref{sectFSA}. The predictive distribution at $n_p$ new space-time points $\{(\boldsymbol{s}^p_1,t^p_1),...,(\boldsymbol{s}^p_{n_p},t^p_{n_p})\}$ is $\boldsymbol{Y}^p | \boldsymbol{Y}\sim\mathcal{N}(\boldsymbol{\mu}^p,\mathbf{\Sigma}^p)$ with predictive mean
\begin{align*}
\boldsymbol{\mu}^p=\boldsymbol{X}^p{\boldsymbol{\beta}}+{\mathbf{\Sigma}}^{\mathrm{T}}_{n{n_p}}\widetilde{\mathbf{\Sigma}}^{-1}(\boldsymbol{Y}-{\boldsymbol{X}}{\boldsymbol{\beta}})
\end{align*}
and predictive covariance 
\begin{align*}
\mathbf{\Sigma}^p={\mathbf{\Sigma}}_{{n_p}} + \sigma^2 \boldsymbol{I}_{n_p}-{\mathbf{\Sigma}}_{n{n_p}}^{\mathrm{T}}\widetilde{\mathbf{\Sigma}}^{-1} {\mathbf{\Sigma}}_{n{n_p}}\in\mathbb{R}^{{n_p}\times {n_p}}, 
\end{align*}
where $\boldsymbol{X}^p_i=\Big({X}\big((\boldsymbol{s}^p_i,t^p_i)\big)_{1}, \ldots, {X}\big((\boldsymbol{s}^p_i,t^p_i)\big)_{p}\big) \in \mathbb{R}^{1\times p}$ is the $i$-th row of $\boldsymbol{X}^p\in\mathbb{R}^{n_p \times p}$ containing predictor variables for prediction $i$, $i=1, \ldots, {n_p}$, ${\mathbf{\Sigma}}_{n{n_p}}=\left[c\left((\boldsymbol{s}_i,t_i), (\boldsymbol{s}^p_j,t^p_j)\right)\right]_{i=1:n, j=1:{n_p}} \in\mathbb{R}^{n\times {n_p}}$ is a cross-covariance matrix, and ${\mathbf{\Sigma}}_{{n_p}}=\left[c\left((\boldsymbol{s}^p_i,t^p_i), (\boldsymbol{s}^p_j,t^p_j)\right)\right]_{i=1:{n_p}, j=1:{n_p}} \in\mathbb{R}^{{n_p}\times {n_p}}$.

\subsection{Extension to non-Gaussian responses for rainfall modeling}\label{sect.nonG}

Precipitation exhibits a mixed discrete-continuous distribution, with a point mass at zero and a skewed positive tail. Existing approaches either model occurrence and amount separately \citep{coe1982fitting,wilks1999multisite} or jointly via censored latent-variable models \citep{bell1987space,sanso2004bayesian,sigrist2012dynamic}, following the Tobit framework of \citet{tobin1958estimation} adapted by \citet{stidd1973estimating}. We follow this latter approach. Specifically, to model precipitation, we use the likelihood
\begin{align*}
p(\boldsymbol{Y}\mid \boldsymbol{\mu},\sigma^2,\lambda)&=\prod_{i=1}^n p\!\left(Y((\boldsymbol{s}_i,t_i)) \mid \mu((\boldsymbol{s}_i,t_i)),\sigma^2,\lambda\right)
=\prod_{i=1}^n 
\Bigg[\Phi\!\left(-\frac{\mu((\boldsymbol{s}_i,t_i))}{\sigma}\right)\Bigg]^{\mathbbm{1}\{Y((\boldsymbol{s}_i,t_i))=0\}}
\\[-2pt]
&\quad\times
\Bigg[
\frac{1}{\sigma}\,
\phi\!\left(\frac{Y((\boldsymbol{s}_i,t_i))^{1/\lambda}-\mu((\boldsymbol{s}_i,t_i))}{\sigma}\right)\,
\frac{1}{\lambda\,Y((\boldsymbol{s}_i,t_i))^{\,1-1/\lambda}}
\Bigg]^{\mathbbm{1}\{Y((\boldsymbol{s}_i,t_i))>0\}},
\end{align*}
where $\boldsymbol{\mu} = \boldsymbol{X}\boldsymbol{\beta} + \boldsymbol{b}$. In such a model, rainfall amounts arise from a censored and power-transformed variable,
\begin{align}
\boldsymbol{Y}
= \max\!\bigl(\boldsymbol{0},\, \boldsymbol{X}\boldsymbol{\beta} + \boldsymbol{b} + \boldsymbol{\epsilon}\bigr)^{\lambda}, 
\qquad 
\boldsymbol{b} \sim \mathcal{N}(\boldsymbol{0}, \mathbf{\Sigma}), 
\qquad 
\boldsymbol{\epsilon} \sim \mathcal{N}\!\left(\boldsymbol{0}, \sigma^2 \boldsymbol{I}_n\right),
\label{Eq:Model2}
\end{align}
where $\lambda>0$ controls the skewness. The censoring at zero models rainfall occurrence, while the transformation accounts for right-skewed positive amounts and increasing variance at higher intensities. Together, these components define a zero-censored power-transformed normal (ZC-PTN) likelihood, linking precipitation occurrence and intensity through a shared latent GP. As the marginal likelihood is intractable, we use a Laplace approximation \citep{tierney1986accurate}; see, e.g., \citet{williams2006gaussian} and \citet{sigrist2022latent}. Appendix \ref{App:Lap} contains a detailed description of the optimization procedure, including derivations of the Laplace approximation.

\section{Approximations and novel methodological contributions}\label{sectFSA}

Large-scale spatio-temporal GP models require approximations to remain computationally feasible. Among these, the Vecchia approximations \citep{vecchia1988estimation} have emerged as one of the most accurate and scalable approaches for spatial and spatio-temporal data  \citep{rambelli2025accuracy}, and have been described as a ``leader among the sea of approximations’’ by \citet{guinness2019gaussian}.

\subsection{Vecchia approximation}

Vecchia approximations can be viewed as structured composite likelihood methods \citep{varin2011overview} in which the full Gaussian likelihood is factorized into a sequence of conditional densities. For a finite-dimensional version of GP $\boldsymbol{b} = \big(b((\boldsymbol{s}_1,t_1)),\ldots,b((\boldsymbol{s}_n,t_n))\big)^{\mathrm{T}}$, the exact likelihood factorizes as $p(\boldsymbol{b} \mid \boldsymbol{\theta})
=
\prod_{i=1}^n p\Big(b((\boldsymbol{s}_i,t_i)) \mid b((\boldsymbol{s}_1,t_1)),\ldots,b((\boldsymbol{s}_{i-1},t_{i-1})),\boldsymbol{\theta}\Big)$. A Vecchia approximation replaces each full conditioning set with a small neighborhood set $N(i)$: $p(\boldsymbol{b} \mid \boldsymbol{\theta})
\approx
\prod_{i=1}^n p\big(b((\boldsymbol{s}_i,t_i)) \mid \boldsymbol{b}_{N(i)}, \boldsymbol{\theta}\big)$, where $N(i)$ typically contains the $m_v$ nearest neighbors among the preceding observations. This yields an approximation
\[
\boldsymbol{b} \stackrel{\text{approx}}{\sim} \mathcal{N}(\mathbf{0}, {\mathbf{\Sigma}}^\mathrm{s}),
\qquad
{\mathbf{\Sigma}}^\mathrm{s} = \boldsymbol{B}^{-1} \boldsymbol{D} \boldsymbol{B}^{-T},
\]
with a sparse precision matrix $({\mathbf{\Sigma}}^\mathrm{s})^{-1} = \boldsymbol{B}^{T}\boldsymbol{D}^{-1}\boldsymbol{B}$. Here, $\boldsymbol{B}\in\mathbb{R}^{n\times n}$ is a unit lower-triangular matrix whose nonzero off-diagonal
entries in row $i$ occur only in the columns indexed by $N(i)$ and are given by $\boldsymbol{A}_i \;=\; \mathbf{\Sigma}_{i,N(i)} \,
        {\mathbf{\Sigma}}_{N(i)}^{-1}$, where $\mathbf{\Sigma}_{i,N(i)}$ denotes the
cross-covariances between $b((\boldsymbol{s}_i,t_i))$ and its Vecchia neighbors
$\{b((\boldsymbol{s}_k,t_k)) : k \in N(i)\}$, and ${\mathbf{\Sigma}}_{N(i)}$
is the covariance submatrix indexed by $N(i)$. The diagonal matrix
$\boldsymbol{D} = \mathrm{diag}(D_1,\ldots,D_n)\in\mathbb{R}^{n\times n}$ is given by $D_i \;=\; {\mathbf{\Sigma}}_{i,i}
        - \boldsymbol{A}_i\, \mathbf{\Sigma}_{i,N(i)}^\mathrm{T}$. %Because the number of neighbors $m_v$ is usually small, $\boldsymbol{B}$ and $\boldsymbol{D}$ are highly sparse, enabling fast log-likelihood evaluations. %This makes Vecchia particularly attractive for daily weather data with thousands of stations and long time series.

\subsection{Low-rank inducing point approximations}

A second class of scalable GP methods is based on low-rank approximations constructed from a set of $m \ll n$ inducing points, or knots, $\mathcal{S}_{\mathcal{T}}^* = \{(\boldsymbol{s}_1^*,t_1^*),\ldots,(\boldsymbol{s}_m^*,t_m^*)\}$. Let $\boldsymbol{b}^* = \big(b((\boldsymbol{s}_1^*,t_1^*)),\ldots,b((\boldsymbol{s}_m^*,t_m^*))\big)^{\mathrm{T}}$ denote the latent GP at the inducing points. The predictive process approximation \citep{banerjee2008gaussian} approximates $b((\boldsymbol{s},t)) \approx \boldsymbol{c}((\boldsymbol{s},t),\mathcal{S}_{\mathcal{T}}^*)^{\mathrm{T}} \mathbf{\Sigma}_m^{-1} \boldsymbol{b}^*$, where $\mathbf{\Sigma}_m\in\mathbb{R}^{m\times m}$ is the $m \times m$ covariance matrix of the inducing points and  
$\boldsymbol{c}((\boldsymbol{s},t),\mathcal{S}_{\mathcal{T}}^*)$ collects their cross-covariances.  
The implied low-rank covariance is $\mathbf{\Sigma}_{mn}^{\mathrm{T}} \mathbf{\Sigma}_m^{-1} \mathbf{\Sigma}_{mn}$, where $\mathbf{\Sigma}_{mn}\in\mathbb{R}^{m\times n}$ is the cross-covariance matrix between inducing points and observation locations. The fully independent training conditional (FITC) approximation, also known as modified predictive process approximation, \citep{quinonero2005unifying, finley2009improving} adds a diagonal variance correction, giving
\[
\boldsymbol{b} \stackrel{\text{approx}}{\sim} \mathcal{N}(0, {\mathbf{\Sigma}}^\mathrm{l}),
\qquad
{\mathbf{\Sigma}}^\mathrm{l}
=
\mathbf{\Sigma}_{mn}^{\mathrm{T}} \mathbf{\Sigma}_m^{-1} \mathbf{\Sigma}_{mn}
+ \mathrm{diag}\!\big({\mathbf{\Sigma}}
- \mathbf{\Sigma}_{mn}^{\mathrm{T}} \mathbf{\Sigma}_m^{-1} \mathbf{\Sigma}_{mn}\big).
\]
Low-rank approaches capture global structures efficiently, but may struggle to represent fine-scale local variation unless many inducing points are used.

\subsection{The Vecchia-inducing-point full-scale approximation}

Vecchia-inducing-point full-scale (VIF) approximations introduced in \citet{gyger2025vecchia} combine the strengths of low-rank and Vecchia methods through a full-scale decomposition. The latent process is decomposed as $b((\boldsymbol{s},t))
=
b_\mathrm{l}((\boldsymbol{s},t)) + b_\mathrm{s}((\boldsymbol{s},t))$, where $b_\mathrm{l}$ is a low-rank predictive process capturing large-scale structure and $b_\mathrm{s}$ is a residual process capturing local dependence not explained by $b_\mathrm{l}$. The covariance of the low-rank component is $\mathbf{\Sigma}^\mathrm{l}
=
\mathbf{\Sigma}_{mn}^{\mathrm{T}} \mathbf{\Sigma}_m^{-1} \mathbf{\Sigma}_{mn}$,
while the residual covariance is $\mathrm{Cov}(\boldsymbol{b}_{s})
=
{\mathbf{\Sigma}}
- \mathbf{\Sigma}_{mn}^{\mathrm{T}} \mathbf{\Sigma}_m^{-1}\mathbf{\Sigma}_{mn}$.
This residual covariance is dense, and it is approximated using a Vecchia approximation $\mathrm{Cov}(\boldsymbol{b}_{s})^{-1}\approx{(\mathbf{\Sigma}}^\mathrm{s})^{-1}
=
\boldsymbol{B}^{\top}\boldsymbol{D}^{-1}\boldsymbol{B}$, where the Vecchia coefficients now use residual covariances adjusted for the low-rank component. Specifically, $D_i = {\mathbf{\Sigma}}_{i,i}
- \mathbf{\Sigma}_{mi}^{\mathrm{T}} \mathbf{\Sigma}_m^{-1}\mathbf{\Sigma}_{mi}
- \boldsymbol{A}_i\big(\mathbf{\Sigma}_{i,N(i)}^\mathrm{T}-\mathbf{\Sigma}_{mN(i)}^{\mathrm{T}} \mathbf{\Sigma}_m^{-1}\mathbf{\Sigma}_{mi}\big)$, and
$\boldsymbol{A}_i
=
\big(\mathbf{\Sigma}_{i,N(i)} - \mathbf{\Sigma}_{mi}^{\mathrm{T}}\mathbf{\Sigma}_m^{-1}\mathbf{\Sigma}_{mN(i)}\big)
\big({\mathbf{\Sigma}}_{N(i)}
- \mathbf{\Sigma}_{mN(i)}^{\mathrm{T}}\mathbf{\Sigma}_m^{-1}\mathbf{\Sigma}_{mN(i)}\big)^{-1}$.
The resulting full-scale approximation is $\mathbf{\Sigma}\approx
\mathbf{\Sigma}^\mathrm{l} + {\mathbf{\Sigma}}^\mathrm{s}$.

%The VIF approach reduces to FITC if the number of Vecchia neighbors is zero and to the pure Vecchia approximation if no inducing points are used. Thus, VIF bridges the two approaches.

%Applying the Sherman-Woodbury-Morrison formula results to a computational complexity of order \(\mathcal{O}\big(n \cdot (m_v^3 + m_v^2\cdot m + m^2)\big)\) for evaluating the negative log-likelihood and its derivatives, and a memory requirement of order \(\mathcal{O}\big(n \cdot (m + m_v)\big)\). Moreover, the computational costs for the predictive means and variances are $\mathcal{O}\big({n_p}\cdot(m_v^3+m_v^2\cdot m) + n\cdot (m_v+m)\big)$ and $\mathcal{O}\big(n_p\cdot(m_v + m\cdot m_v) + n\cdot(m\cdot m_v + m^2)\big)$, respectively. 

\subsection{Selecting Vecchia neighbors and inducing points}\label{sec5a}

The performance of Vecchia and inducing-points-based GP approximations depends critically on the choice of conditioning sets and inducing point locations. Correlation-based neighbor selection can substantially improve Vecchia accuracy, especially under anisotropy or non-stationarity \citep{kang2023correlation}, but existing approaches are either restricted to gridded data or do not readily extend to general non-separable space-time covariances \citep{datta2016nonseparable}. For inducing-point methods, common strategies include random subsampling, kMeans++ \citep{arthur2007k}, and tree-based approaches \citep{terenin2022numerically}, with kMeans++ offering a favorable accuracy-cost trade-off in spatial FITC and full-scale settings \citep{gyger2024iterative}. Extending such strategies to non-separable space-time models remains an open challenge.

\subsubsection{Correlation-based Vecchia neighbor selection}
Classical Vecchia approximations select neighbors via Euclidean distances, but \citet{kang2023correlation} demonstrate that correlation-based searches often yield higher accuracy. Depending on the covariance structure, this may be reduced to using the Euclidean distance after an appropriate coordinate transformation. In our setting, however, the Gneiting covariance in \eqref{Gneiting} does not admit such a transformation. Correlation-based neighbor search uses the following correlation metric:
\[
d_c((\boldsymbol{s}_i,t_i),(\boldsymbol{s}_j,t_j)) = \sqrt{1- \left|\frac{[\mathbf{\Sigma}]_{ij}}{\sqrt{[\mathbf{\Sigma}]_{ii}[\mathbf{\Sigma}]_{jj}}}\right|}.
\]
We use an extension of the cover tree algorithm \citep{beygelzimer2006cover} to enable efficient correlation-based neighbor selection for general, non-separable space-time covariance functions. In contrast to standard constructions that randomly select candidate nodes during tree building, we insert points sequentially according to their index order, following the procedure outlined in Algorithm~\ref{alg:covertree}. This modification restricts candidate neighbors to observations with smaller indices, which is consistent with the Vecchia factorization and simplifies the neighbor search, leading to improved runtime performance, see Algorithm~\ref{alg:kNN} for more details. \citet{gyger2025vecchia} have also used such a modified cover tree algorithm for correlation-based neighbor search in VIF approximations. Because covariance parameters evolve during optimization, we follow \citet{kang2023correlation} and update the neighbor sets at iterations indexed by powers of two and once more upon convergence. If an update alters the likelihood, the optimization is resumed until
convergence. %Since the correlation-based neighbor sets depend on the covariance parameters, the underlying Vecchia approximation changes during optimization, making a formal convergence analysis difficult. To the best of our knowledge, theoretical guarantees are currently unavailable for this type of adaptive Vecchia approximation. However, related iterative neighbor-update strategies have previously been considered in scaled and correlation-based Vecchia approaches \citep{kang2023correlation}. %To the best of our knowledge, this is the first use of the modified cover-tree-based correlation search in a pure Vecchia setting.  %elkin2023new

\begin{algorithm}[!ht]
\caption{Construction of cover tree for correlation-based Vecchia neighbor search}\label{alg:covertree}
\begin{algorithmic}[1]
\Require Set of data points $\mathcal{D} = \{(\boldsymbol{s}_1,t_1),\ldots,(\boldsymbol{s}_n,t_n)\}$
\Ensure Cover tree $\mathcal{CT}$

\State Initialize:
Maximal distance: $R_\text{max} = 1$ \par
Empty tree: $\mathcal{CT}$ \par
Knot set of $\mathcal{CT}$: $\mathcal{V} \gets \{\}$ \par
Number of levels in $\mathcal{CT}$: $l \gets 0$ \par
Covered data points per level and knot: $\mathcal{C}_{0,0} \gets \mathcal{S}$
\State
Set root in $\mathcal{CT}$: $\mathcal{CT}_{1,1} \gets (\boldsymbol{s}_1,t_1)$, $\mathcal{V} \gets \{(\boldsymbol{s}_1,t_1)\}$, $l \gets 1$, $\mathcal{C}_{l,(\boldsymbol{s}_1,t_1)} \gets \mathcal{C}_{0,0} \setminus \{(\boldsymbol{s}_1,t_1)\}$ 
\While {size($\mathcal{V}$) $<$ $n$}
  \State $R_l \gets R_\text{max}/2^l$ \Comment{New maximal distance between new knots in tree $\mathcal{CT}$ and descendants}
  \State $\mathcal{K} \gets \mathcal{CT}_{l,:}$ \Comment{Set of knots of tree $\mathcal{CT}$ in level $l$}
  \For{$k \in \mathcal{K}$}
  \State $\text{ind}\gets 0$ \Comment{Index of knot in tree $\mathcal{CT}$ at level $l + 1$}
  \While {$\mathcal{C}_{l,k} \neq \{\}$}
        \State $\text{ind}\gets \text{ind} + 1$
        \State $\mathcal{CT}_{l+1,\text{ind}} \gets \{(\boldsymbol{s}_i,t_i) \in \mathcal{C}_{l,k} \mid i=\min\{j\mid (\boldsymbol{s}_j,t_j) \in \mathcal{C}_{l,k}\}\}$ \Comment{Point with smallest index in $\mathcal{C}_{l,k}$}
        \State $\mathcal{V}\gets \mathcal{V} \cup \{\mathcal{CT}_{l+1,\text{ind}}\}$
        \State $\mathcal{C}_{l,k}\gets \mathcal{C}_{l,k}\setminus\{\mathcal{CT}_{l+1,\text{ind}}\}$
         \State $\mathcal{C}_{l+1,\mathcal{CT}_{l+1,\text{ind}}}\gets \{(\boldsymbol{s}_i,t_i) \in \mathcal{C}_{l,k} \mid d_c((\boldsymbol{s}_i,t_i),\mathcal{CT}_{l+1,\text{ind}}) \leq R_l\}$
         \State $\mathcal{C}_{l,k}\gets \mathcal{C}_{l,k}\setminus\mathcal{C}_{l+1,\mathcal{CT}_{l+1,\text{ind}}}$
  \EndWhile      
  \EndFor
  \State $l \gets l + 1$
\EndWhile
\end{algorithmic}
\end{algorithm}

\begin{algorithm}[!ht]
\caption{Find $m_v$ nearest Vecchia neighbors with respect to the metric $d_c$}\label{alg:kNN}
\begin{algorithmic}[1]
\Require Query point $(\boldsymbol{s}_i,t_i)\in\mathcal{D}$, cover tree $\mathcal{CT}$, number of neighbors $m_v$
\Ensure $\mathcal{N}_{m_v}$ set of $m_v$ nearest Vecchia neighbors of $(\boldsymbol{s}_i,t_i)$

\State Initialize: Maximal distance: $R_\text{max} = 1$ \par
Number of levels in tree $\mathcal{CT}$: $l \gets \text{depth}(\mathcal{CT})$ \par
Set of potential nearest neighbors: $\mathcal{Q} \gets \{\mathcal{CT}_{1,1}\}$ \par
Distance to $m_v$ closest point in $\mathcal{Q}$: $D_{m_v} \gets 1$ \Comment{Set to 1 if $|\mathcal{Q}| < m_v$} 
\For {$j = 1 \text{ to } l$}
  \State $\mathcal{C} \gets \{(\boldsymbol{s}_k,t_k) \in \text{Children}((\boldsymbol{s},t))\mid(\boldsymbol{s},t)\in\mathcal{Q} \cap k < i\} \cup \mathcal{Q}$ \Comment{Children of $\mathcal{Q}$ with index $<i$}
  \State $D_{m_v} \gets m_v\text{-th} \text{ smallest distance }d_c((\boldsymbol{s},t),(\boldsymbol{s}_i,t_i)) \text{ for } (\boldsymbol{s},t) \in \mathcal{C}$ \Comment{Set to 1 if $|\mathcal{C}| < m_v$}
  \State $\mathcal{Q} \gets \{(\boldsymbol{s},t) \in \mathcal{C}\mid d_c((\boldsymbol{s},t), (\boldsymbol{s}_i,t_i)) \leq D_{m_v} + 1/2^{j-1}\}$ 
\EndFor
\State $\mathcal{N}_{m_v} \gets \text{Find $m_v$ nearest points to $(\boldsymbol{s}_i,t_i)$ in $\mathcal{Q}$ by brute-force search}$
\end{algorithmic}
\end{algorithm}

Moreover, a Vecchia approximation depends on the ordering of observations, since each conditional density is formed using only a subset of previously ordered points. Proposed strategies include simple coordinate-based orderings, maximum-minimum (max-min), and random orderings \citep{guinness2018permutation}. For space-time data, however, the problem becomes more delicate: ordering solely by spatial coordinates can lead to conditioning sets drawn from irrelevant times, whereas ordering only by time may overlook essential spatial structure within each snapshot. \citet{idir2025improving} systematically compared ordering schemes for space-time Vecchia approximations and found that random ordering and max-min ordering generally provided the best accuracy. Motivated by these results and the computational efficiency of a random ordering, we adopt a time-stratified random ordering: observations are first ordered by time to preserve temporal causality, and within each time point, locations are randomly permuted. We have also compared this approach with a complete random ordering using simulated data and found nearly identical results; see Appendix \ref{App:Sim}.

\subsubsection{Space-time separated kMeans++ algorithm for inducing points selection}

For non-separable space-time covariance functions such as the one in \eqref{Gneiting}, spatial and temporal correlations may each be strong while their joint dependence decays quickly. As a result, if clustering is applied directly in the full space-time domain, inducing points can be selected that are near in space and time yet weakly correlated when considered jointly, leading to an inefficient low-rank representation. This is particularly relevant for weather forecasting, where the main objective is short lead time prediction, often at locations with sparse or no observational history. In such settings, inducing points are unlikely to be time-wise close to the prediction sites, and forecast accuracy depends critically on how well the temporal evolution of the field is represented. Thus, capturing temporal structure can be more influential than maximizing the domain coverage of inducing points. To address this, we propose a space-time separated kMeans++ (sts-kMeans++) inducing-point selection strategy: spatial and temporal centers are selected independently via kMeans++, and their Cartesian product defines a structured but irregular grid. This approach yields inducing points that may share locations at multiple times (enhancing temporal representation) and occupy distinct locations at a single time point (preserving spatial support). The numbers of spatial and temporal centers, $m_s$ and $m_t$, are chosen proportional to the number of unique time points and the average number of stations per time; see Algorithm~\ref{algo:kMean_spacetime}. The resulting method has computational cost $\mathcal{O}(n(m_s + m_t))$. %In Section 2 of the Supplementary Material \citep{SupplementGyger26}, we visualize the joint and sts-kMeans++ inducing point selection.

\begin{algorithm}[H]
\caption{Space-time separated kMeans++ inducing points}\label{algo:kMean_spacetime}
\footnotesize
\begin{algorithmic}[1]
\Require Data $\mathcal{D}= \{(\boldsymbol{s}_1,t_1),\ldots,(\boldsymbol{s}_n,t_n)\} \subset \mathcal{S}\times \mathcal{T}$ with $\mathcal{T}$ the set of unique time points and $\mathcal{S}$ the set of unique space points, total inducing points $m$%, iterations $i_\text{max}$
\Ensure $\mathcal{S}_\mathcal{T}^*=\{(\boldsymbol{s}_1^*,t_1^*),\ldots,(\boldsymbol{s}_{\hat{m}}^*,t_{\hat{m}}^*)\}$ with $\hat{m}\approx m$
\State  \quad $m_s \gets \mathrm{round}\!\big(\sqrt{\frac{m n}{|\mathcal{T}|^2}}\big)$, \quad $m_t \gets \mathrm{round}\!\big(\sqrt{\frac{m |\mathcal{T}|^2}{n}}\big)$ \Comment{$\hat{m} = m_s\cdot m_t \approx m$}
\State $\bar{\mathcal{T}} \gets \texttt{kMeans++}(\mathcal{T},m_t)$, \quad $\bar{\mathcal{S}} \gets \texttt{kMeans++}(\mathcal{S},m_s)$

\State $\mathcal{S}_{\mathcal{T}}^* \gets \{({\boldsymbol{s}}, {t})|({\boldsymbol{s}}, {t})\in \bar{\mathcal{S}}\times \bar{\mathcal{T}} \}$
\end{algorithmic}
\end{algorithm}

\subsection{GPU acceleration}\label{sect_GPU}

Gaussian process computations rely heavily on repeated matrix multiplications, linear solves, and neighbor searches. Even with Vecchia and low-rank approximations, these operations remain the main computational bottlenecks for large space-time datasets, motivating targeted acceleration. Modern platforms combine central processing units (CPUs) and GPUs, whose complementary architectures are well suited to this workload. In this work, we accelerate the dominant computations by offloading dense linear algebra involving cross-covariance matrices (e.g., products and solves with $\boldsymbol{\Sigma}_{mn}$ and $\boldsymbol{\Sigma}_{m}$), sparse matrix-matrix multiplications involving the Vecchia factors $\mathbf{B}$ and $\mathbf{D}$, and Vecchia neighbor searches to the GPU. All methods are implemented in the \texttt{GPBoost} software library using optimized NVIDIA \texttt{CUDA} routines, including \texttt{cuBLAS}, \texttt{cuSPARSE}, and \texttt{cuSOLVER}. Currently, GPU support for the Gneiting \eqref{Gneiting} and Mat\'ern covariance is provided for smoothness values $\nu \in \{0.5,1.5,2.5\}$, for which these kernels admit closed-form expressions involving only elementary functions, thereby avoiding the need to evaluate the modified Bessel function $K_\nu$ on the GPU. The remaining sequential or low-dimensional computations are handled on the CPU. This hybrid approach enables substantial speedups without rewriting the full code base, while preserving portability and reproducibility. Note that calculating the matrices $\mathbf{B}$ and $\mathbf{D}$ in Vecchia approximations on the GPU did not result in a speed-up compared to the CPU-based Vecchia implementation in the \texttt{GPBoost} library. This is because the construction of $\mathbf{B}$ and $\mathbf{D}$ is already highly efficient on the CPU due to optimized multi-threading, while the GPU implementation incurs substantial overhead from memory transfers and synchronization for many small computations, outweighing potential parallelization gains.

\section{Simulated experiments}\label{sect4}

We first compare the predictive and inferential performance of the above-introduced GP approximations using simulated experiments. Specifically, we evaluate Vecchia approximations with either Euclidean or correlation-based nearest-neighbor selection, both using $m_v = 30$ neighbors. For low-rank inducing-point methods, we consider FITC approximations with inducing points selected via both the standard kMeans++ and the proposed sts-kMeans++ algorithm, in both cases using $m = 500$ inducing points. Finally, we evaluate the hybrid VIF approximation, which combines correlation-based Vecchia neighbor selection with sts-kMeans++-selected inducing points, using $m_v = 30$ and $m = 500$. All experiments are conducted on a system equipped with an AMD EPYC~7763 processor using 16 CPU threads, 512~GB of system memory, and an NVIDIA L40S GPU with 48~GB of VRAM. GPU-based computations refer to executions that utilize the L40S GPU in addition to the CPU, whereas CPU-based computations are performed exclusively on the AMD EPYC~7763 processor. All methods presented in this article are implemented in the \texttt{GPBoost} library (\url{https://github.com/fabsig/GPBoost}), and we use version 1.6.6 to perform the simulated experiments and the real-world application. %The code and datasets required to reproduce all experiments are publicly available at \url{https://github.com/TimGyger/SpaceTimeGPApprox}.
%\subsection{Synthetic space-time fields}

\begin{figure}[ht!]
\centering
\includegraphics[scale=0.35]{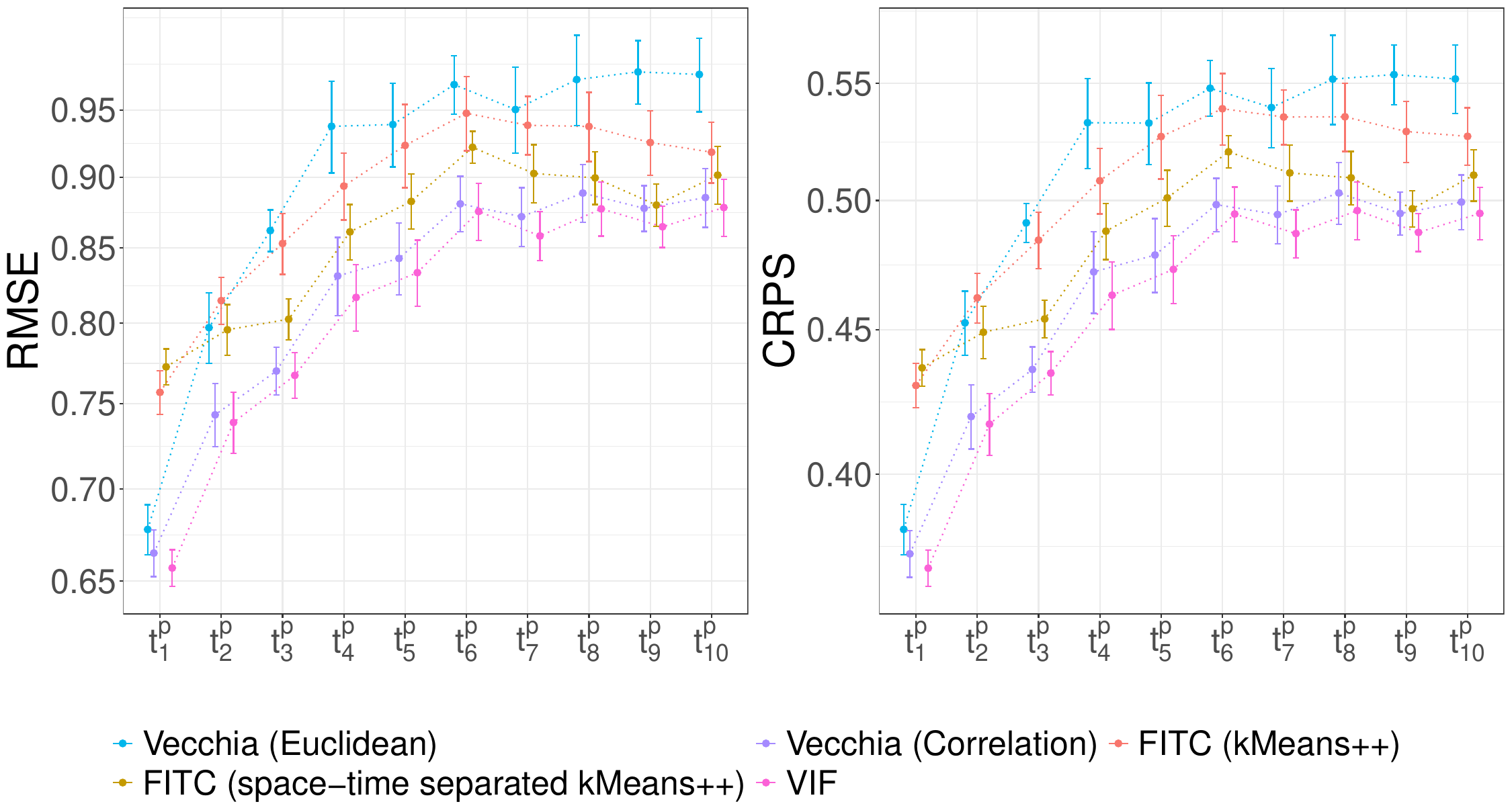}
\caption{Average RMSE and CRPS (log scale) $\pm$ one standard error vs. lead time for the Vecchia, FITC, and VIF approximations.}\label{fig:SimR}
\end{figure}

First, we generate $10$ zero-mean spatio-temporal GP samples of size $15,000$ using the non-separable covariance~\eqref{Gneiting} with nugget $\sigma^2 = 0.01$ and parameters $(\sigma_1^2, a, c, \alpha, \nu, \gamma, \delta) = (1, 0.5, 20, 0.4, 1.5, 0.4, 0.2)$. A total of $500$ spatial locations are sampled uniformly over $[0,1]^2$ and observed at 30 equally spaced time points. Observations with $t \leq 20$ form the training set ($n=10,000$), and those with $t>20$ form the test set ($n_p=5,000$). Parameters are estimated by maximizing the marginal likelihood using the limited-memory BFGS algorithm. Predictive performance is evaluated using the RMSE and the continuous ranked probability score (CRPS) given by $
- \frac{1}{n_p} \sum_{i=1}^{n_p} {{\sigma}_{i}^p}(\frac{1}{\sqrt{\pi}}-2\cdot \phi(\frac{{y}^*_i-{ \mu}^p_{i}}{{{\sigma}_{i}^p}})-\frac{{y}^*_i-{ \mu}^p_{i}}{{{\sigma}_{i}^p}}(2\cdot \Phi(\frac{{y}^*_i-{ \mu}^p_{i}}{{{\sigma}_{i}^p}})-1))$ for Gaussian likelihoods,
where $\phi(x)=\frac{1}{\sqrt{2 \pi}}\exp(-\frac{x^2}{2})$ and $\Phi(x)$ are the density and cumulative distribution functions (CDFs), respectively, of a standard normal distribution, ${\boldsymbol{y}}^*$ is the test response, and $\boldsymbol{\mu}^p$ and ${\boldsymbol{\sigma}}^p$ are the predictive means and variances, respectively. Figure~\ref{fig:SimR} summarizes the RMSE and CRPS across different lead times. The results show that the correlation-based neighbor selection leads to a clear accuracy gain for the Vecchia approximation, yielding uniformly lower errors than Euclidean-based neighbors at every forecast horizon. For the FITC approximation, choosing inducing points via the proposed sts-kMeans++ procedure consistently improves the prediction accuracy relative to kMeans++. Across all settings, the VIF approximation provides the most accurate predictions and the most stable uncertainty quantification, outperforming both standalone Vecchia and FITC approaches.

\begin{figure}[ht!]
    \centering        
    \includegraphics[scale=0.5]{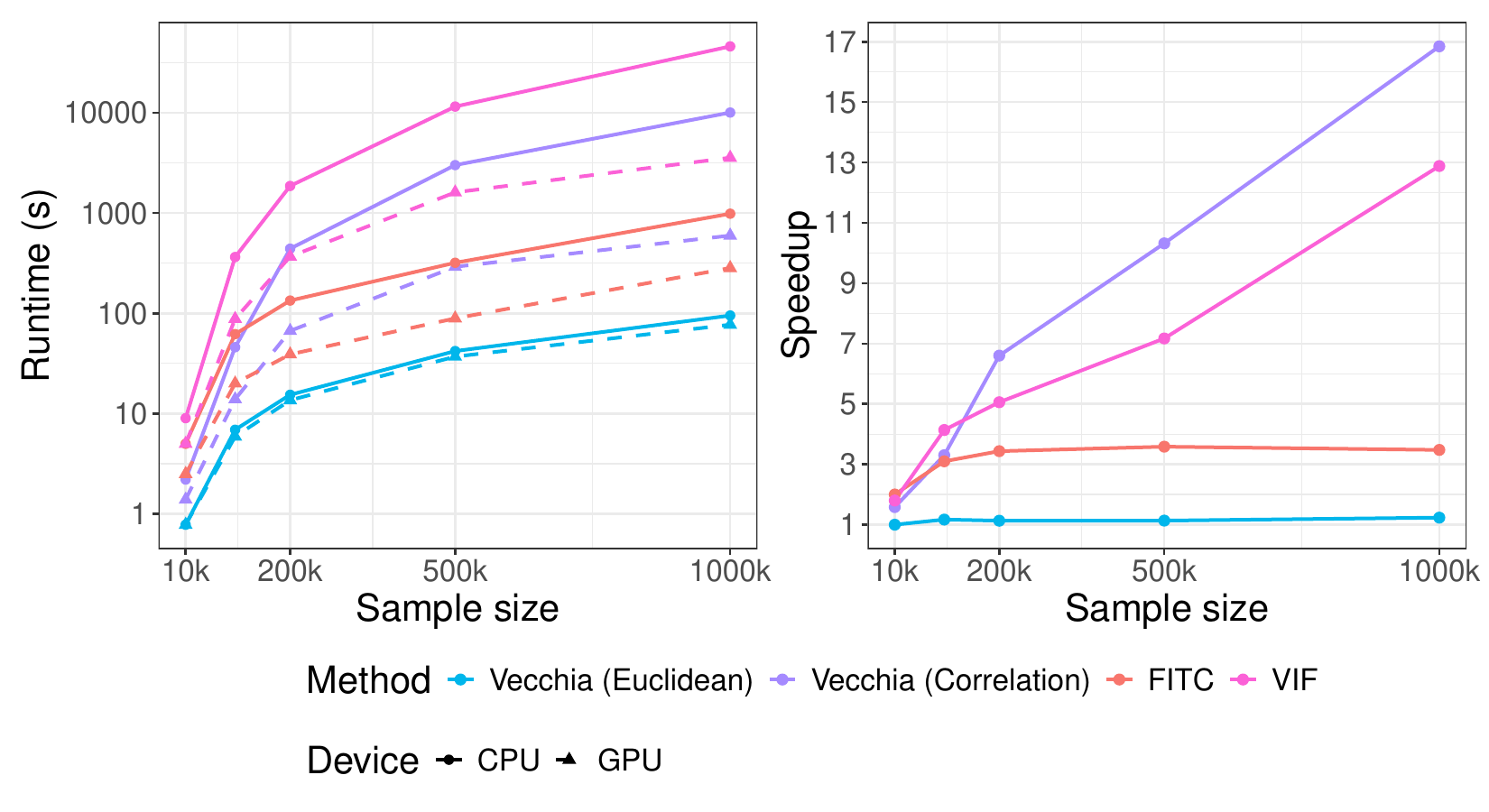}    
    \caption{Runtimes in seconds (log-scale) and GPU speedups per optimization step across sample sizes for the Vecchia, FITC, and VIF approximations.}\label{fig:RWRGPU}
\end{figure}

Next, we analyze the speedup provided by GPU acceleration. Figure~\ref{fig:RWRGPU} reports runtimes and speedups for a single optimization iteration (two likelihood evaluations, gradient computation, and two Vecchia neighbor searches) across varying sample sizes and for different approximation methods. GPU acceleration yields substantial gains, particularly for methods with correlation-based neighbor searches (Vecchia and VIF). For instance, GPU acceleration provides a speedup of a factor of close to $20$ for the correlation-based Vecchia approximation for a sample size of $1,000,000$. FITC and VIF approximations also benefit from GPU acceleration due to matrix multiplications involving $\boldsymbol{\Sigma}_{mn}$. 

Additional simulation results and implementation details are provided in Appendix \ref{App:Sim}. There, we visualize the simulated datasets, compare estimated covariance parameters with their true values, and assess the sensitivity of the proposed correlation-based Vecchia approximation to the ordering strategy. In Appendix \ref{App:Sim}, we also evaluate the methods on the large-scale benchmark datasets from sub-competitions~\textit{2a} and~\textit{2b} of \citet{abdulah2022second}, including comparisons with exact GP inference and with the top-performing competition entries. Furthermore, we report additional runtime experiments comparing CPU and GPU implementations, including operation-specific GPU speedups for likelihood, gradient, and neighbor-search computations. %These experiments additionally consider the large benchmark data, non-Gaussian likelihoods fitted via Laplace approximation, and high-dimensional ARD Matérn covariance models. 
Overall, the additional results provide further evidence for the high prediction accuracy, scalability, and computational benefits of the proposed approximations.

\section{Application to short-term weather prediction}\label{sect5}

In the following, we apply the proposed methodology to generating short-term forecasts of daily maximum temperature and precipitation across the conterminous United States using the datasets and models described in Sections \ref{sect_data_sq} and \ref{sect2}. We additionally compare the resulting forecasts with those from the NWP HRRR model. 

After excluding stations with observations available for fewer than 95\% of the study period to ensure sufficient temporal coverage and stable estimation, the observational network comprises 3,057 stations with highly heterogeneous spatial coverage. Such variability directly motivates scalable space-time modeling, as accurate prediction must transfer information across regions with widely differing station density. The dataset contains 608 consecutive days from 2024-01-01 to 2025-08-31. Forecasts are produced using expanding-window training datasets with observations from the year 2024 serving as the initial training set. To evaluate the prediction accuracy, a random subset of approximately 25\% of the stations is withheld entirely during training, resulting in training datasets containing between approximately $800,000$ and $1.3$ million space-time points. Moreover, this spatio-temporal test data design ensures that every model is evaluated not only for temporal forecasting at observed sites but also for its ability to transfer information to previously unmonitored locations. In practice, forecast users often require predictions in areas with sparse observational histories, at newly installed sensors, or in regions without any monitoring infrastructure (e.g., agricultural zones, remote mountain terrain, coastal transition areas). Beginning on January 1, 2025, we sequentially generate forecasts on every day for daily temperatures and rainfall with lead times of five and three days, respectively. The covariance parameters are re-estimated at the end of each month which allows for adapting to temporal non-stationarity (e.g., shifts in variability and correlation structure). Compared to a daily re-estimation, this reduces computation times while retaining operational realism. We use the same computational environment including GPU acceleration and the same number of Vecchia neighbors and inducing points as in Section \ref{sect4}. 

%For comparison with HRRR, we use archived operational forecasts from the NOAA HRRR archive, accessed via the public AWS S3 bucket. Forecasts are initialized once daily, and hourly forecasts up to 48 hours ahead are extracted at the station locations using values from the nearest HRRR grid cell. Daily maximum temperature forecasts are obtained by taking the maximum of the hourly temperature forecasts over each forecast day, while daily precipitation forecasts are obtained by summing the corresponding hourly accumulated precipitation forecasts. Since archived HRRR forecasts are only available up to 48 hours ahead, the HRRR comparison is restricted to two-day-ahead forecasts.

%The spatio-temporal models introduced in Section~\ref{sect2} are fit using the scalable GP approximations developed in Section~\ref{sectFSA} and evaluated in an operational forecasting setup designed to mirror real-time use. 

 % The \texttt{Python} code used to run all experiments is publicly accessible at \url{https://github.com/TimGyger/SpaceTimeGPApprox}.

 %A scalable GP model that performs well only at instrumented sites but deteriorates when predicting elsewhere has limited operational utility.
 
%By evaluating both temporal and spatial generalization, the forecasting framework therefore directly tests whether scalable GP approximation strategies preserve the underlying space-time dependence and can robustly transfer learned structure to unobserved parts of the weather field.

\subsection{Short-term prediction of daily maximum temperature}\label{sect501}

Daily maximum temperature is modeled using the specification in \eqref{Eq:Model} with the space-time covariance function given in \eqref{Gneiting}. We fix the Matérn smoothness parameter to $\nu = 1.5$. Estimating $\nu$ on a GPU is currently not implemented due to the presence of the modified Bessel function in the covariance, but estimating it on subsets of the data using CPU gives $\nu \approx 1.5$ (results not tabulated). In addition, experiments on a subset of the data indicated that the predictive accuracy is not sensitive to the choice of $\nu$; see Section~\ref{sect503}. The model is fitted using the scalable approximation strategies introduced in Section~\ref{sectFSA}. We generate rolling one- to five-day-ahead forecasts throughout 2025 and compare the GP-based approaches with two non-GP baselines: a simple persistence forecast using the previous day’s maximum temperature and a fixed-effects linear regression model without a GP. In addition, we compare against the operational NOAA HRRR model using archived forecasts accessed through the public AWS S3 archive. Forecasts are initialized once daily, and hourly predictions are extracted from the nearest HRRR grid cell at each station location. Daily maximum temperature forecasts are then obtained by taking the maximum of the hourly forecasts over each forecast day. Since archived HRRR forecasts are only available up to 48 hours ahead, the HRRR comparison is restricted to one- and two-day-ahead forecasts. 

\begin{figure}[ht!]
    \centering
    \includegraphics[scale=0.5]{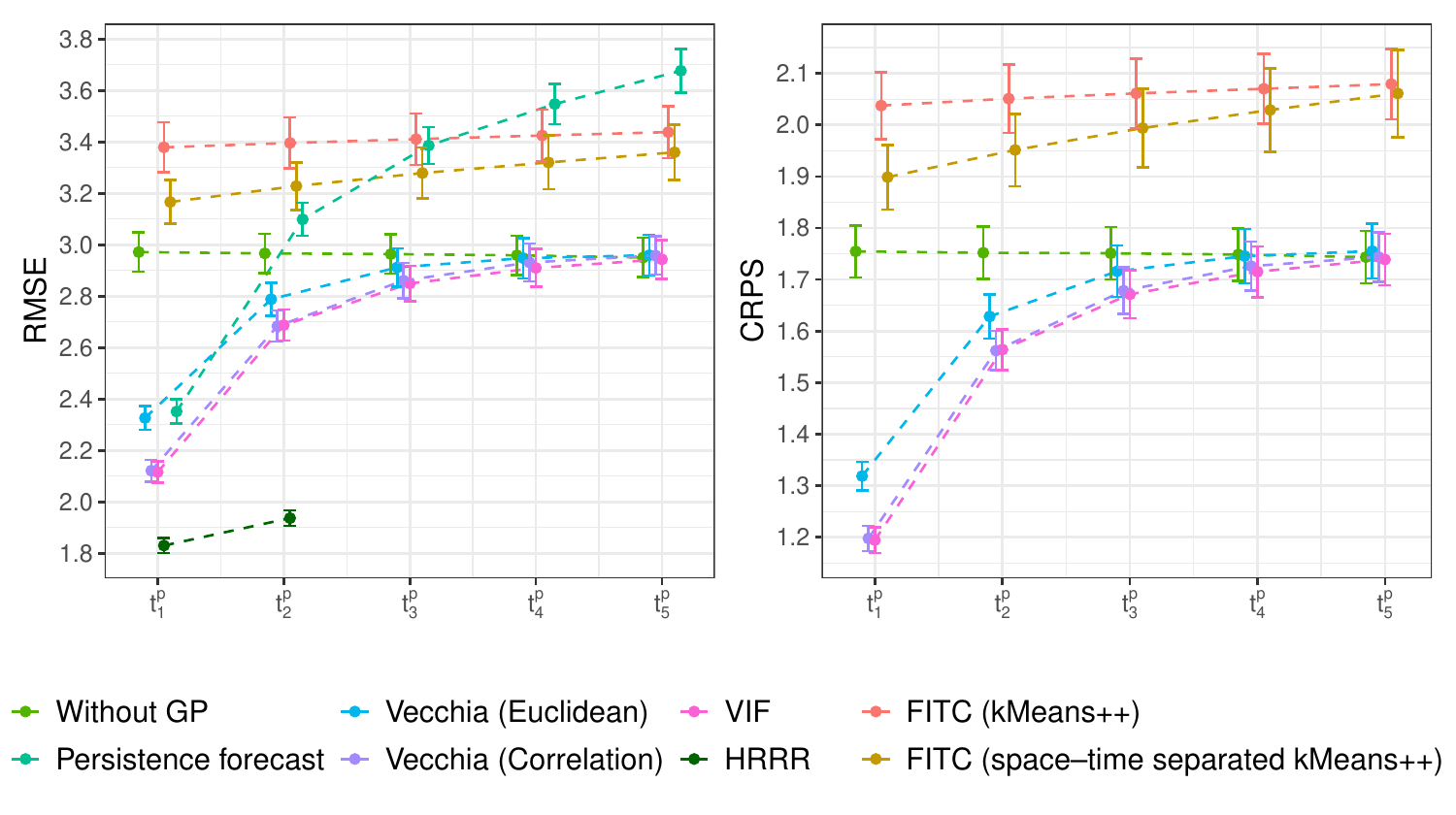}
    \caption{Temperature modeling results: average RMSE and CRPS $\pm$ one standard error vs. lead time (days) for one- to five-day spatio-temporal forecasts using the Vecchia, FITC, and VIF approximations. Persistence and fixed-effects-only baselines, as well as HRRR forecasts for the lead times where archived forecasts are available, are included for comparison.}
    \label{fig:RWR}
\end{figure}

Figure~\ref{fig:RWR} summarizes the forecast performance for spatio-temporal prediction using the RMSE and CRPS. The correlation-based Vecchia and VIF approximations yield the best overall performance among all GP methods and statistical baselines. In particular, a correlation-based neighbor selection improves the Vecchia approximation relative to Euclidean-distance-based neighbors. The FITC approximation is highly sensitive to the placement of the inducing points, and the proposed sts-kMeans++ strategy substantially enhances the FITC approximation's accuracy compared with standard kMeans++. The operational HRRR forecasts achieve the lowest overall RMSE at short lead times, where historical NWP forecast data are available. We note, however, that NWP models use much more data and computational resources compared to the statistical models considered in this article. Regarding the remaining benchmarks, both VIF and correlation-based Vecchia clearly outperform the persistence forecast and the fixed-effects linear model at every horizon. This demonstrates the importance of explicitly modeling space-time dependence even for very short-range forecasting, where relying solely on past observations or covariates is insufficient. 
%As an illustrative example, Figure~\ref{fig:RWRes} in Appendix~\ref{App:PandT} shows residuals and predictive variances from the first five-day forecast in August~2025 under the VIF approximation. Predictive uncertainty increases systematically with the forecast horizon, reflecting that longer lead times yield less information about the evolving  state. The widening intervals are consistent with the typical lifetime of synoptic-scale systems, indicating that uncertainty estimates arise from physically meaningful space-time dependence rather than ad-hoc adjustments.

\ifdefined\ARXIV
  \begin{table}[ht!]
\footnotesize
 \centering
 %\begin{threeparttable}
 \begin{tabular}{ |p{2.5cm}|p{1.5cm}||p{1.6cm}|p{2.cm}|p{2.cm}|p{2.cm}|p{1.cm}|}
  \hline
   \multicolumn{2}{|c||}{}&\makecell[c]{\textbf{Vecchia}\\Euclidean}& \makecell[c]{\textbf{Vecchia}\\Correlation}& \makecell[c]{\textbf{FITC}\\kMeans++}& \makecell[c]{\textbf{FITC}\\Space-time\\kMeans++}& \makecell[c]{\textbf{VIF}}\\
  \hline 
  \makecell[l]{Temporal\\forecasting} & \makecell[l]{RMSE\\CRPS} & \makecell[l]{2.808\\1.626}& \makecell[l]{2.749\\1.578}& \makecell[l]{3.422\\2.046}& \makecell[l]{3.292\\1.978}& \makecell[l]{2.737\\1.573}\\
  \hline 
  \makecell[l]{Spatio-temporal\\forecasting}& \makecell[l]{RMSE\\CRPS}& \makecell[l]{2.829\\1.633}& \makecell[l]{2.759\\1.581}& \makecell[l]{3.448\\2.060}& \makecell[l]{3.311\\1.987}& \makecell[l]{2.749\\1.577}\\
  %\hline 
  %\makecell[l]{Combined \\ forecasting} & \makecell[l]{RMSE\\CRPS}& \makecell[l]{2.813\\1.628}& \makecell[l]{2.751\\1.579}& \makecell[l]{3.428\\2.050}& \makecell[l]{3.297\\1.980}& \makecell[l]{2.740\\1.574}\\
  \Xhline{3\arrayrulewidth}
Runtime & \makecell[l]{Train\\Prediction} & \makecell[l]{835 s\\33 s} & \makecell[l]{1541 s\\46 s} & \makecell[l]{1234 s\\20 s} & \makecell[l]{1292 s\\19 s} & \makecell[l]{6543 s\\113 s}\\
  \hline 
 \end{tabular}
 \caption{ Average five-day RMSE, CRPS, and average computational runtimes for temperature modeling. Training time corresponds to parameter estimation; prediction time corresponds to prediction of means and variances.}\label{ExtraIntro}
 %\end{threeparttable}
\end{table}
\else
  \begin{table}[ht!]
%\footnotesize
 \centering
 %\begin{threeparttable}
 \caption{ Average five-day RMSE, CRPS, and average computational runtimes for temperature modeling. The results distinguish between temporal forecasting and spatio-temporal forecasting. Average runtimes in seconds (s) are shown for parameter estimation (Train) and computation of predictive mean and variances (Prediction). The training datasets contain between approximately $800,000$ and $1.3$ million space-time points.}\label{ExtraIntro}
 
 \begin{tabular}{ |p{2.1cm}|p{1.5cm}||p{1.5cm}|p{1.7cm}|p{1.7cm}|p{1.8cm}|p{0.9cm}|}
  \hline
   \multicolumn{2}{|c||}{}&\makecell[c]{\textbf{Vecchia}\\Euclidean}& \makecell[c]{\textbf{Vecchia}\\Correlation}& \makecell[c]{\textbf{FITC}\\kMeans++}& \makecell[c]{\textbf{FITC}\\sts kMeans++}& \makecell[c]{\textbf{VIF}}\\
  \hline 
  \makecell[l]{Temporal\\forecasting} & \makecell[l]{RMSE\\CRPS} & \makecell[l]{2.808\\1.626}& \makecell[l]{2.749\\1.578}& \makecell[l]{3.422\\2.046}& \makecell[l]{3.292\\1.978}& \makecell[l]{2.737\\1.573}\\
  \hline 
  \makecell[l]{Spatio-temporal\\forecasting}& \makecell[l]{RMSE\\CRPS}& \makecell[l]{2.829\\1.633}& \makecell[l]{2.759\\1.581}& \makecell[l]{3.448\\2.060}& \makecell[l]{3.311\\1.987}& \makecell[l]{2.749\\1.577}\\
  %\hline 
  %\makecell[l]{Combined \\ forecasting} & \makecell[l]{RMSE\\CRPS}& \makecell[l]{2.813\\1.628}& \makecell[l]{2.751\\1.579}& \makecell[l]{3.428\\2.050}& \makecell[l]{3.297\\1.980}& \makecell[l]{2.740\\1.574}\\
  \Xhline{3\arrayrulewidth}
Runtime & \makecell[l]{Train\\Prediction} & \makecell[l]{835 s\\33 s} & \makecell[l]{1541 s\\46 s} & \makecell[l]{1234 s\\20 s} & \makecell[l]{1292 s\\19 s} & \makecell[l]{6543 s\\113 s}\\
  \hline 
 \end{tabular}
%\end{threeparttable}
\end{table}
\fi

Table~\ref{ExtraIntro} reports the average RMSE and CRPS of the approximated GP models across the full five-day forecast horizon for both spatio-temporal forecasting and temporal-only predictions at stations with observed historical data, together with average computational runtimes. Overall, the predictive performance in the spatio-temporal and temporal-only settings is qualitatively similar. The VIF approximation provides the highest prediction accuracy, albeit at a higher training cost than the Vecchia methods. The FITC approximation is computationally inexpensive but substantially less accurate. Across all scalable approximations, forecast generation is inexpensive: computing predictive means and variances requires only a few tens of seconds. Thus, the dominant computational expense arises from parameter estimation rather than daily forecasting. Even for the most accurate method (VIF), a full re-estimation takes on the order of one hour, while the Vecchia variants complete training within minutes. These runtimes remain practical even without GPU acceleration. Even on standard CPU hardware, monthly parameter re-estimation can be performed offline (e.g., overnight) after which operational 5-day forecasts can be issued almost instantly. Consequently, scalable GP models can be deployed in routine forecasting workflows without specialized hardware, while still delivering high-accuracy, uncertainty-aware predictions at continental scale.

\begin{figure}[ht!]
    \centering
    \ifdefined\ARXIV
  \includegraphics[width=\linewidth]{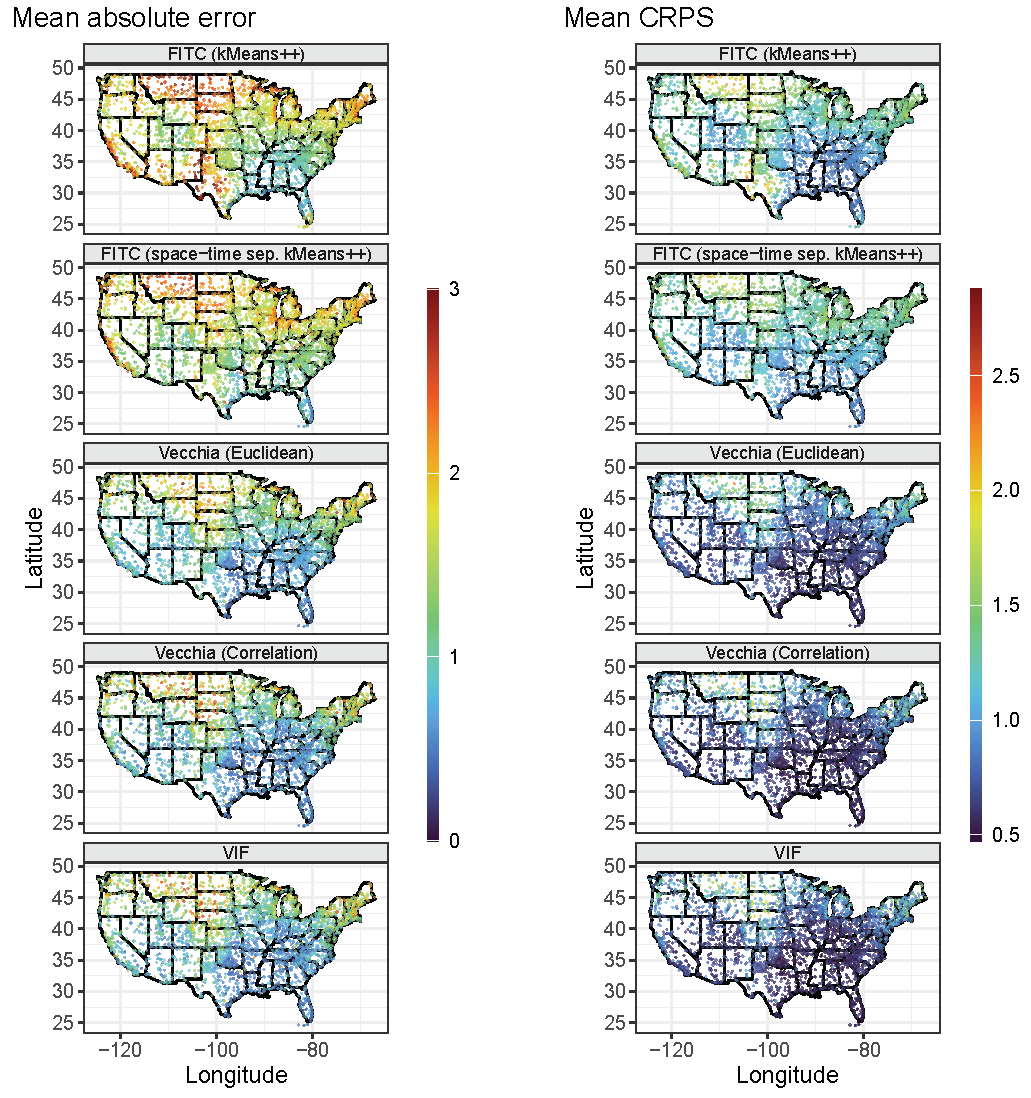}
\else
  \includegraphics[width=\linewidth]{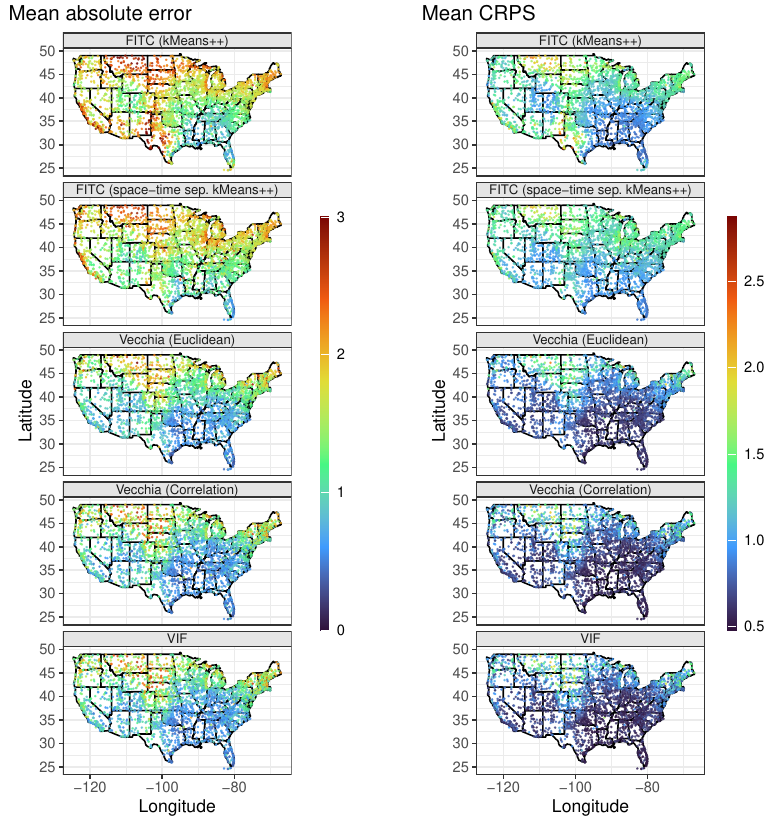}
\fi
    \caption{Mean absolute error and mean CRPS of one-day-ahead forecasts for August 2025, computed per station for the Vecchia, FITC, and VIF approximations.}
    \label{fig:RWRA}
\end{figure}

 Furthermore, spatial variation in the station-wise averaged one-day-ahead forecast accuracy is shown in Figure \ref{fig:RWRA} for August 2025. The FITC approximations, particularly with standard kMeans++ inducing points, perform uniformly worse than the other methods across all regions. In contrast, differences between the VIF approximation and the Vecchia methods (with either Euclidean or correlation-based neighbor selection) are visually small. For the VIF and Vecchia approaches, errors are lowest throughout the Southeast, where summer temperature fields vary smoothly and station density is high. Higher errors occur in regions with strong synoptic variability or complex terrain, including the northern Rockies, the Upper Midwest, and portions of the Northeast, where sharp frontal transitions and local land-atmosphere interactions lead to more abrupt temperature changes. 

Table~\ref{covparsT} reports the monthly estimated covariance parameters by the VIF model. The spatial and temporal effective ranges remain remarkably stable across months, indicating that short-lead temperature forecasts are governed by persistent large-scale dependence rather than rapid structural changes. Such parameter stability is desirable in practice, as it implies that models need not be re-tuned frequently to maintain forecasting skill. The estimates show a spatial effective range of roughly 1,400~km, reflecting broad continental-scale coherence in daily maximum temperatures, and a temporal range of about seven days, consistent with the lifespan of synoptic systems. The non-separability parameter ($\gamma \approx 1$) suggests strong space-time interaction, in line with slowly evolving air-mass advection that shapes temperature dynamics across much of the United States. 
%\begin{table}[ht!]
%\footnotesize
 %\centering
 %\begin{threeparttable}
 %\begin{tabular}{ |p{0.9cm}|p{.9cm}|p{.9cm}|p{0.9cm}|p{.9cm}|p{.9cm}|p{.9cm}|p{2.cm}|p{2.cm}|}
  %\hline
  %{$\sigma^2$}&{$\sigma_1^2$} &{$a$} &{$c\cdot 10^5$} & {$\alpha$}& {$\gamma$}  & {$\delta$} & \textbf{Time range} & \textbf{Space range}\\
  %\hline 
%1.539 & 6.193 & 0.090 & 0.337 & 0.831 &  0.999 & 1.667 & 6.7 days & 1,408 km\\
  %\hline 
 %\end{tabular}
 %\caption{\label{covparsT01} VIF covariance estimates (July); ranges defined by correlation decay to 0.05.}
 %\end{threeparttable}
%\end{table}

\ifdefined\ARXIV
  \begin{table}[ht!]
 \centering
 \begin{threeparttable}
 \begin{tabular}{ |p{1.2cm}||p{0.9cm}|p{0.9cm}|p{0.9cm}|p{1.3cm}|p{0.9cm}|p{0.9cm}|p{0.9cm}|p{1.3cm}|p{1.5cm}|}
  \hline
  Date&{$\sigma^2$}&{$\sigma_1^2$} &{$a$} &{$c$} & {$\alpha$}& {$\gamma$}  & {$\delta$} & \textbf{Time range} & \textbf{Space range}\\
  \hline 
Dec 24&1.379 & 5.497 & 0.087 & 3.47e-06 & 0.842 &  0.998 & 1.643 & 6.6 days & 1,367 km\\
\hline
Jan 25&1.387 & 5.635 & 0.087 & 3.47e-06 & 0.840 &  0.999 & 1.647 & 6.7 days & 1,367 km\\
\hline
Feb 25&1.414 & 5.775 & 0.087 & 3.48e-06 & 0.839 &  0.998 & 1.650 & 6.7 days & 1,367 km\\
\hline
Mar 25&1.464 & 6.058 & 0.088 & 3.50e-06 & 0.831 &  0.999 & 1.656 & 6.7 days & 1,355 km\\
\hline
Apr 25&1.540 & 6.307 & 0.089 & 3.39e-06 & 0.823 &  0.998 & 1.662 & 6.8 days & 1,399 km\\
\hline
May 25&1.573 & 6.407 & 0.090 & 3.35e-06 & 0.825 &  0.997 & 1.668 & 6.7 days & 1,416 km\\
\hline
Jun 25&1.561 & 6.366 & 0.089 & 3.37e-06 & 0.826 &  0.998 & 1.671 & 6.7 days & 1,408 km\\
\hline
 Jul 25& 1.539 & 6.193 & 0.090 & 3.37e-06 & 0.831 &  0.999 & 1.667 & 6.7 days & 1,408 km\\
  \hline 
 \end{tabular}
 \caption{\label{covparsT} Estimated covariance parameters for the VIF model at each re-estimation, with effective ranges reported as the time or distance at which correlation decays to 0.05.}
 \end{threeparttable}
\end{table}
\else
  \begin{table}[ht!]
 \centering
 \begin{threeparttable}
 \caption{\label{covparsT} Estimated covariance parameters for the VIF model at each re-estimation, with effective ranges reported as the time or distance at which correlation decays to 0.05.}
 \begin{tabular}{ |p{1.cm}||p{0.8cm}|p{0.8cm}|p{0.8cm}|p{0.8cm}|p{0.8cm}|p{0.8cm}|p{0.7cm}|p{1.6cm}|p{1.7cm}|}
  \hline
  Date&{$\sigma^2$}&{$\sigma_1^2$} &{$a$} &{$c\cdot 10^5$} & {$\alpha$}& {$\gamma$}  & {$\delta$} & \textbf{Time range} & \textbf{Space range}\\
  \hline 
Dec 24&1.379 & 5.497 & 0.087 & 0.347 & 0.842 &  0.998 & 1.643 & 6.6 days & 1,367 km\\
\hline
Jan 25&1.387 & 5.635 & 0.087 & 0.347 & 0.840 &  0.999 & 1.647 & 6.7 days & 1,367 km\\
\hline
Feb 25&1.414 & 5.775 & 0.087 & 0.348 & 0.839 &  0.998 & 1.650 & 6.7 days & 1,367 km\\
\hline
Mar 25&1.464 & 6.058 & 0.088 & 0.350 & 0.831 &  0.999 & 1.656 & 6.7 days & 1,355 km\\
\hline
Apr 25&1.540 & 6.307 & 0.089 & 0.339 & 0.823 &  0.998 & 1.662 & 6.8 days & 1,399 km\\
\hline
May 25&1.573 & 6.407 & 0.090 & 0.335 & 0.825 &  0.997 & 1.668 & 6.7 days & 1,416 km\\
\hline
Jun 25&1.561 & 6.366 & 0.089 & 0.337 & 0.826 &  0.998 & 1.671 & 6.7 days & 1,408 km\\
\hline
 Jul 25& 1.539 & 6.193 & 0.090 & 0.337 & 0.831 &  0.999 & 1.667 & 6.7 days & 1,408 km\\
  \hline 
 \end{tabular}
 
 \end{threeparttable}
\end{table}
\fi

%The QQ-plot~\ref{fig:QQ} Appendix \ref{App:PandT} shows that standardized residuals from the VIF approximation are close to Gaussian over the central quantiles, indicating that the model adequately captures day-to-day temperature variability. Departures emerge mainly in the upper and lower tails, corresponding to rare synoptic extremes that evolve more abruptly than the proposed model can represent. These deviations are modest and confined to infrequent events, suggesting that a Gaussian observation model remains adequate for short-lead operational forecasting. Although heavier-tailed alternatives such as Student-t likelihoods could, in principle, better accommodate extreme departures, the magnitude of the observed tail behavior is unlikely to materially affect predictive performance.

To assess potential spatial non-stationarity in the covariance structure, we re-estimated the VIF model at the final recalibration date on three broad geographic subsets (West, Central, East). The resulting parameter estimates for the Central and Eastern subsets are very similar to those obtained from the full dataset (results not tabulated). Spatial effective ranges varied only modestly, from about $1,477$ km to $1,597$ km, slightly bigger than the full-domain estimate, while temporal effective ranges remained between $8.6$ and $11$ days. The estimated non-separability parameter was also close to $\gamma \approx 1$ in both regions. In contrast, the Western subset exhibited a noticeably larger spatial effective range together with a reduced non-separability parameter, reflecting the broader-scale temperature coherence and weaker space-time coupling over the western United States. Overall, stationarity appears adequate at the continental scale, with only moderate regional deviations. %An important direction for future work is the development of nonstationary or covariate-dependent space-time covariance functions that explicitly accommodate such regional differences.

%Because spatial coordinates are expressed in meters, the spatial range parameter is numerically much smaller than the other covariance parameters, which can hinder numerical optimization by creating poorly scaled gradients. To check robustness, we re-estimated the VIF-approximated model after rescaling locations to kilometers. As expected for stationary covariances, this changes only the spatial range parameter by a factor of approximately 1,000, while all other parameters and predictive performance remain essentially unchanged.

\subsection{Short-term prediction of precipitation}\label{sect502}

For precipitation, we use the model described in Section \ref{sect.nonG} and the covariance in \eqref{Gneiting} with the Matérn smoothness fixed to $\nu = 0.5$. This corresponds to an exponential covariance, producing a rougher latent process that more realistically captures the sharp spatial and temporal variability characteristic of rainfall. Estimating the smoothness parameter using only CPU on subsets of the data gives $\nu \approx 0.5$. In addition, experiments on a subset of the data showed that the predictive accuracy is not sensitive to the choice of $\nu$; see Section~\ref{sect503}. Daily precipitation exhibits a mixed distribution with a discrete component (occurrence) and a continuous component (amount).  To account for this, we use the zero-censored power-transformed normal (ZC-PTN) likelihood introduced in \eqref{Eq:Model2} in Section~\ref{sect.nonG}. A Laplace approximation is used for parameter estimation and prediction. To reduce the computational complexity, we use the iterative methods of \citet{kundig2025iterative} and \citet{gyger2025vecchia}, which replace explicit matrix factorizations with iterative methods that only require matrix-vector operations.
To evaluate the predictive accuracy, we use separate metrics for occurrences and intensities. Because skill in precipitation forecasting deteriorates more rapidly with lead time than for temperature, we focus on one- to three-day-ahead forecasts, an operationally relevant horizon at which meaningful predictability can still be obtained. We also examined up to five-day-ahead forecasts; however, accuracy remained similar or only slightly degraded. For rainfall occurrence, we use the Brier score (BS) and the log score (LS), which assess both sharpness and calibration of the implied occurrence probabilities. Conditional on rainfall occurrence, positive-amount predictions are evaluated using the mean absolute error (MAE), which is robust to the heavy right tail of precipitation amounts. To assess probabilistic predictions, we additionally consider the CRPS, computed by sampling from the ZC-PTN predictive distributions. For comparison, we include a fixed-effects linear regression model without a GP and a simple persistence benchmark based on the previous day’s precipitation occurrence and amount. We additionally compare against the operational NOAA HRRR model. As for temperature, forecasts are initialized once daily and extracted from the nearest HRRR grid cell at each station location. Daily precipitation forecasts are obtained by summing the hourly accumulated precipitation forecasts.  

\begin{figure}[ht!]
    \centering
    \includegraphics[scale=0.5]{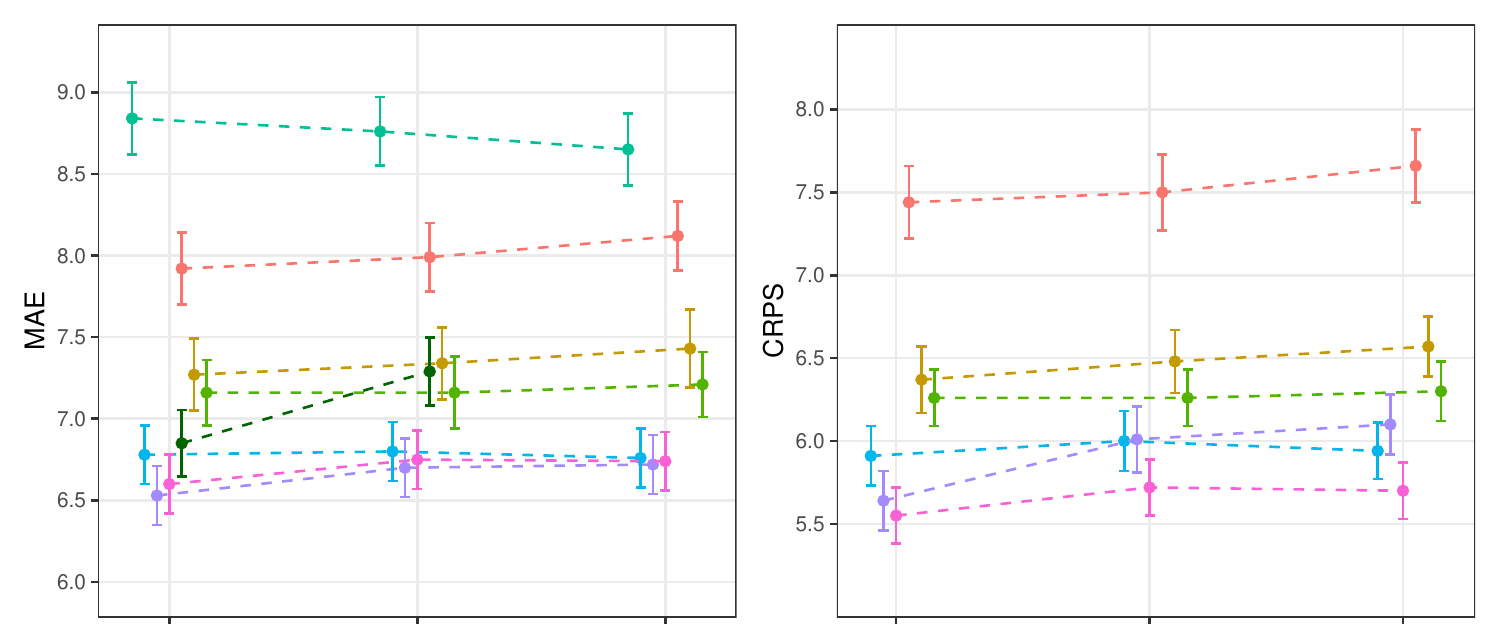}
    \includegraphics[scale=0.5]{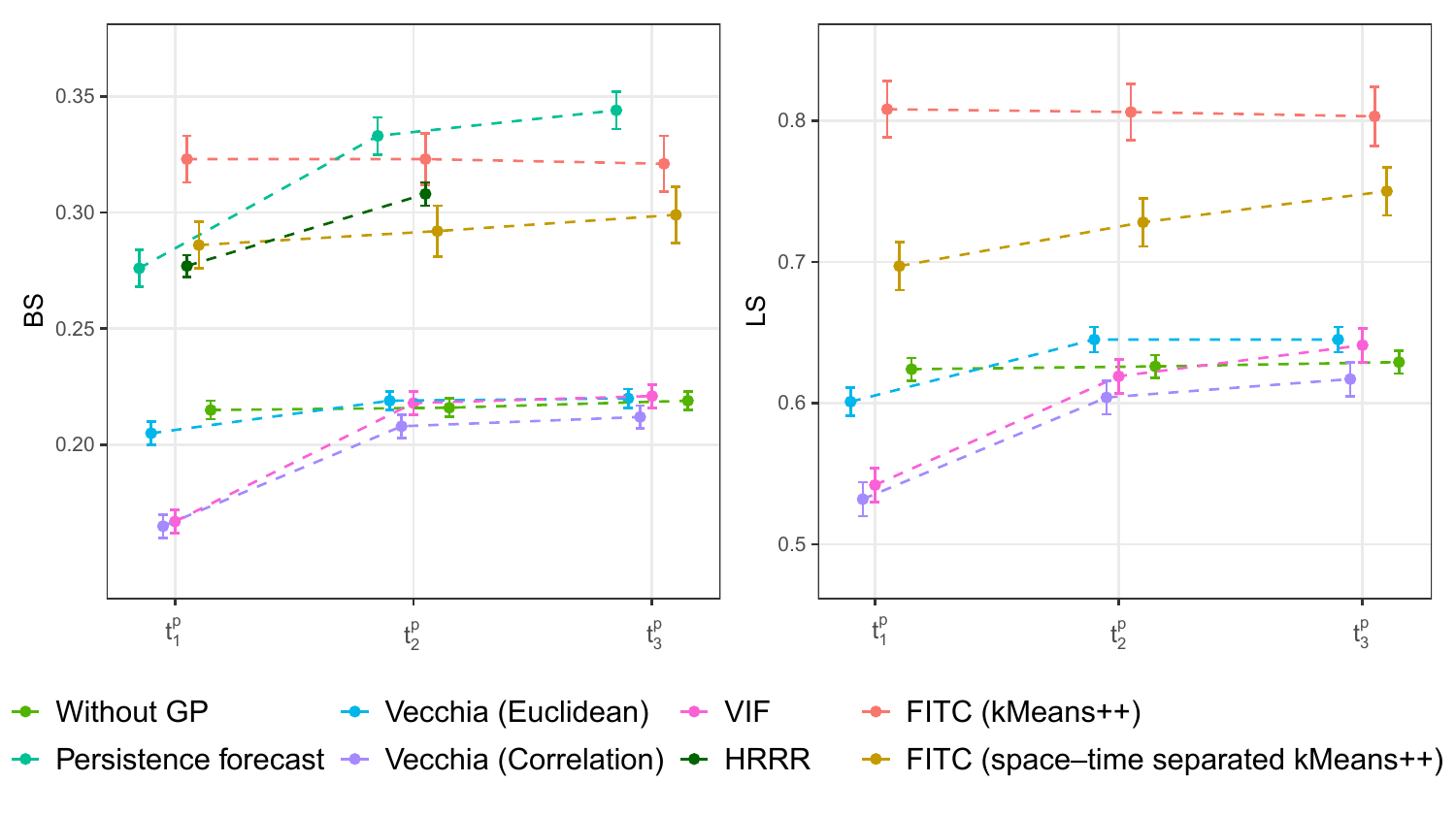}
    \caption{Precipitation modeling results: Average MAE and CRPS for precipitation amounts and Brier score (BS) and log score (LS) for precipitation occurrence $\pm$ one standard error vs. lead time (days) for one- to three-day spatio-temporal forecasts using the Vecchia, FITC, and VIF approximations.  Persistence and fixed-effects-only baselines, as well as HRRR forecasts for the lead times where archived forecasts are available, are included for comparison.}
    \label{fig:RWRnG01}
\end{figure}

Figure~\ref{fig:RWRnG01} reports the accuracy measures for precipitation amounts (MAE, CRPS) and occurrences (BS, LS) versus lead times. Concerning precipitation amounts, we first observe that the correlation-based neighbor selection improves the Vecchia approximation for one-day ahead forecasts. As in the temperature application, the FITC approximation is sensitive to the inducing-point placement, with the proposed sts-kMeans++ strategy substantially outperforming standard kMeans++. The VIF approximation achieves the lowest CRPS across all horizons and all metrics. Moreover, the persistence forecast performs consistently inferior to the GP approximations, underscoring both the value of relevant covariates and the importance of explicitly modeling space-time dependence in short-term precipitation forecasting. Concerning the fixed-effects linear regression model without a GP component, both the VIF and Vecchia approximations yield superior performance. In contrast to the temperature application, the operational HRRR forecasts yield less accurate forecasts than the best the GP-based approaches. A likely reason is that precipitation exhibits substantially stronger local variability and intermittency than temperature, making point-level station observations more difficult to represent accurately using precipitation values from the nearest HRRR grid cell. 

For precipitation occurrence, we find that a Vecchia approximation with correlation-based neighbor selection consistently outperforms its Euclidean-based counterpart. The improvements are most pronounced at short horizons, where the correlation-based Vecchia and VIF approximations yield the lowest BS and LS. Within the FITC family, the choice of inducing-point placement has a modest effect, although the proposed sts-kMeans++ strategy offers better performance than standard kMeans++. Nonetheless, both FITC variants remain inferior to the other GP approximations, reflecting their difficulty in capturing highly localized, rapidly evolving precipitation processes. While persistence forecasts perform reasonably well at short lead times, the best GP-based models consistently achieve superior predictive accuracy. Moreover, for one-day-ahead forecasts, both the VIF and correlation-based Vecchia approximations clearly outperform the fixed-effects linear regression model without a GP component. 

Table~\ref{ExtraIntro2} reports the average MAE, CRPS, BS, and LS across the full three-day forecast horizon for both spatio-temporal forecasting and temporal-only predictions at stations with observed historical data, together with average computational runtimes. Similar to the temperature forecasting results, the performance differences between temporal and spatio-temporal forecasting remain relatively small for rainfall occurrence, indicating that the GP models successfully transfer information from monitored to unmonitored stations with only limited loss in predictive accuracy. Although the gap between temporal and spatio-temporal forecasting is somewhat larger for precipitation than for temperature, it nevertheless remains modest despite the inherently localized and rapidly evolving nature of rainfall. In particular, convective storms, orographic lifting, and frontal passages would generally be expected to make prediction at unobserved stations substantially more difficult. The comparatively small deterioration in predictive skill suggests that the GP models capture relevant fine-scale spatial dependencies and, together with the covariates, effectively propagate information to new locations through the joint space-time covariance structure.

Compared with the Gaussian likelihood used for temperature modeling, all methods exhibit higher runtimes under the non-Gaussian ZC-PTN likelihood. Nevertheless, the methods remain computationally feasible even in the most demanding setting: the most accurate model (VIF) requires only a few hours for parameter estimation, whereas the computationally cheaper Vecchia approximations require on the order of tens of minutes. As in the Gaussian case, prediction itself remains inexpensive, requiring only tens of seconds, implying that the computational burden is dominated by model fitting rather than forecasting.

\ifdefined\ARXIV
  \begin{table}[ht!]
\footnotesize
 \centering
 \begin{threeparttable}
 \begin{tabular}{ |p{2.2cm}|p{1.4cm}||p{1.5cm}|p{1.8cm}|p{2.cm}|p{2.cm}|p{1.3cm}|}
  \hline
   \multicolumn{2}{|c||}{}&\makecell[c]{\textbf{Vecchia} \\ Euclidean}& \makecell[c]{\textbf{Vecchia} \\ Correlation}& \makecell[c]{\textbf{FITC} \\ kMeans++ }& \makecell[c]{\textbf{FITC} \\ Space-time \\kMeans++ }& \makecell[c]{\textbf{VIF} } \\
  \hline 
  \makecell[l]{Temporal \\ forecasting}& \makecell[l]{MAE \\ CRPS \\ BS \\ LS} & \makecell[l]{6.634 \\ 5.780 \\ 0.217 \\ 0.630}& \makecell[l]{6.551 \\ 5.790 \\ 0.197 \\ 0.579}& \makecell[l]{7.941 \\7.412 \\ 0.324 \\ 0.806}& \makecell[l]{7.280 \\6.404 \\ 0.291 \\ 0.719}& \makecell[l]{6.571 \\5.560 \\ 0.207 \\ 0.600}\\
  \hline 
  \makecell[l]{Spatio-temporal \\ forecasting}& \makecell[l]{MAE \\ CRPS \\ BS \\ LS}& \makecell[l]{6.801 \\ 5.942 \\ 0.218 \\ 0.631}& \makecell[l]{6.664 \\ 5.901 \\ 0.198 \\ 0.581}& \makecell[l]{8.014 \\ 7.533 \\ 0.324 \\ 0.817}& \makecell[l]{7.341 \\ 6.474 \\ 0.293 \\0.724} & \makecell[l]{6.694 \\5.653 \\ 0.207 \\ 0.601}\\
  %\hline 
  %\makecell[l]{Combined \\forecasting}& \makecell[l]{MAE \\ CRPS \\ BS \\ LS}& \makecell[l]{6.761 \\ 5.912 \\ 0.217 \\ 0.630}& \makecell[l]{6.659 \\ 5.880 \\ 0.197 \\ 0.579} & \makecell[l]{7.974 \\ 7.510 \\ 0.324 \\ 0.812}& \makecell[l]{7.334 \\ 6.451 \\ 0.291 \\0.720} & \makecell[l]{6.670 \\ 5.634 \\ 0.207 \\ 0.600}\\
  \Xhline{3\arrayrulewidth}
Runtime& \makecell[l]{Train \\ Prediction} & \makecell[l]{3,321 s \\ 201 s}  &\makecell[l]{5,618 s \\ 278 s} & \makecell[l]{2,201 s \\ 28 s} &\makecell[l]{2,187 s \\ 29 s} &\makecell[l]{10,728 s \\ 413 s}\\
  \hline 
 \end{tabular}
 \caption{\label{ExtraIntro2} Average three-day MAE, CRPS, BS, and LS, and average computational runtimes for precipitation modeling. The results distinguish between temporal forecasting and spatio-temporal forecasting. Average runtimes in seconds (s) are shown for parameter estimation (Train) and computation of predictive mean and variances (Prediction). The training datasets contain between approximately $800,000$ and $1.3$ million space-time points.}
     \end{threeparttable}
 \end{table}
\else
  \begin{table}[ht!]
\footnotesize
 \centering
 \begin{threeparttable}
 \caption{\label{ExtraIntro2} Average three-day MAE, CRPS, BS, and LS, and average computational runtimes for precipitation modeling. The results distinguish between temporal forecasting and spatio-temporal forecasting. Average runtimes in seconds (s) are shown for parameter estimation (Train) and computation of predictive mean and variances (Prediction). The training datasets contain between approximately $800,000$ and $1.3$ million space-time points.}
 
 \begin{tabular}{ |p{2.cm}|p{1.5cm}||p{1.6cm}|p{1.8cm}|p{1.8cm}|p{1.8cm}|p{1.3cm}|}
  \hline
   \multicolumn{2}{|c||}{}&\makecell[c]{\textbf{Vecchia} \\ Euclidean}& \makecell[c]{\textbf{Vecchia} \\ Correlation}& \makecell[c]{\textbf{FITC} \\ kMeans++ }& \makecell[c]{\textbf{FITC} \\ sts kMeans++ }& \makecell[c]{\textbf{VIF} } \\
  \hline 
  \makecell[l]{Temporal \\ forecasting}& \makecell[l]{MAE \\ CRPS \\ BS \\ LS} & \makecell[l]{6.634 \\ 5.780 \\ 0.217 \\ 0.630}& \makecell[l]{6.551 \\ 5.790 \\ 0.197 \\ 0.579}& \makecell[l]{7.941 \\7.412 \\ 0.324 \\ 0.806}& \makecell[l]{7.280 \\6.404 \\ 0.291 \\ 0.719}& \makecell[l]{6.571 \\5.560 \\ 0.207 \\ 0.600}\\
  \hline 
  \makecell[l]{Spatio-temporal \\ forecasting}& \makecell[l]{MAE \\ CRPS \\ BS \\ LS}& \makecell[l]{6.801 \\ 5.942 \\ 0.218 \\ 0.631}& \makecell[l]{6.664 \\ 5.901 \\ 0.198 \\ 0.581}& \makecell[l]{8.014 \\ 7.533 \\ 0.324 \\ 0.817}& \makecell[l]{7.341 \\ 6.474 \\ 0.293 \\0.724} & \makecell[l]{6.694 \\5.653 \\ 0.207 \\ 0.601}\\
  %\hline 
  %\makecell[l]{Combined \\forecasting}& \makecell[l]{MAE \\ CRPS \\ BS \\ LS}& \makecell[l]{6.761 \\ 5.912 \\ 0.217 \\ 0.630}& \makecell[l]{6.659 \\ 5.880 \\ 0.197 \\ 0.579} & \makecell[l]{7.974 \\ 7.510 \\ 0.324 \\ 0.812}& \makecell[l]{7.334 \\ 6.451 \\ 0.291 \\0.720} & \makecell[l]{6.670 \\ 5.634 \\ 0.207 \\ 0.600}\\
  \Xhline{3\arrayrulewidth}
Runtime& \makecell[l]{Train \\ Prediction} & \makecell[l]{3,321 s \\ 201 s}  &\makecell[l]{5,618 s \\ 278 s} & \makecell[l]{2,201 s \\ 28 s} &\makecell[l]{2,187 s \\ 29 s} &\makecell[l]{10,728 s \\ 413 s}\\
  \hline 
 \end{tabular}
     \end{threeparttable}
 \end{table}
\fi

\begin{figure}[ht!]
    \centering
    \ifdefined\ARXIV
  \includegraphics[width=\linewidth]{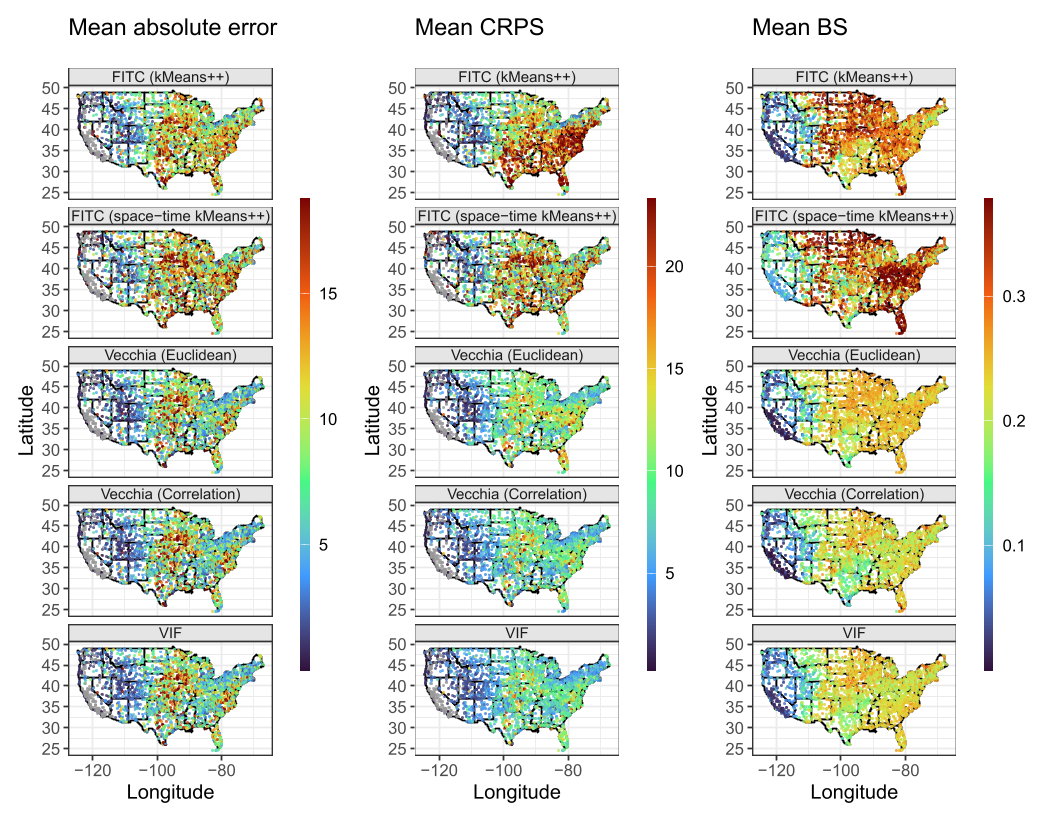}
\else
  \includegraphics[width=\linewidth]{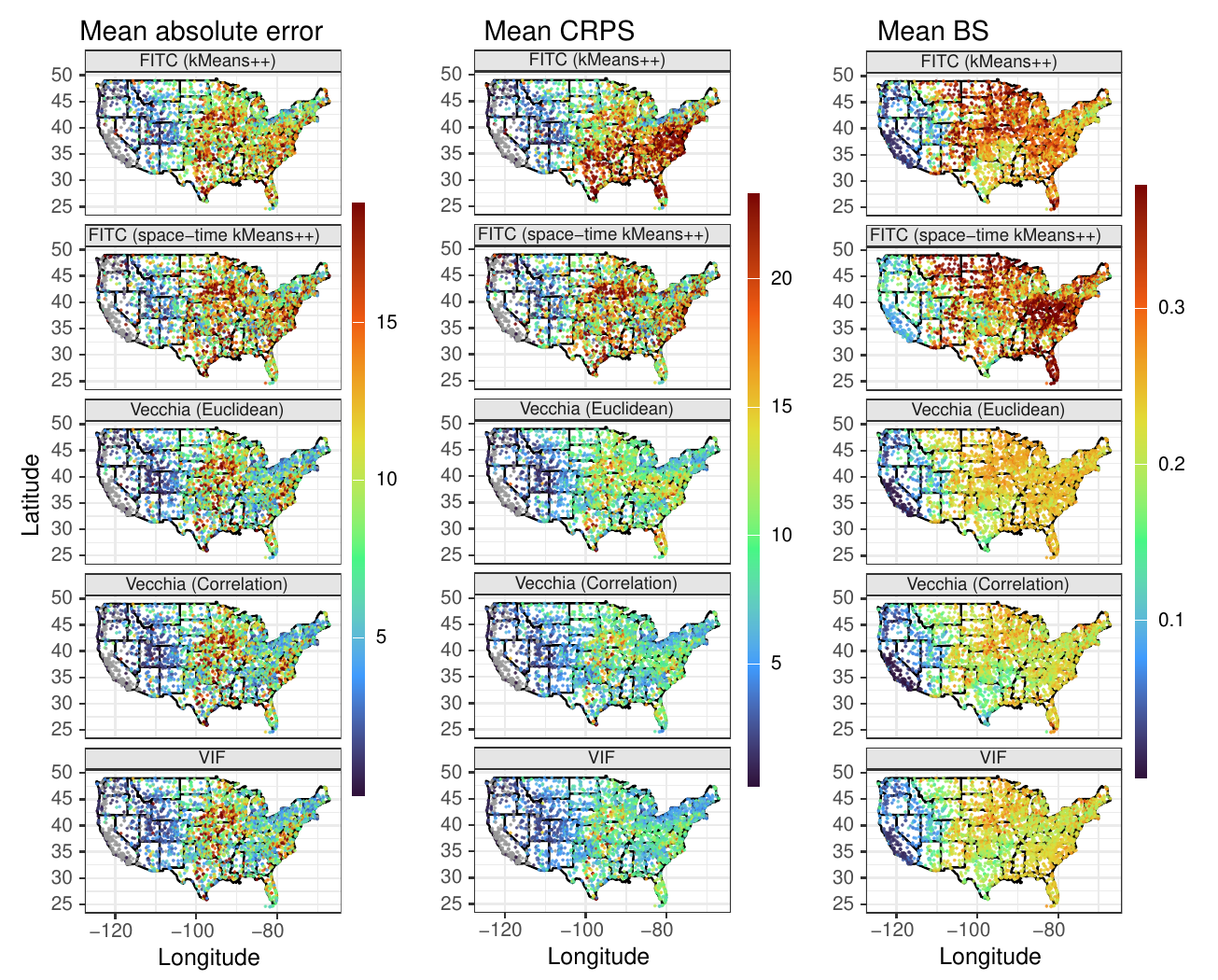}
\fi
    \caption{Mean absolute error, mean CRPS, and mean BS of one-day-ahead forecasts for August 2025, computed per station for the Vecchia, FITC, and VIF approximations.
}
    \label{fig:RWRA01}
\end{figure}

Spatial variability in the averaged station-wise one-day-ahead forecast accuracy is shown in Figure \ref{fig:RWRA01} for August 2025. The benefits of correlation-based Vecchia neighbor selection are most evident in the CRPS and BS fields, where it yields lower error patterns than Euclidean neighbors, particularly in the middle and eastern parts of the country. %Differences in mean absolute error are comparatively small as the MAE ignores uncertainty quantification. %%By contrast, the FITC approximation shows substantial sensitivity to the inducing-point selection strategy, particularly for CRPS and BS, and remains clearly inferior to the other methods across most regions.
For the VIF and Vecchia approaches, forecast errors generally increase from west to east. Much of the western United States experiences dry, spatially smooth summer conditions, which are easier to model and, therefore, yield lower uncertainty. In contrast, the eastern half of the country, particularly the Southeast and the Mid-Atlantic, exhibits frequent, localized rainfall driven by convective storms, tropical moisture intrusions, and frontal boundaries, making both occurrence and intensity harder to predict. Higher errors also appear across the Central and Upper Midwest (including states such as Iowa, Minnesota, and Missouri). This region sits along common summertime storm tracks and is strongly influenced by mesoscale convective systems that can produce large rainfall gradients over short distances. These systems evolve rapidly overnight and propagate long distances, introducing substantial spatial and temporal variability that challenges short-term prediction, especially at unobserved stations. 

Table~\ref{covparsT101} reports the monthly estimated covariance parameters by the VIF model. Across all monthly re-estimations, the parameters remain stable, indicating that the large-scale space-time dependence structure of precipitation is persistent over the year, with only modest seasonal adjustments. Between January and July 2025, the temporal effective range increases slightly (3.3 to 3.7 days), while the spatial range decreases (3,300 km to 2,770 km), reflecting the transition from winter’s broad synoptic precipitation to more localized warm-season convection. The non-separability parameter increases only marginally (0.26 to 0.265), implying that the overall strength of space-time interaction remains essentially unchanged. In comparison with the temperature application, the precipitation model exhibits both a shorter temporal range and weaker non-separability, underscoring the fundamentally faster and more localized dynamics of precipitation. %Despite these differences, the smooth month-to-month evolution of all parameters demonstrates that the monthly re-estimation strategy effectively tracks gradual seasonal changes while avoiding overfitting to short-lived weather events. 
The power-transformation parameter $\lambda$ also increases modestly, consistent with heavier-tailed rainfall amounts during the warm season.
% in the dependence structure at the temporal resolution considered, and indicate that a single global covariance model provides an adequate description of the data.

\begin{table}[ht!]
%\footnotesize
 \centering
 \begin{threeparttable}
 \caption{\label{covparsT101} Estimated auxiliary and covariance parameters for the VIF model at each re-estimation, with effective ranges reported as the time or distance at which correlation decays to 0.05 .}
 \begin{tabular}{ |p{1.3cm}||p{0.7cm}|p{0.7cm}|p{0.9cm}|p{.7cm}|p{0.8cm}|p{.7cm}|p{.7cm}|p{.7cm}|p{1.4cm}|p{1.2cm}|}
\hline Date & $\sigma$ & $\lambda$ & $\sigma_1^2$ & $a$ & $c\cdot 10^5$ & $\alpha$ & $\gamma$ & $\delta$ & \textbf{Time range} & \textbf{Space range} \\
\hline Dec 24 & 3.111 & 1.415 & 16.212 & 0.232 & 0.099 & 0.905 & 0.260 & 2.489 & 3.3 days & $3,263 \mathrm{~km}$ \\
\hline Jan 25 & 3.105 & 1.422 & 15.450 & 0.232 & 0.100 & 0.889 & 0.261 & 2.484 & 3.3 days & $3,338 \mathrm{~km}$ \\
\hline Feb 25 & 3.101 & 1.418 & 16.234 & 0.231 & 0.100 & 0.885 & 0.262 & 2.436 & 3.4 days & $3,419 \mathrm{~km}$ \\
\hline Mar 25 & 3.118 & 1.421 & 12.471 & 0.231 & 0.101 & 0.879 & 0.267 & 2.408 & 3.5 days & $2,966 \mathrm{~km}$ \\
\hline Apr 25 & 3.229 & 1.429 & 16.465 & 0.230 & 0.107 & 0.872 & 0.263 & 2.385 & 3.6 days & $2,800 \mathrm{~km}$ \\
\hline May 25 & 3.234 & 1.453 & 18.197 & 0.230 & 0.107 & 0.867 & 0.264 & 2.381 & 3.6 days & $2,800 \mathrm{~km}$ \\
\hline Jun 25 & 3.276 & 1.464 & 18.863 & 0.230 & 0.107 & 0.863 & 0.266 & 2.361 & 3.6 days & $2,800 \mathrm{~km}$ \\
\hline Jul 25 & 3.415 & 1.508 & 18.286 & 0.230 & 0.108 & 0.860 & 0.265 & 2.353 & 3.7 days & $2,774 \mathrm{~km}$ \\
\hline
\end{tabular}

\end{threeparttable}
\end{table}

To further assess potential spatial non-stationarity in the covariance structure, we re-estimated the VIF model at the final recalibration date on three geographic subsets: West, Central, and East. The resulting parameter estimates (results not tabulated) were highly similar to those obtained from the full dataset. Spatial effective ranges varied only modestly - from about $2,463$ km to $2,627$ km, slightly smaller than the full-domain estimate - while temporal effective ranges remained between $3.6$ and $3.9$ days. The estimated non-separability parameter was stable across all subsets, with $\gamma \approx 0.26$ in each case. These findings suggest no evidence of large-scale spatial non-stationarity.

\subsection{Additional results}\label{sect503}

Appendix \ref{App:PandT} presents additional analyses for temperature and precipitation forecasting, including temporal-only prediction at observed stations, comparisons with a simpler Bernoulli GP model for rainfall occurrence, visualizations of estimated GP random effects, residual diagnostics, and calibration assessments. The results show that temporal and spatio-temporal forecasting perform similarly, with only a modest loss in predictive skill for precipitation at unobserved locations despite the localized nature of rainfall. Residual analyses indicate some underestimation for extreme precipitation events but no strong systematic bias overall. Calibration analyses further show that the VIF and correlation-based Vecchia approximations are substantially better calibrated than the FITC and Euclidean-based Vecchia approximations. Experiments with fixed smoothness parameters $\nu \in {0.5,1.5,2.5}$ further indicate that predictive accuracy is largely insensitive to the choice of $\nu$.

\section{Conclusion}

This study demonstrates that scalable GP models can deliver accurate short-term forecasts of temperature and precipitation across the continental United States while remaining computationally feasible at operational scale. Using a nonseparable space-time covariance, the models capture realistic weather dependence, with spatial correlation varying across time lags. Two methodological advances underpin this performance: a scalable correlation-based Vecchia neighbor selection method and the sts-kMeans++ method for selecting inducing points. Together, these ideas are integrated into the VIF approximation, which provides the strongest predictive accuracy and most reliable uncertainty quantification in both Gaussian and non-Gaussian settings, while Vecchia methods alone offer an attractive alternative when training cost is paramount. GPU acceleration further reduces computation time, enabling scaling to millions of observations.

The case study highlights several scientifically relevant findings. Temperature fields exhibit long-range spatial coherence and persistent anomalies, whereas precipitation shows shorter temporal memory and weaker space-time coupling. Despite these differences, the scalable GP approaches generalize well to unobserved locations, an essential property for forecasting systems with nonuniform station coverage. Comparisons with the operational NOAA HRRR forecasting system further illustrate both the strengths and limitations of station-based GP forecasting. For temperature, HRRR achieves the overall best predictive performance at the short lead times where comparisons are available, highlighting the strength of large-scale operational NWP systems. For precipitation, however, the GP-based methods outperform the HRRR NWP system, likely because localized precipitation variability is more difficult to capture accurately using coarse gridded forecasts extracted at station locations. Beyond predictive accuracy, the probabilistic nature of GP models provides calibrated uncertainty estimates, which are increasingly recognized as essential for modern weather forecasting and forecast evaluation \citep{bulte2026uncertainty,pohle2026uncertainty}. The methods and software developed here also extend naturally to other large-scale, uncertainty-aware prediction tasks, including ecology, environmental monitoring, and transportation networks.

An important direction for future work is the development of nonstationary or covariate-dependent space-time covariance structures that better capture regional variability in dependence strength and persistence. Correlation-based Vecchia neighbor selection is likely to remain valuable in this setting, as it adapts to local dependence and may help stabilize inference when dependence varies across the domain. The estimation of the covariance smoothness parameter $\nu$ could further enhance flexibility and allow the data to inform spatial and temporal regularity. Recent computational advances for estimating Matérn smoothness parameters include automatic differentiation approaches \citep{geoga2023fitting} and stable Fisher scoring optimization methods \citep{hong2026fisher}. Additional extensions include alternative temporal resolutions (e.g., intraday or weekly), machine-learning-based mean models combined with spatio-temporal correlation \citep{sigrist2022gaussian,sigrist2022latent,saha2023random,zhan2025neural}, the modeling of additional meteorological variables such as wind speed, humidity, or radiation, and the joint modeling of multiple variables through multivariate covariance functions. Furthermore, future work could investigate the integration of scalable GP approaches into emerging benchmark frameworks for weather forecasting, such as WeatherBench2 \citep{rasp2024weatherbench}, enabling more systematic comparisons with modern AI-based and operational forecasting systems. Finally, while we focus on a GPU implementation within a single ecosystem, extending the approach to and systematically comparing across different hardware platforms (e.g., AMD, Intel, or Apple GPUs) represents an interesting direction for future research.
 %Finally, we have used fixed smoothness parameters

% In summary, scalable GP approximations, when aligned with the underlying space-time structure and supported by modern computation, offer a powerful and interpretable framework for real-time weather forecasting.

%\section*{Disclosure statement}

%The authors report there are no competing interests to declare.
\ifdefined\ARXIV
\section*{Acknowledgments}
This research was partially supported by the Swiss Innovation Agency Innosuisse (grant no. 57667.1 IP-ICT).
\fi

%\bibliographystyle{abbrvnat}\bibliographystyle{IEEEtran}

% This creates an appendix chapter, comment if not needed.
%\appendix
%\section{First Appendix Chapter Title}

%% file: appendix.tex
\appendix
\begin{bibunit}

\begin{appendices}
\setcounter{figure}{0}
\renewcommand{\thefigure}{A\arabic{figure}}
\setcounter{table}{0}
\renewcommand{\thetable}{A\arabic{table}}

\section{Dataset}\label{App:DATA}

The following figures provide visualizations and fully reproducible code of the dataset.

\begin{figure}[ht!]
\centering
\ifdefined\ARXIV
  \includegraphics[scale=0.5]{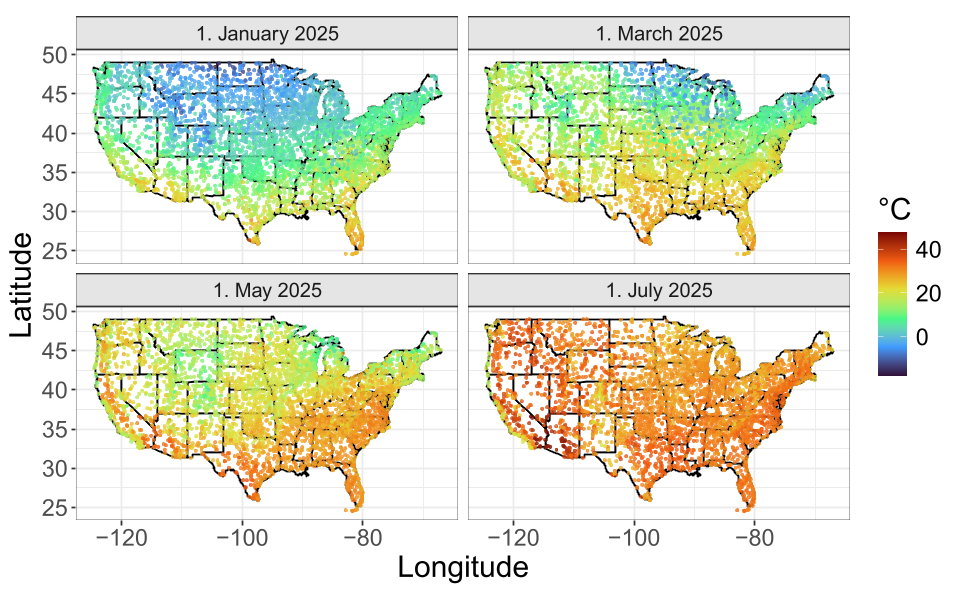}
\else
  \includegraphics[scale=0.5]{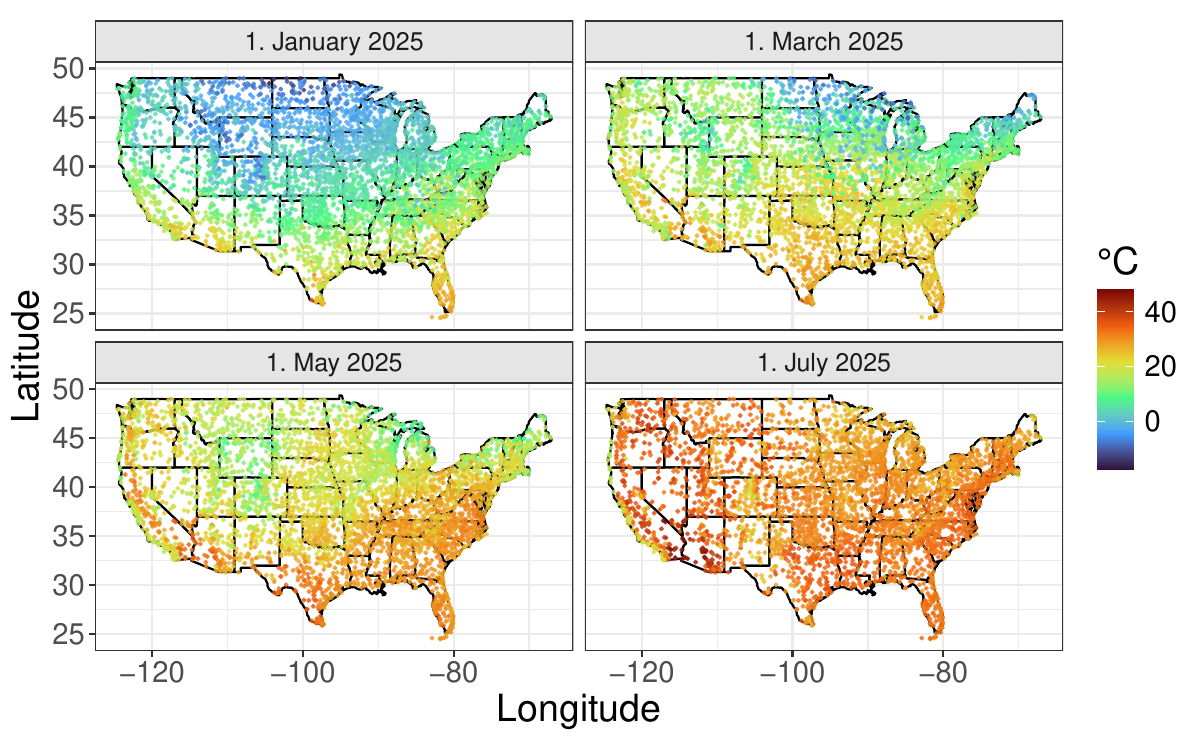}
\fi
\caption{Daily maximum temperature (°C with resolution: 0.1) across the conterminous United States, shown for four selected dates in 2025.}\label{fig:RWTemp}
\end{figure}

\begin{figure}[ht!]
\centering
\ifdefined\ARXIV
  \includegraphics[scale=0.5]{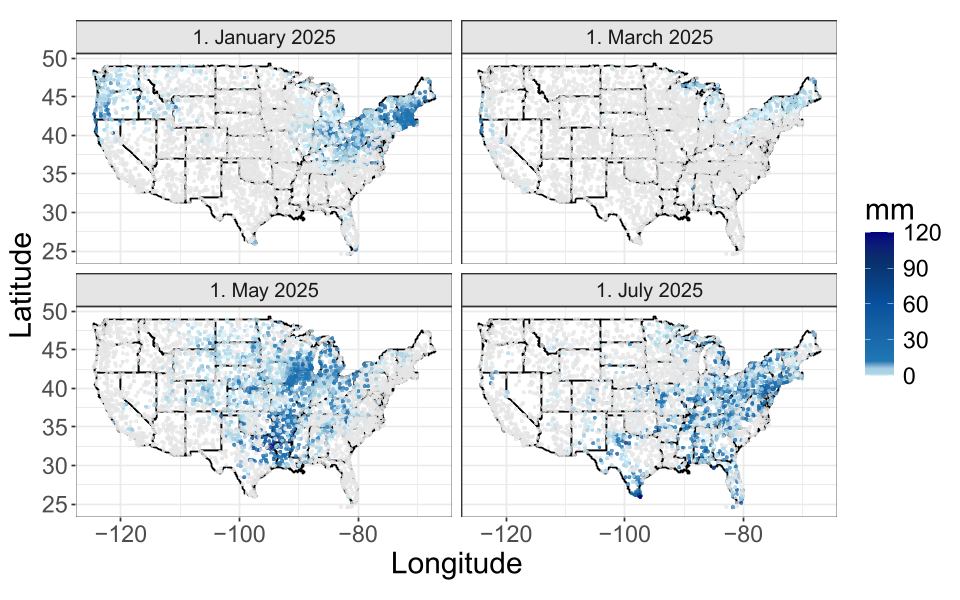}
\else
  \includegraphics[scale=0.5]{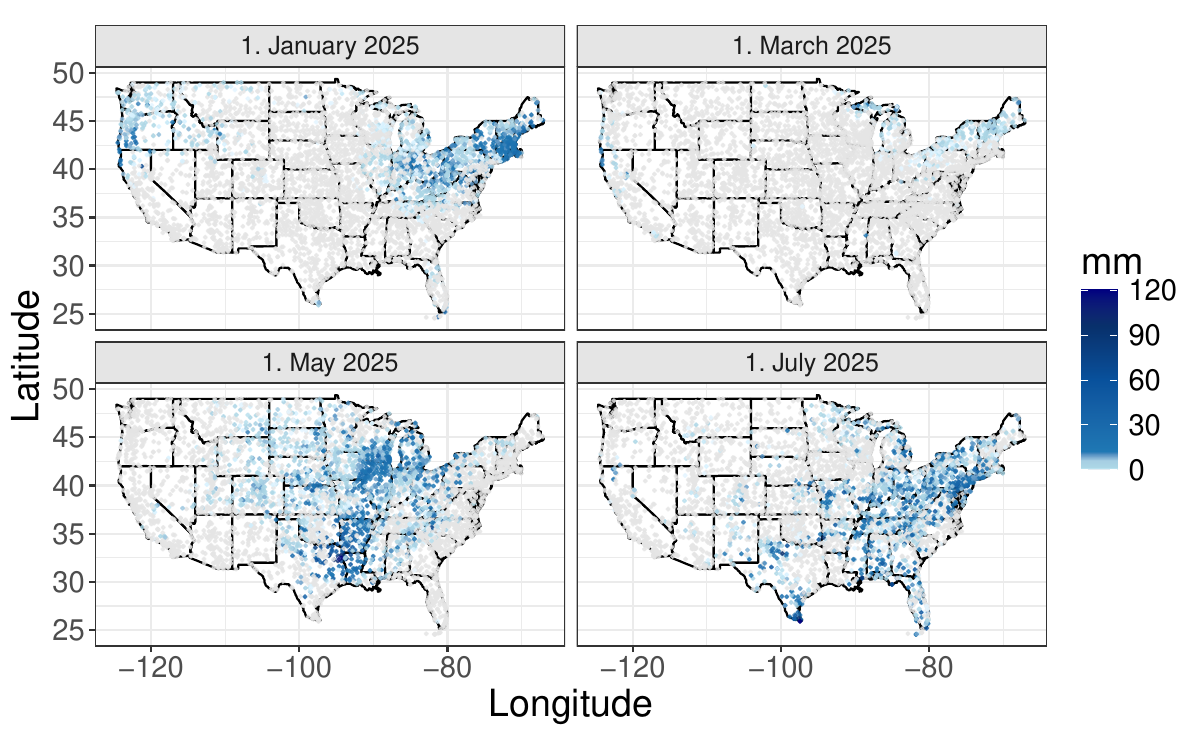}
\fi
\caption{Daily rainfall in millimeters (mm) across the conterminous United States at four selected days in 2025.}\label{fig:RWT}
\end{figure}

%\begin{figure}[ht!]
%\centering
%\includegraphics[scale=0.5]{figures/Real_World_data_examples.pdf}
%\caption{Daily maximum temperature (°C) and precipitation (mm) at five representative GHCNd stations across the full study period, with annotated average temperature, total rain amount (mm), and counts of rainy days (precipitation > 0 mm). Note the different scales in the panels.}\label{fig:RWT123}
%\end{figure}

\begin{figure}[H]
\centering
\ifdefined\ARXIV
  \includegraphics[scale=0.5]{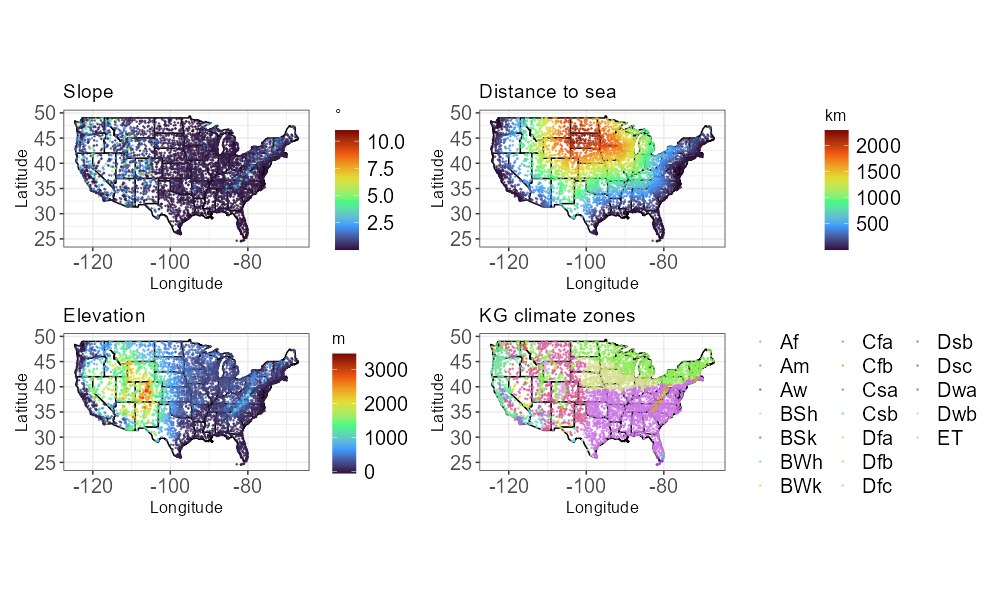}
\else
  \includegraphics[scale=0.5]{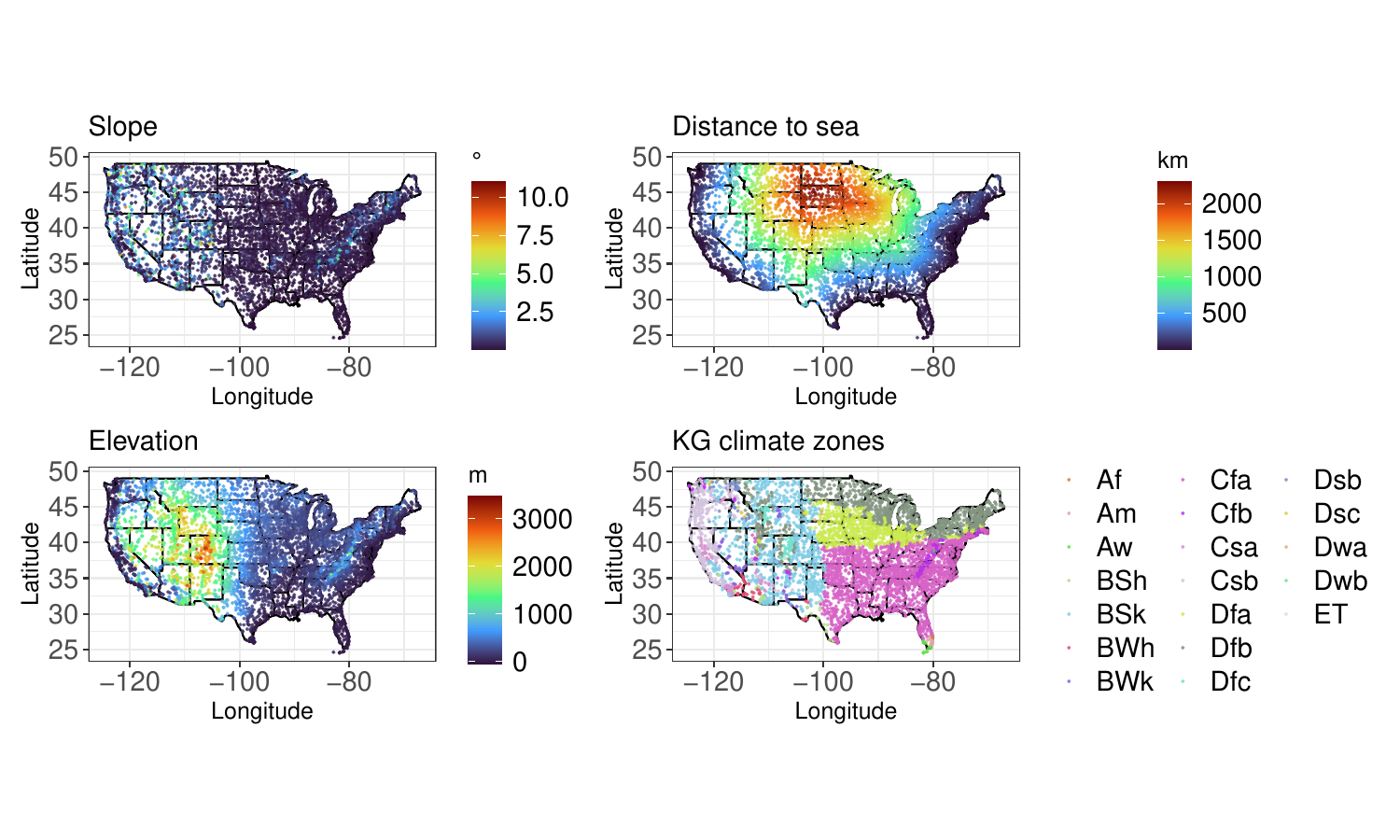}
\fi
%{figures/Real_World_Cov.pdf}
\caption{Example geographic and climatic covariates used in the model.}\label{fig:RWR_cov}
\end{figure}

\section{Laplace approximation and latent Gaussian process models}\label{App:Lap}

In the following, we assume that the response variable $\boldsymbol{y}=\left(y((\boldsymbol{s}_1,t_1)), \ldots, y((\boldsymbol{s}_n,t_n))\right)^\mathrm{T} \in \mathbb{R}^n$ follows a parametric distribution with a density of $p(\boldsymbol{y} \mid \boldsymbol{\mu}, \boldsymbol{\xi}) = \prod_{i=1}^n p(y_i|\mu_i,\xi)$ with respect to a sigma finite product measure conditional on $\boldsymbol{\mu} = F(\boldsymbol{X}) + \boldsymbol{b}$. This density has, potentially link function-transformed, parameters $\boldsymbol{\mu} \in \mathbb{R}^n$ and additional auxiliary parameters $\boldsymbol{\xi} \in \Xi \subset \mathbb{R}^r$. For instance, $\mu_i$ and $\xi$ can be the mean and variance of a Gaussian distribution, the log-mean and the shape parameter of a gamma likelihood, or $\mu_i$ can be the logit-transformed success probability of a Bernoulli distribution for which there is no additional parameter $\xi$. The parameter $\boldsymbol{\mu}$ is modeled as the sum of fixed effects $F(\boldsymbol{X})$ and a finite-dimensional version of a GP $\boldsymbol{b}$, $\boldsymbol{\mu} = F(\boldsymbol{X}) + \boldsymbol{b}$, where $\boldsymbol{X}\in\mathbb{R}^{n\times p}$ contains predictor variables and $\boldsymbol{b} \sim \mathcal{N}(\boldsymbol{0}, \mathbf{\Sigma})$. We again assume for simplicity that $F(\boldsymbol{X})=\boldsymbol{0}$, but this assumption can be easily relaxed. Estimation of the covariance $\boldsymbol{\theta} \in \mathbb{R}^{q}$ and auxiliary parameters $\boldsymbol{\xi}$ is done by minimizing the negative log-marginal likelihood
 \begin{align}
 \begin{split}
-\log \big(p(\boldsymbol{y} \mid \boldsymbol{\theta}, \boldsymbol{\xi})\big)=-\log \Big(\int p(\boldsymbol{y} \mid \boldsymbol{b}, \boldsymbol{\xi}) p(\boldsymbol{b} \mid \boldsymbol{\theta}) d\boldsymbol{b}\Big).
\end{split}\label{NEGLL} 
\end{align}

% For applying the Laplace approximation, we assume that $p\left(y(\boldsymbol{s}_i) \mid b(\boldsymbol{s}_i), \boldsymbol{\xi}\right)$ is three times differentiable in $b(\boldsymbol{s}_i)$.
For non-Gaussian likelihoods, there is typically no analytic expression for $p(\boldsymbol{y} \mid \boldsymbol{b}, \boldsymbol{\xi})$ and an approximation has to be used. In this article, we use the Laplace approximation \citep{tierney1986accurate}, additionally assuming that $p(\boldsymbol{y} \mid \boldsymbol{\mu}, \boldsymbol{\xi})$ is log-concave in $\boldsymbol{\mu}$. The Laplace approximation for (\ref{NEGLL}) is given by
\begin{align}
L^\textit{LA}(\boldsymbol{y} , \boldsymbol{\theta} , \boldsymbol{\xi})=-\log p\left(\boldsymbol{y}  \mid \Tilde{\boldsymbol{b}}, \boldsymbol{\xi} \right)+\frac{1}{2} \Tilde{\boldsymbol{b}}^{ T} \mathbf{\Sigma}^{-1} \Tilde{\boldsymbol{b}}+\frac{1}{2} \log \operatorname{det}\left(\mathbf{\Sigma} \boldsymbol{W}+\boldsymbol{I}_n\right),\label{LA} 
\end{align}
where $\Tilde{\boldsymbol{b}}=\operatorname{argmax}\limits_{\boldsymbol{b}} \log p(\boldsymbol{y} \mid \boldsymbol{b}, \boldsymbol{\xi})-\frac{1}{2} \boldsymbol{b}^\mathrm{T} \mathbf{\Sigma}^{-1} \boldsymbol{b}$ is the mode of $p(\boldsymbol{y} \mid \boldsymbol{b}, \boldsymbol{\xi}) {p}(\boldsymbol{b} \mid \boldsymbol{\theta})$, and $\boldsymbol{W} \in \mathbb{R}^{n \times n}$ is diagonal with 
\begin{align}
    W_{i i}=-\left.\frac{\partial^2 \log p\left(y(\boldsymbol{s}_i) \mid b(\boldsymbol{s}_i), \boldsymbol{\xi}\right)}{\partial b(\boldsymbol{s}_i)^2}\right|_{\boldsymbol{b}=\Tilde{\boldsymbol{b}}}. \label{W_diag}
\end{align}

The mode $\Tilde{\boldsymbol{b}}$ is usually found with Newton's method \citep{williams2006gaussian}. Gradients of $L^\textit{LA}(\boldsymbol{y} ,\boldsymbol{\theta} , \boldsymbol{\xi})$ for parameter estimation can be found, e.g., in \citet{williams2006gaussian} and \citet{sigrist2022latent}.

\clearpage
\section{Simulation study: additional figures and results}\label{App:Sim} 
\begin{figure}[ht!]
\centering
     \begin{subfigure}[b]{0.32\linewidth}
         \centering
      \includegraphics[width=\linewidth]{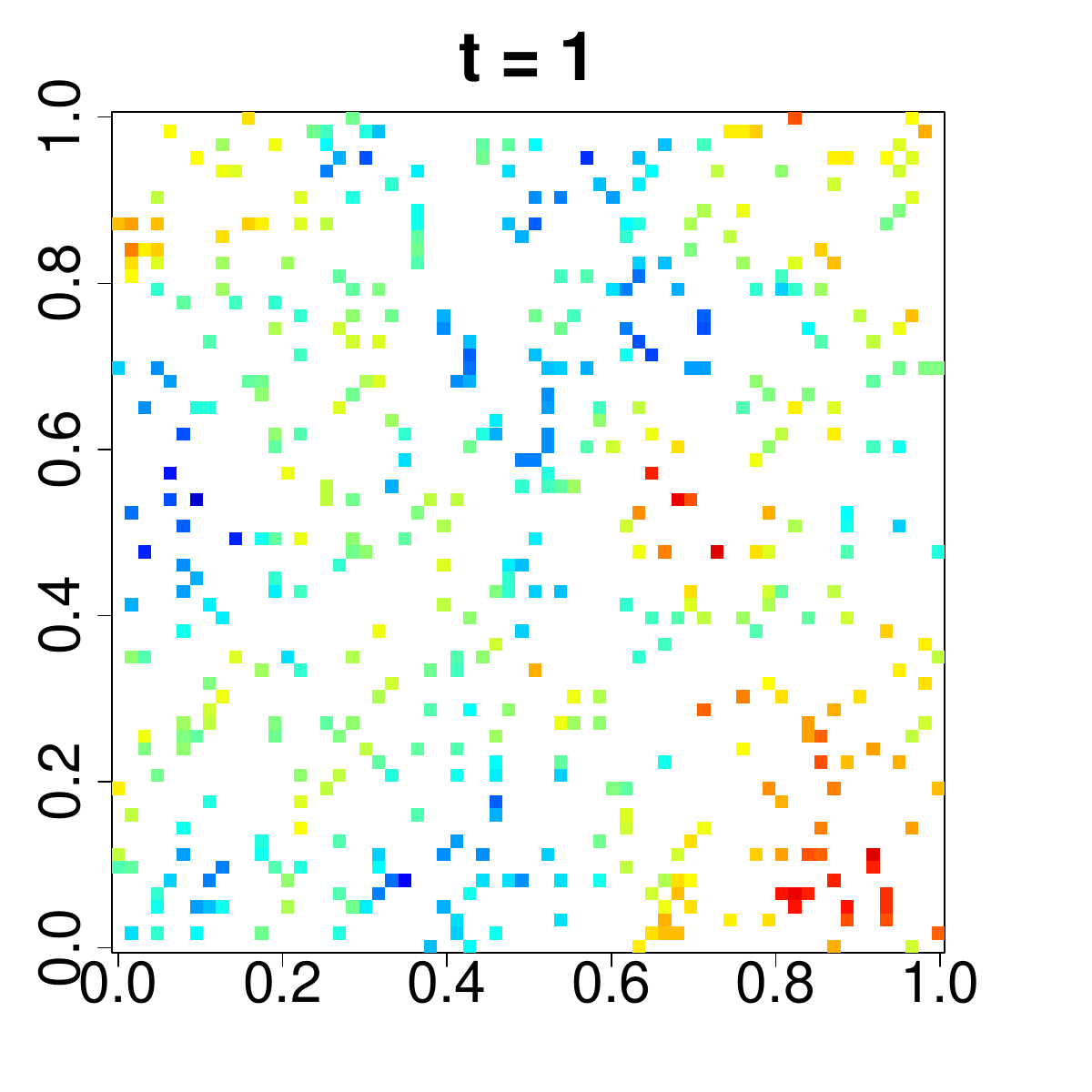}
     \end{subfigure}
     \hfill
     \centering
     \begin{subfigure}[b]{0.32\linewidth}
         \centering
      \includegraphics[width=\linewidth]{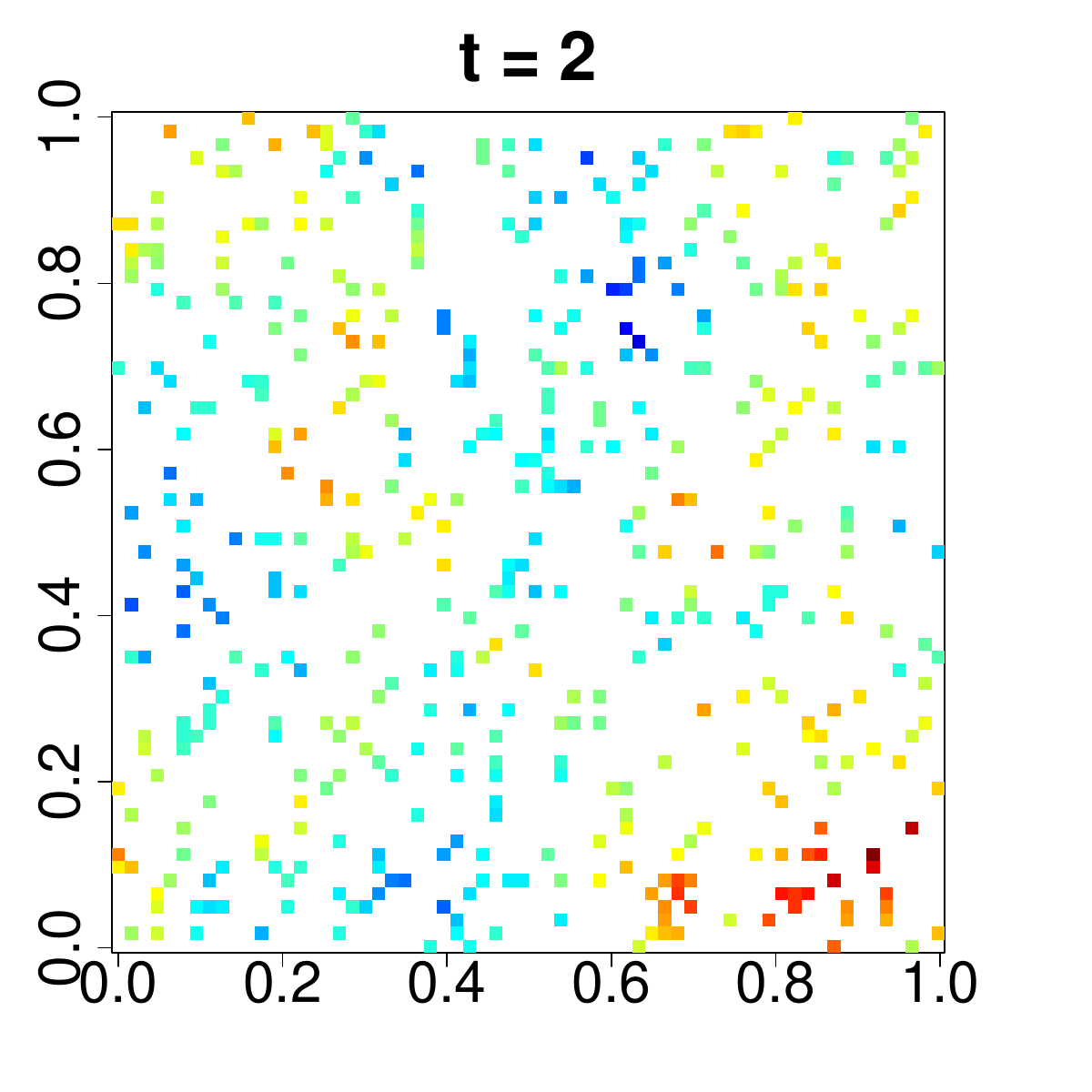}
     \end{subfigure}
     \hfill
     \begin{subfigure}[b]{0.32\linewidth}
         \centering
         \includegraphics[width=\linewidth]{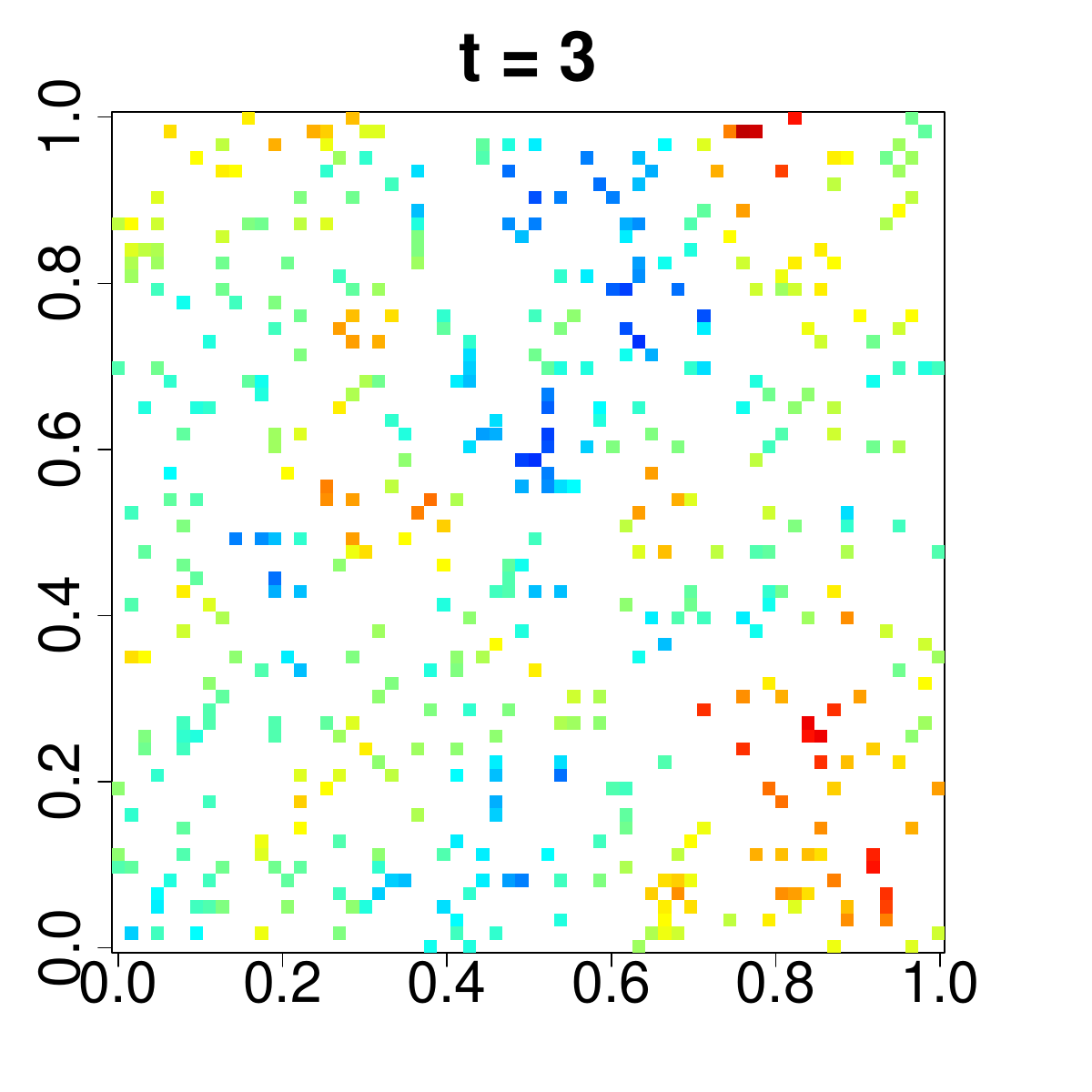}
     \end{subfigure}
     \begin{subfigure}[b]{0.32\linewidth}
         \centering
         \includegraphics[width=\linewidth]{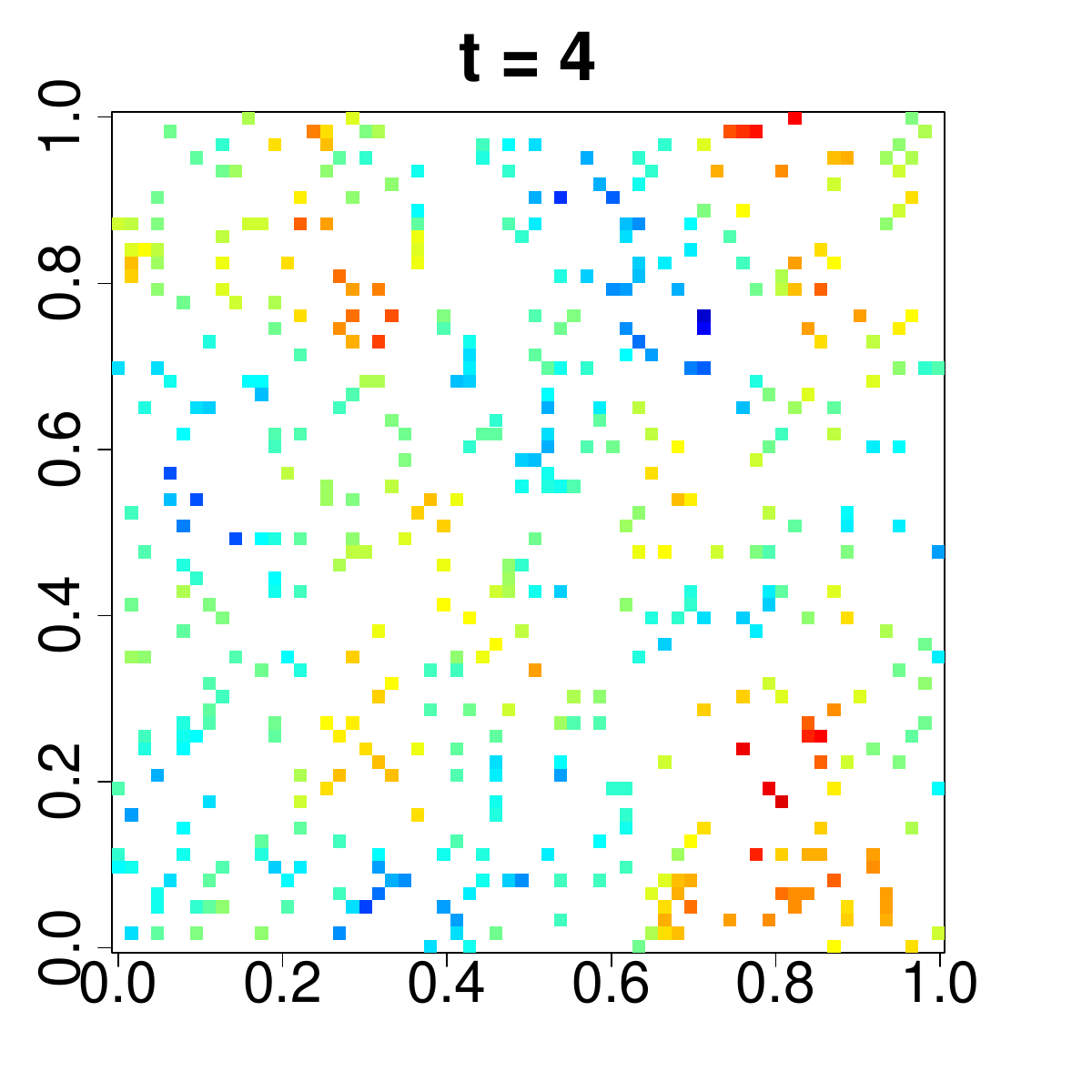}
     \end{subfigure}
     \hfill
     \begin{subfigure}[b]{0.32\linewidth}
         \centering
         \includegraphics[width=\linewidth]{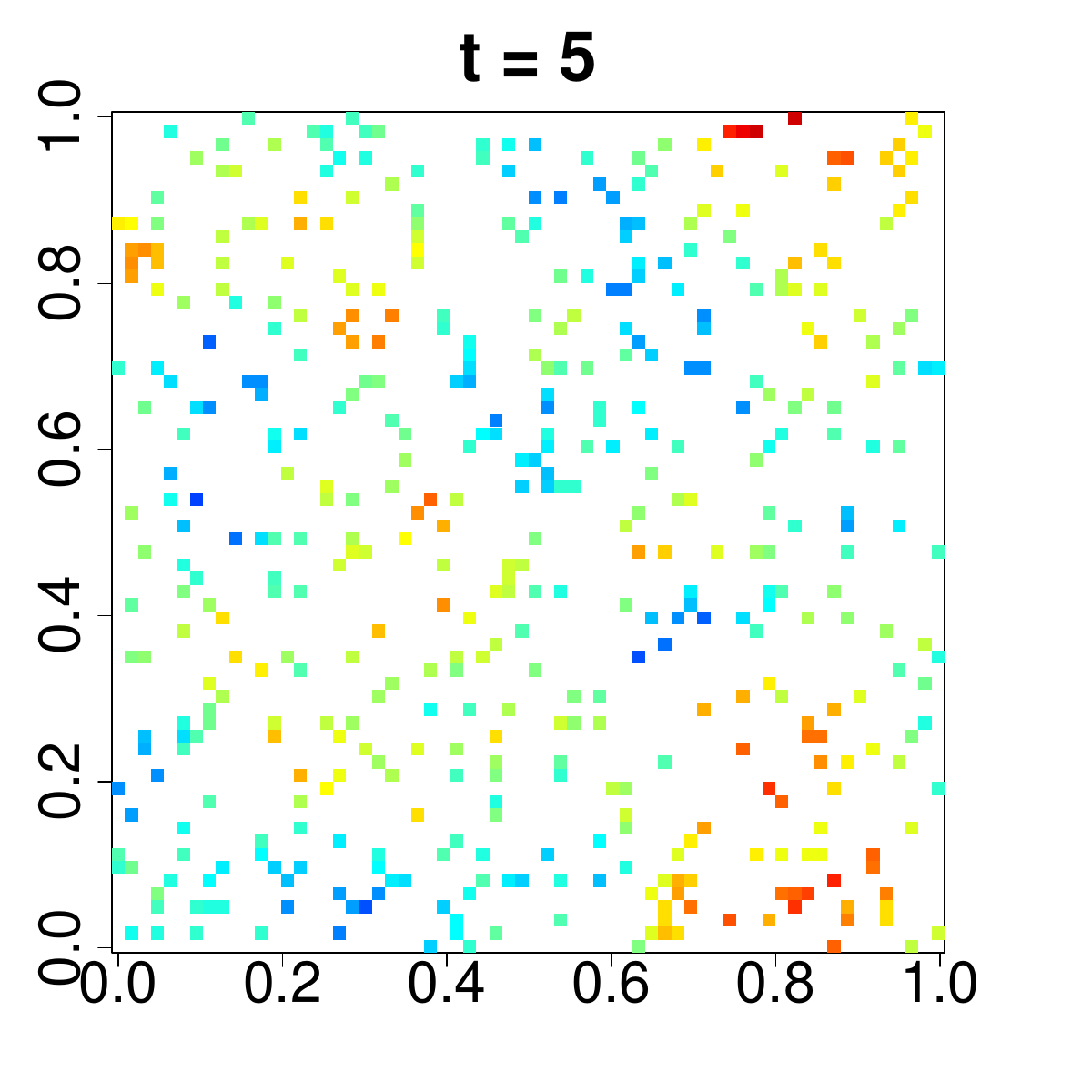}
     \end{subfigure}
     \hfill
     \begin{subfigure}[b]{0.32\linewidth}
         \centering
         \includegraphics[width=\linewidth]{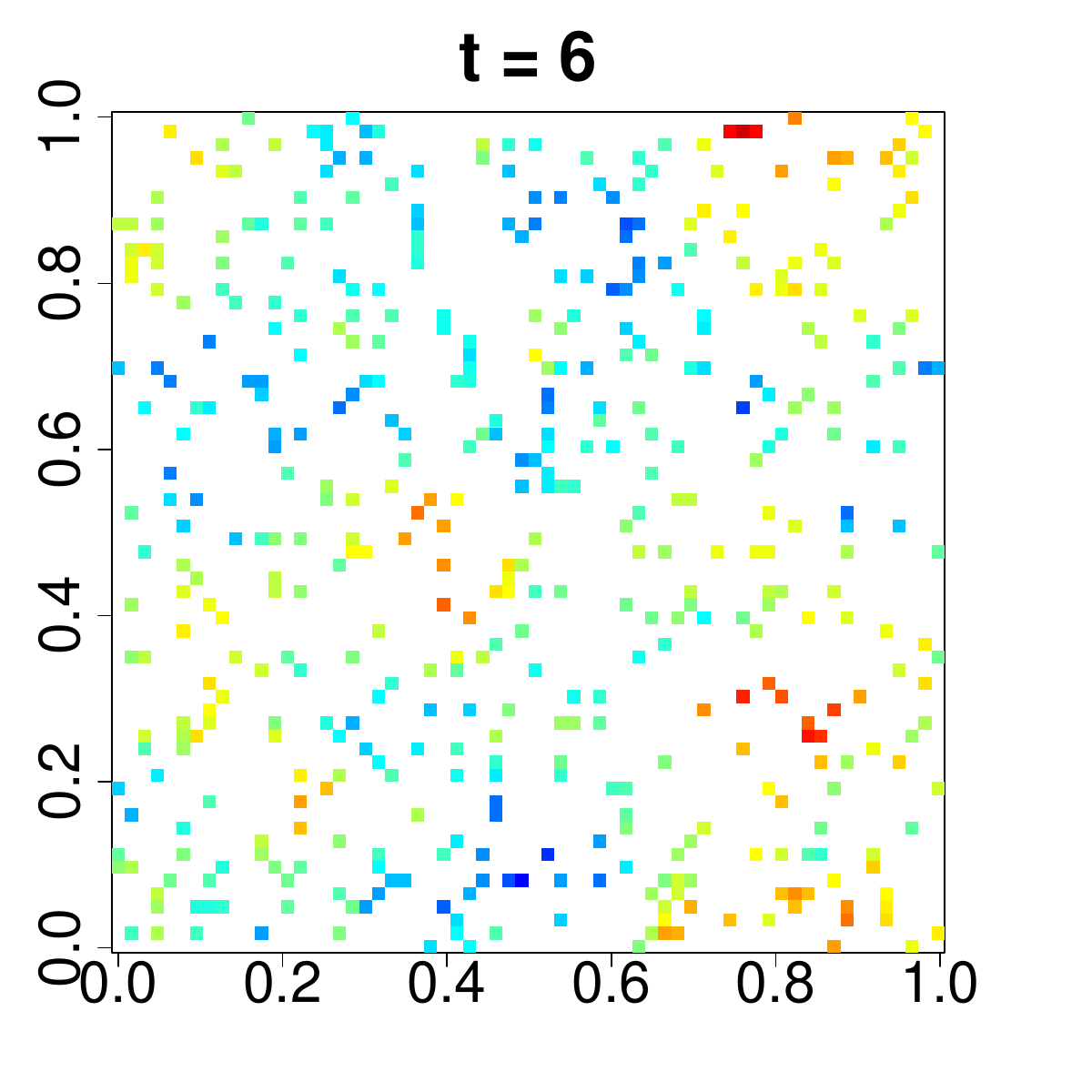}
     \end{subfigure}
     \begin{subfigure}[b]{0.32\linewidth}
         \centering
         \includegraphics[width=\linewidth]{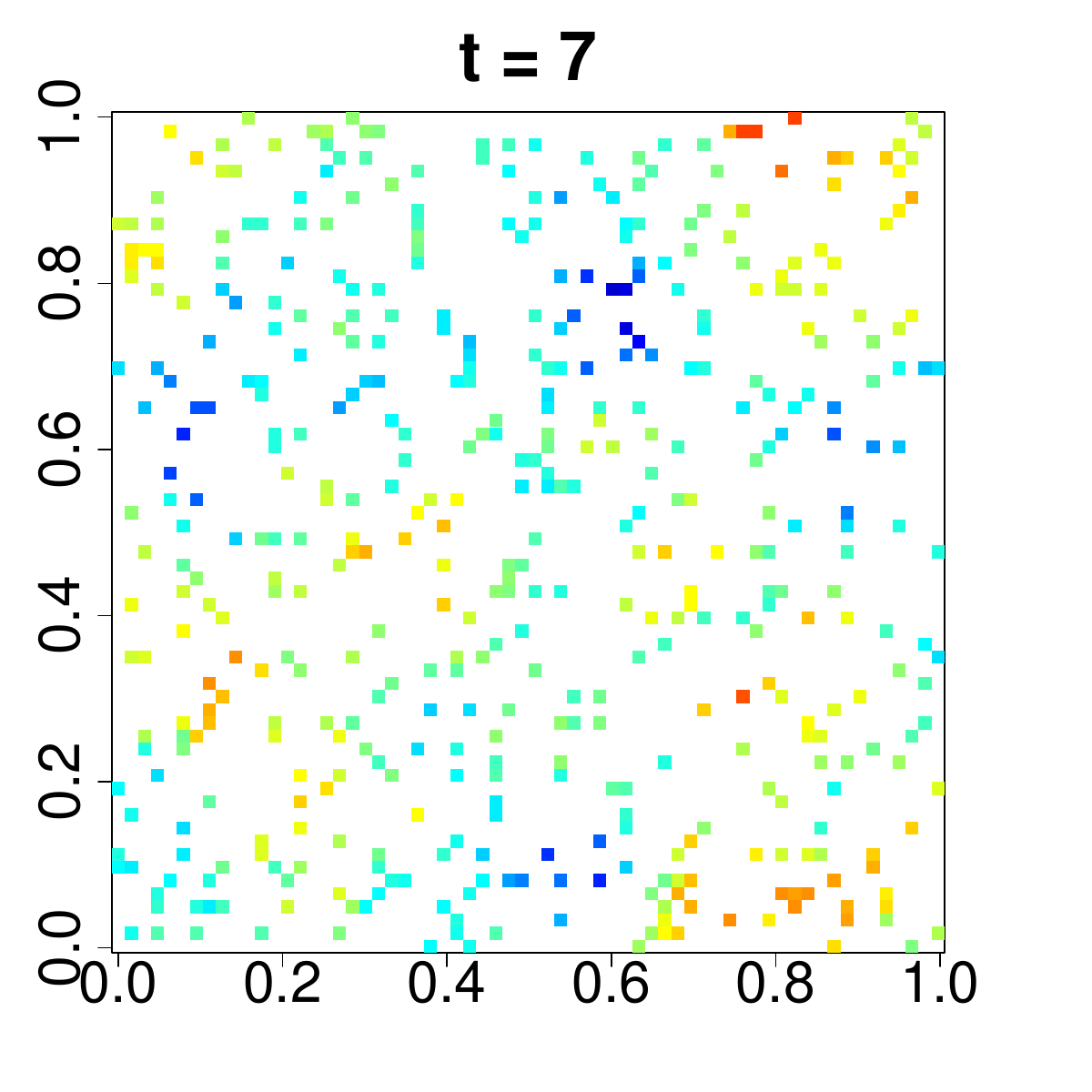}
     \end{subfigure}
     \hfill
     \begin{subfigure}[b]{0.32\linewidth}
         \centering
         \includegraphics[width=\linewidth]{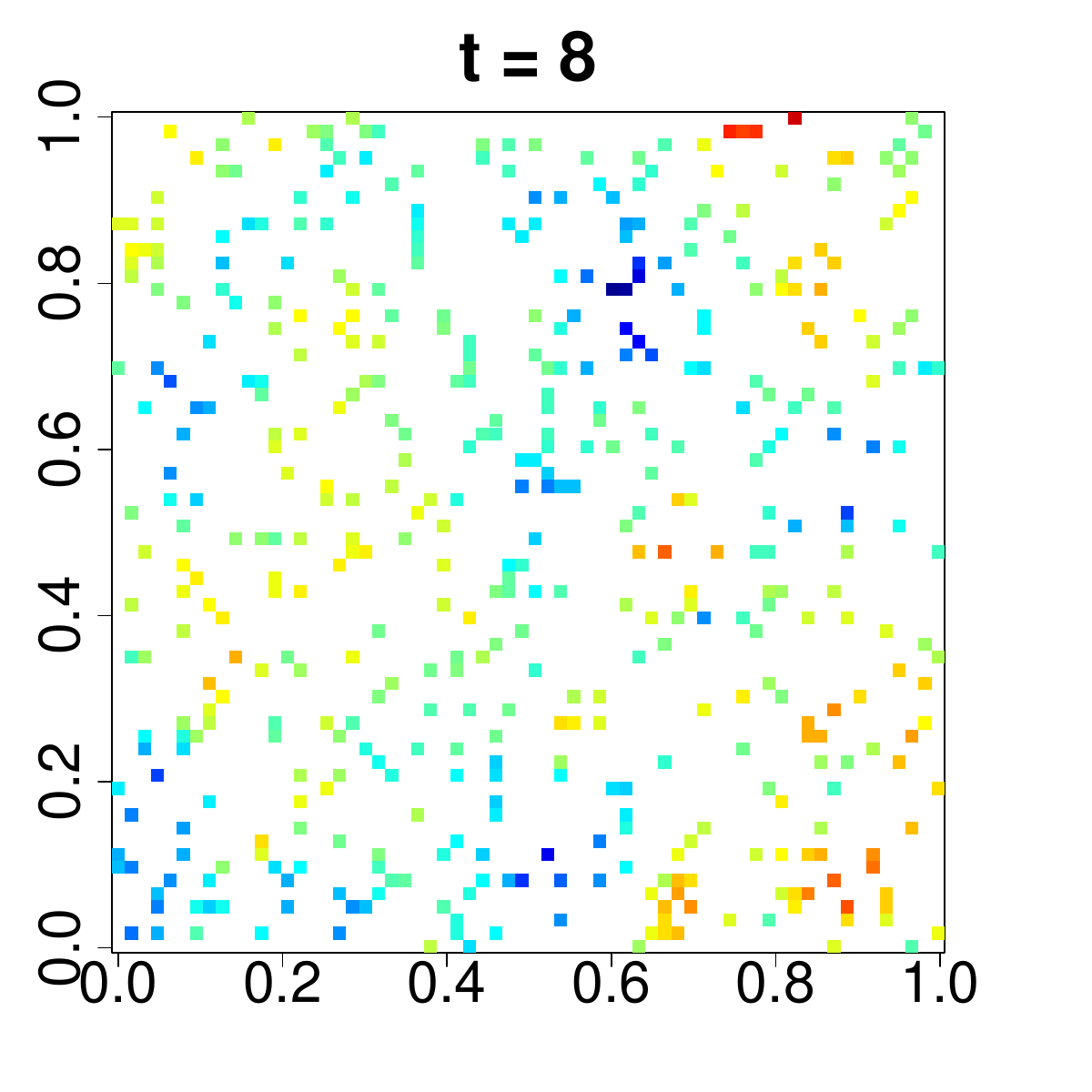}
     \end{subfigure}
     \hfill
     \begin{subfigure}[b]{0.32\linewidth}
         \centering
         \includegraphics[width=\linewidth]{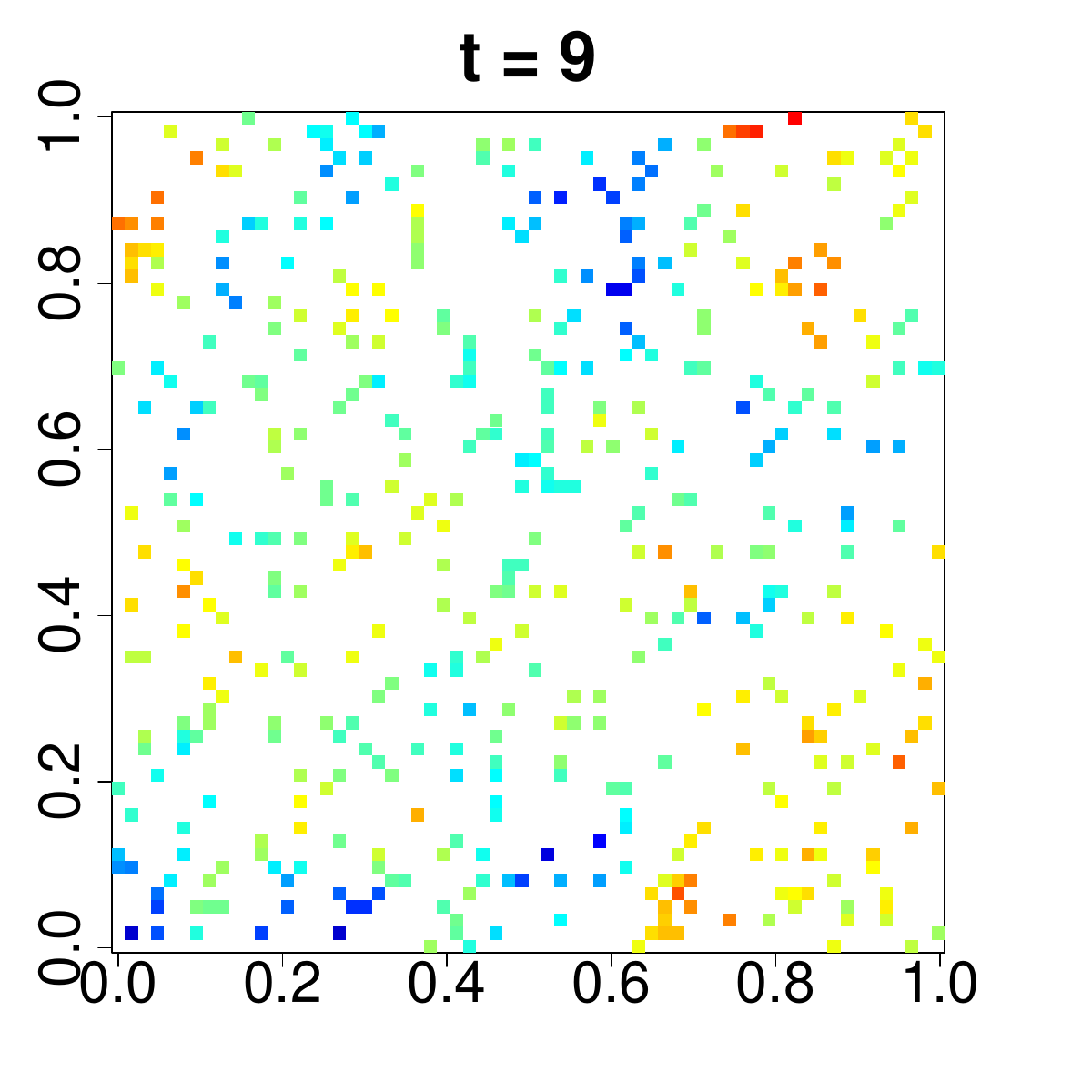}
     \end{subfigure}
     \caption{Synthetic space-time dataset for first nine time points $t \in \{1,2,3,4,5,6,7,8,9\}$ with 500 spatial locations per time point.}
        \label{fig:Sim_dat}
\end{figure}

Figure \ref{fig:Sim_dat} illustrates the synthetic dataset used in the simulation study.

%\clearpage

%\section{Parameter estimation}

\begin{figure}[ht!]
\centering
\includegraphics[width=0.8\linewidth]{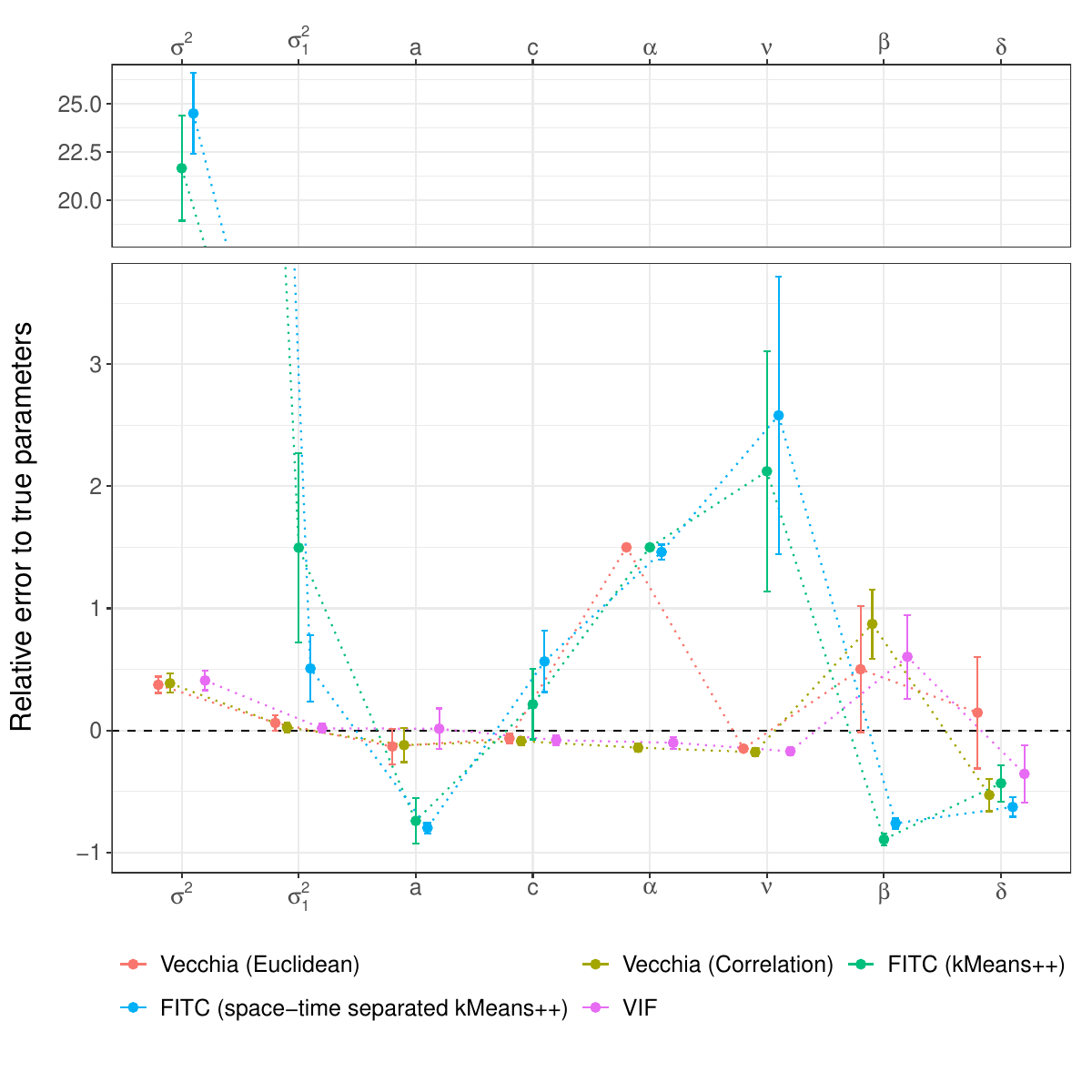}
\caption{Relative error with error bars (mean $\pm$ two standard errors) of covariance parameters for Vecchia, FITC, and VIF approximations.}\label{fig:SimP}
\end{figure}

Figure \ref{fig:SimP} presents the results of a comparison between estimated parameters and their true values. The Vecchia approximation with Euclidean neighbors struggles to identify the temporal smoothness ($\alpha$), whereas the Vecchia approximation with a correlation-based neighbor selection and the VIF approximation yield accurate estimates for all covariance parameters. The FITC approximation exhibits marked biases, especially in the nugget parameter, indicating difficulty in capturing fine-scale variability.

%\section*{Vecchia ordering} 
Table \ref{ordering_comp} presents a comparison of the predictive accuracy obtained with a time-stratified random ordering and a fully random ordering on simulated data, showing nearly identical results.

\begin{table}[ht!]
%\footnotesize
 \centering
 \begin{threeparttable}
 \caption{\label{ordering_comp} Comparison of RMSE (± one standard error) between random ordering and time-stratified random ordering for the correlation-based Vecchia approximation.}
 \begin{tabular}{ |p{1.3cm}|p{4cm}|p{4.5cm}|}
  \hline
  & Random ordering & Time-stratified random ordering\\
  \hline 
\textbf{RMSE}&0.805 $\pm$ 0.008 &0.804 $\pm$ 0.008\\
  \hline 
 \end{tabular}
 \end{threeparttable}
\end{table}

%\section*{Scalability on large benchmark data}

To assess performance at realistic scales, we apply the suggested methods to the synthetic benchmarks from the Second Spatial Statistics Competition \textit{2a} and \textit{2b} \citep{abdulah2022second}, which provide exact GP samples from the Gneiting covariance function used in this work. The datasets contain $100{,}000$ and $1{,}000{,}000$ observations, respectively, with $100$ time points and up to $10{,}000$ spatial locations per time point. Using these benchmarks avoids cubic simulation costs and enables direct comparison with exact methods such as \textit{ExaGeoStat} \citep{abdulah2018exageostat}. We consider three regimes of space-time dependence (strong, moderate, and weak), defined by different covariance parameter settings. The corresponding parameter values and effective spatial and temporal ranges are reported in Table~\ref{Scenarios}. We consider three prediction tasks, namely withheld spatial locations (RS), withheld space time locations (RST), and temporal extrapolation (T10), using 90\% of the data for training and RMSE for evaluation. The results for sub-competition~\textit{2a} reported in Table~\ref{Scenarios_Results2a} show that the VIF and Vecchia with correlation-based neighbor selection approximations consistently achieve RMSE values close to the exact model across all scenarios, whereas Euclidean neighbor selection is less accurate under rapidly decaying dependence. FITC approximations perform poorly in interpolation settings but become competitive for temporal extrapolation, particularly when using sts-kMeans++ for selecting inducing points. This highlights the importance of targeted temporal coverage when forecasting beyond the observation window, where spatial proximity to inducing points is limited. 

\begin{table}[ht!]
%\footnotesize
\centering
\begin{threeparttable}
\caption{\label{Scenarios_Results2a}RMSE for sub-competition~\textit{2a} \citep{abdulah2022second}; last column from Tables S4–S6.}
{
\begin{tabular}{|c|c||c|c|c|c|c|c|}
\hline
\multicolumn{2}{|c||}{\textbf{RMSE}} &
\makecell[c]{Vecchia \\ Euc.} &
\makecell[c]{Vecchia \\ Corr.} &
\makecell[c]{FITC \\ kMeans++ }&
\makecell[c]{FITC \\ sts-kMeans++ } &
VIF &
ExaGeoStat \\
\hline
\multirow{3}{*}{RS}
 & $\mathcal D_1$ & 0.707073 & 0.701579 & 0.888506 & 0.888751 & 0.700932 & 0.696989 \\
 & $\mathcal D_2$ & 0.227223 & 0.227249 & 0.816669 & 0.907848 & 0.227232 & 0.227171 \\
 & $\mathcal D_3$ & 0.046763 & 0.046765 & 0.333600 & 0.700298 & 0.046770 & 0.046780 \\
\Xhline{2\arrayrulewidth}
\multirow{3}{*}{RST}
 & $\mathcal D_1$ & 0.708225 & 0.615680 & 0.894176 & 0.888547 & 0.615148 & 0.610297 \\
 & $\mathcal D_2$ & 0.222245 & 0.222248 & 0.805316 & 0.889083 & 0.222247 & 0.221473 \\
 & $\mathcal D_3$ & 0.046240 & 0.046239 & 0.354850 & 0.703996 & 0.046239 & 0.040588 \\
\Xhline{2\arrayrulewidth}
\multirow{3}{*}{T10}
 & $\mathcal D_1$ & 0.923049 & 0.904621 & 0.926686 & 0.918036 & 0.904412 & 0.901485 \\
 & $\mathcal D_2$ & 0.939343 & 0.938799 & 0.962824 & 0.937738 & 0.938799 & 0.932938 \\
 & $\mathcal D_3$ & 0.913873 & 0.913874 & 0.751174 & 0.744102 & 0.909280 & 0.777060 \\
\hline
\end{tabular}
}

\end{threeparttable}
\end{table}

Table~\ref{Scenarios_Results2b} reports the RMSE of the VIF approximation across all datasets and scenarios in sub-competition~\textit{2b} \citep{abdulah2022second}, compared with the four top competitors.

\begin{table}[ht!]
 \centering
 \begin{threeparttable}
 \caption{\label{Scenarios} The parameter choices for the experiments in the sub-competition \textit{2a} and \textit{2b} in \citep{abdulah2022second}.}
 \begin{tabular}{ |p{1.cm}||p{.6cm}|p{.6cm}|p{.6cm}|p{.6cm}|p{.6cm}|p{.6cm}|p{.6cm}|p{2.5cm}|p{2.5cm}|}
  \hline
  \textbf{Data set}&{$\sigma_1^2$} &{$a$} &{$c$} & {$\alpha$}& {$\nu$} & {$\beta$}  & {$\delta$} & \textbf{Time effective range} & \textbf{Space effective range}\\
  \hline 
  $\mathcal{D}_1$&0.9 &1 &50 & 0.6 & 1 & 0.9  & 0.1  &  11.63 & 0.08 \\
  \hline 
  $\mathcal{D}_2$&0.9 &0.24 &12.5 & 0.6 & 1 & 0.9  & 0.1  &  38.20 & 0.32 \\
  \hline 
  $\mathcal{D}_3$&0.9 &1 &2.5 & 0.08 & 1 & 0.9  & 0.1  &  $\infty$ & 1.6 \\
  \hline 
 \end{tabular}

     \end{threeparttable}
 \end{table}

%\textit{ExaGeoStat} is a high-performance computing framework designed specifically for GP models at massive scale. It implements covariance construction, simulation, and exact likelihood evaluation using distributed linear algebra routines and task-based parallelism, allowing it to circumvent the usual $\mathcal{O}(n^3)$ computational bottleneck on modern multicore, GPU, and exascale architectures.

The \texttt{Envstat.ai} team employed a Deep Neural Network (DNN) for spatio-temporal prediction, extending the spatial model of \citet{chen2020deepkriging}. They incorporated basis functions to capture spatio-temporal dependence. For interpolation tasks, they applied regression, while for forecasting at new locations they adopted a two-stage strategy: (i) interpolate values at the new locations for observed time points, and (ii) train a Long Short-Term Memory (LSTM) network on these interpolated values to obtain forecasts for future time points.  

The \texttt{GpGp} team fitted a Matérn space-time covariance function using the \texttt{GpGp} package to estimate the model parameters. The two estimated range parameters were then used to rescale the coordinates, which guided the ordering and neighbor selection for Vecchia’s approximation. They employed $m_v = 30$ neighbors to fit the model and $m_v = 60$ for prediction.  

The \texttt{RESSTE} team began with an exploratory data analysis using a space-time covariance function and identified that the data were generated from a positively non-separable space-time kernel. Based on this finding, they selected a covariance function from the Gneiting class, combining a Matérn spatial covariance with a Cauchy temporal covariance. Given the large size of the datasets, they used a block-composite likelihood to separately estimate the spatial and temporal parameters, which were then combined to fit the full space-time model. For prediction, they applied ordinary kriging, conditioning on a specified number of nearest spatial neighbors as well as preceding, current, and subsequent time slots.

 The \texttt{GeoModels} team  applied the nearest neighbor weighted composite likelihood method to model both datasets, using the \texttt{GeoModels} package \citep{bevilacqua2022geomodels}.

Table~\ref{Scenarios_Results2b} presents the RMSE of the VIF approximation across all datasets and prediction scenarios from sub-competition \textit{2b} compared to the four best competitors: \texttt{Envstat.ai}, \texttt{GpGp}, \texttt{GeoModels}, \texttt{RESSTE} described above. Overall, the VIF approximation performs strongly across all datasets and scenarios. For the time-extrapolation scenario T10, \texttt{Envstat.ai} is clearly superior to all competing methods.

\begin{table}[ht!]
 \centering
 \begin{threeparttable}
 \caption{\label{Scenarios_Results2b}RMSE for the sub-competition \textit{2b} of \citet{abdulah2022second}. }
 \begin{tabular}{ |p{.8cm}|p{0.6cm}||p{1.8cm}|p{2cm}|p{1.8cm}|p{2cm}|p{1.8cm}|}
  \hline
   \multicolumn{2}{|c||}{\textbf{RMSE}}& \makecell[c]{\textbf{VIF}}& \makecell[c]{\textit{Envstat.ai}}& \makecell[c]{\textit{GpGp} }& \makecell[c]{\textit{GeoModels} } &\makecell[c]{\textit{RESSTE} }\\
  \hline 
  \makecell[l]{RS}& \makecell[l]{$\mathcal{D}_1$ \\ $\mathcal{D}_2$ \\ $\mathcal{D}_3$}& \makecell[l]{0.276685 \\ 0.071518 \\ 0.014095}& \makecell[l]{0.536665 \\ 0.232439 \\ 0.050975}& \makecell[l]{0.277007 \\ 0.071519 \\ 0.014096} & \makecell[l]{0.276670 \\ 0.071517 \\ 0.014098} &  \makecell[l]{0.276730 \\ 0.071523 \\ 0.014097} \\
  \Xhline{3\arrayrulewidth}\makecell[l]{RST}& \makecell[l]{$\mathcal{D}_1$ \\ $\mathcal{D}_2$ \\ $\mathcal{D}_3$}& \makecell[l]{0.257556 \\ 0.070873 \\ 0.014260}& \makecell[l]{0.538314 \\0.232796 \\0.050681} & \makecell[l]{0.274177 \\0.070873 \\0.014260}& \makecell[l]{0.276957 \\ 0.070880 \\ 0.014269} & \makecell[l]{0.264053 \\ 0.071254 \\ 0.013380}\\
  \Xhline{3\arrayrulewidth}
  \makecell[l]{T10}& \makecell[l]{$\mathcal{D}_1$ \\ $\mathcal{D}_2$ \\ $\mathcal{D}_3$}& \makecell[l]{0.890374 \\ 0.947473 \\ 0.677872}& \makecell[l]{0.522757 \\ 0.206519 \\ 0.047245} & \makecell[l]{0.904975 \\ 0.949491 \\ 0.651425}& \makecell[l]{0.888390 \\ 0.948918 \\ 0.728364} & \makecell[l]{0.903714 \\ 0.964134 \\ 0.688624}\\
  \Xhline{3\arrayrulewidth}
  \multicolumn{2}{|c||}{Time}& \makecell[l]{69,063 s}& \makecell[c]{-} & \makecell[c]{-}& \makecell[c]{-} & \makecell[c]{-}\\
  \hline 
 \end{tabular}
     \end{threeparttable}
 \end{table}

\begin{figure}[ht!]
    {\raggedright \Large \bfseries $\mathcal{D}_1:$ \par}
    \vspace{-0.2cm}
     \centering
     \begin{subfigure}[b]{0.32\linewidth}
         \centering
         \includegraphics[width=\linewidth]{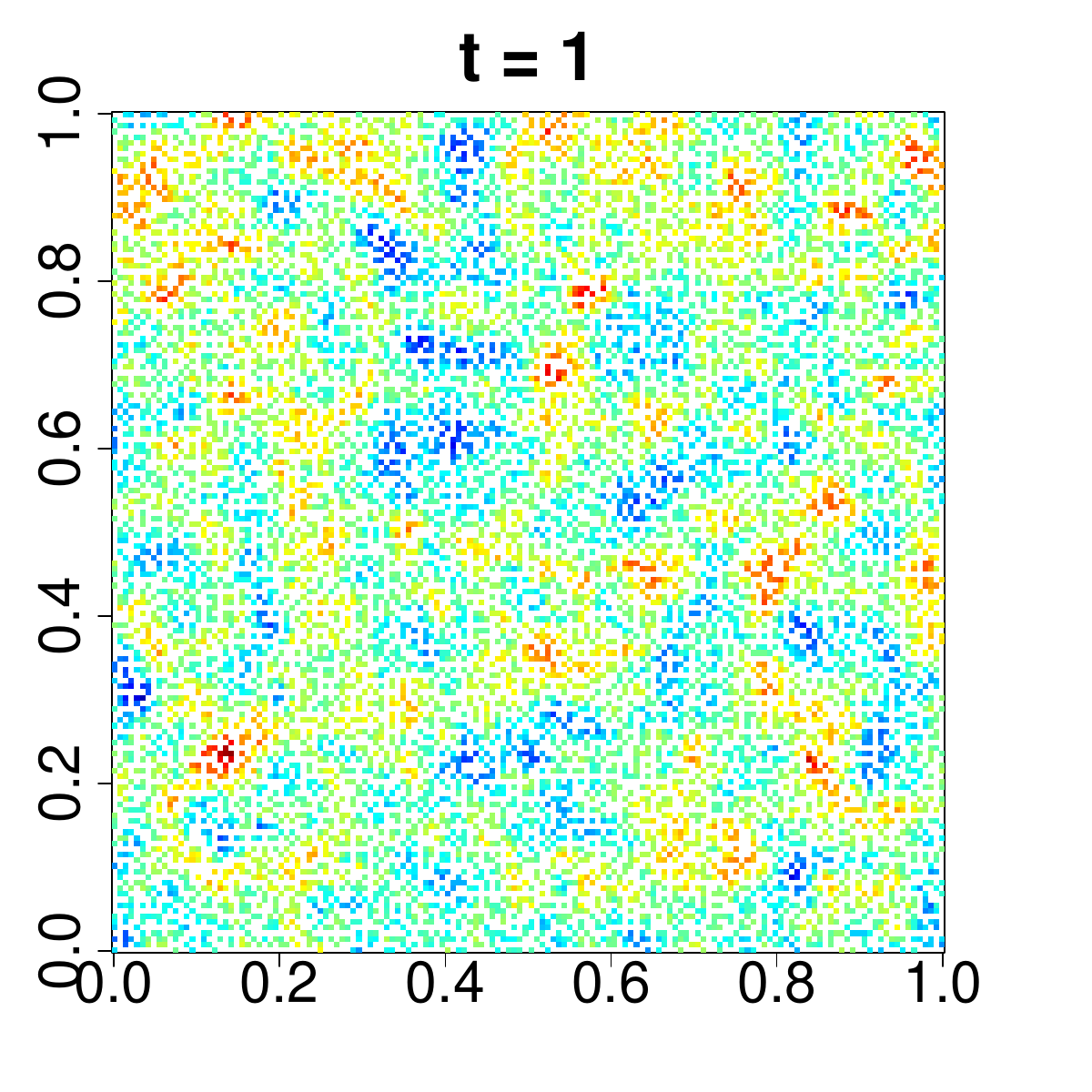}
     \end{subfigure}
     \hfill
     \begin{subfigure}[b]{0.32\linewidth}
         \centering
         \includegraphics[width=\linewidth]{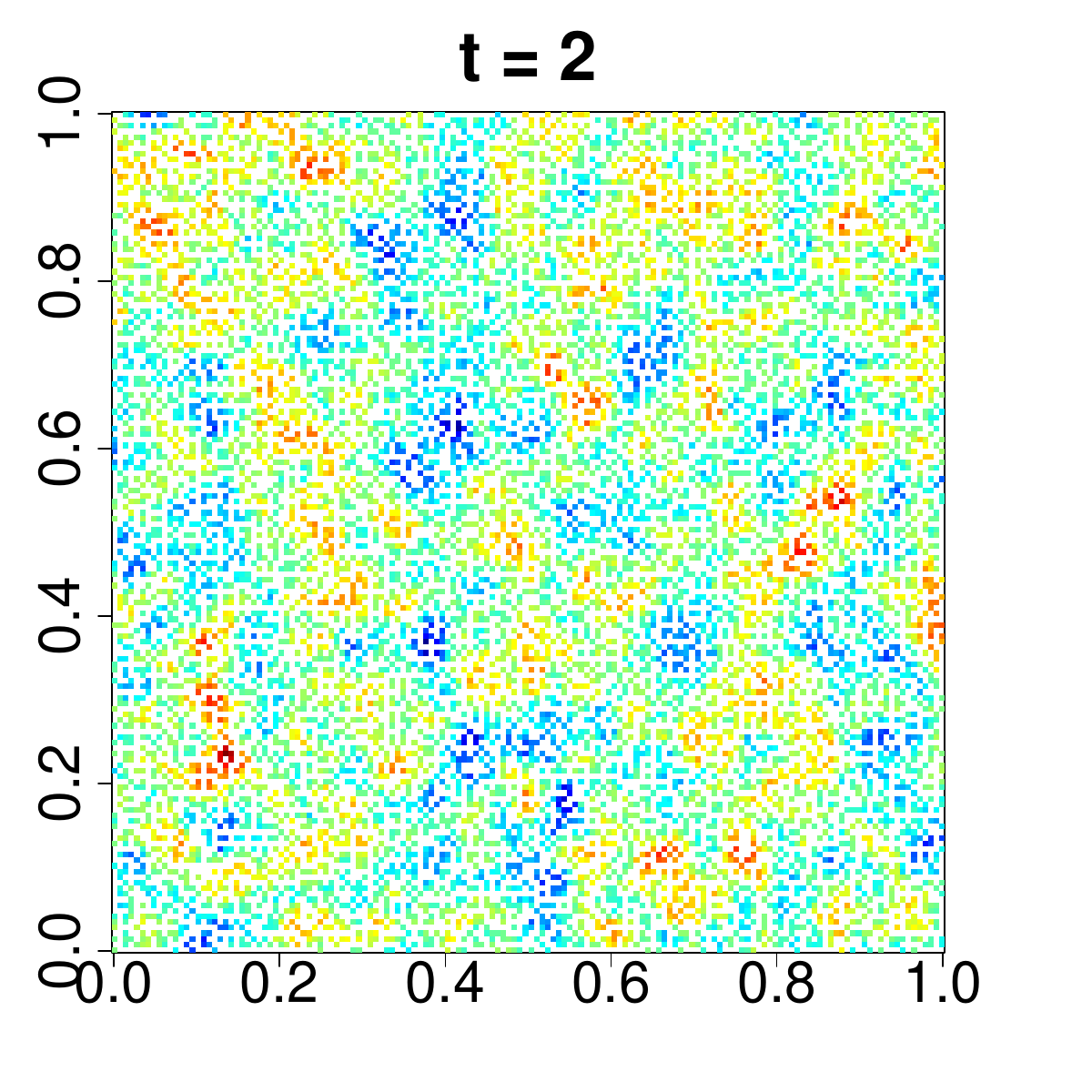}
     \end{subfigure}
     \hfill
     \begin{subfigure}[b]{0.32\linewidth}
         \centering
         \includegraphics[width=\linewidth]{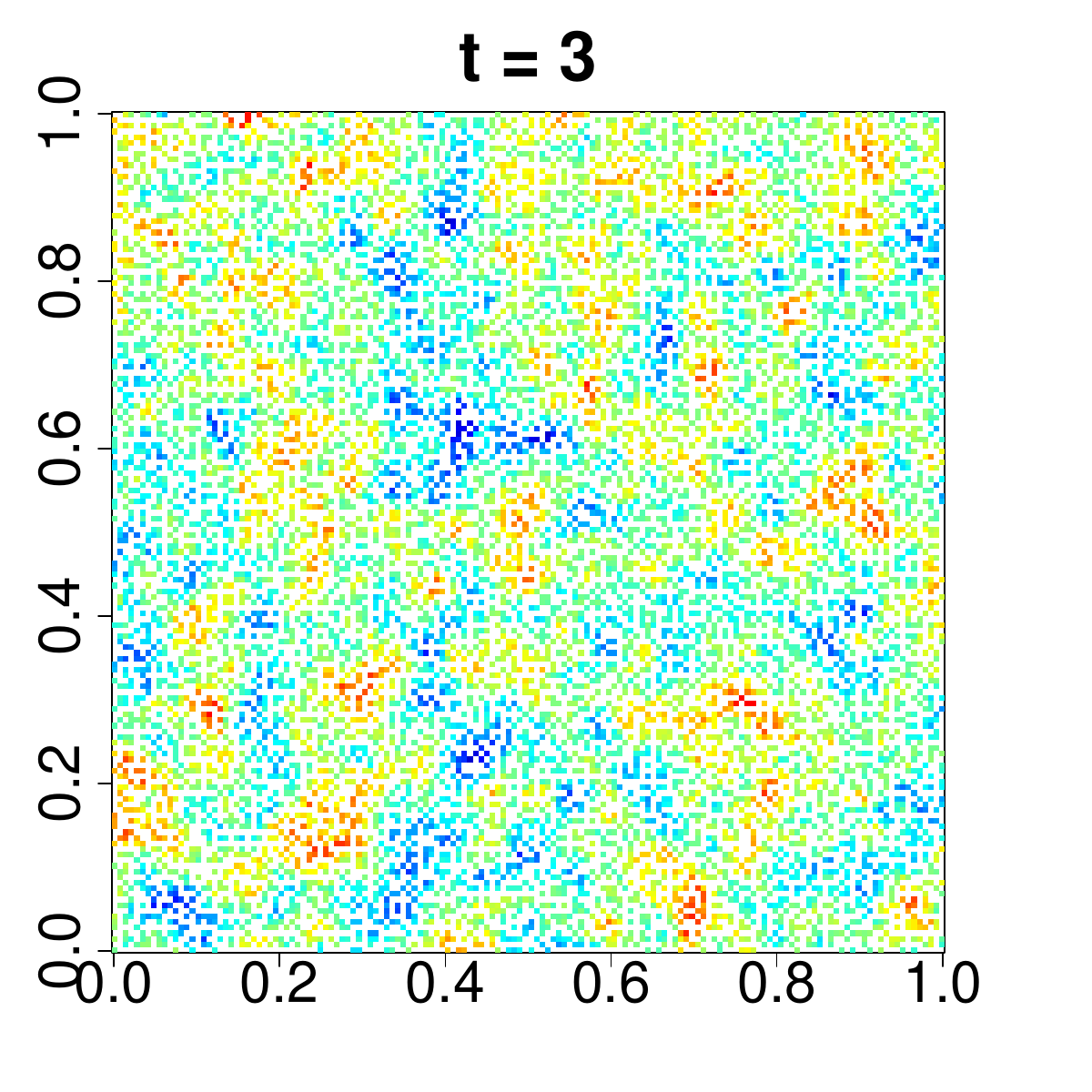}
     \end{subfigure}
     \hfill
     \vspace{-0.4cm}
    {\raggedright \Large \bfseries $\mathcal{D}_2:$ \par}
    \vspace{-0.2cm}
     \begin{subfigure}[b]{0.32\linewidth}
         \centering
         \includegraphics[width=\linewidth]{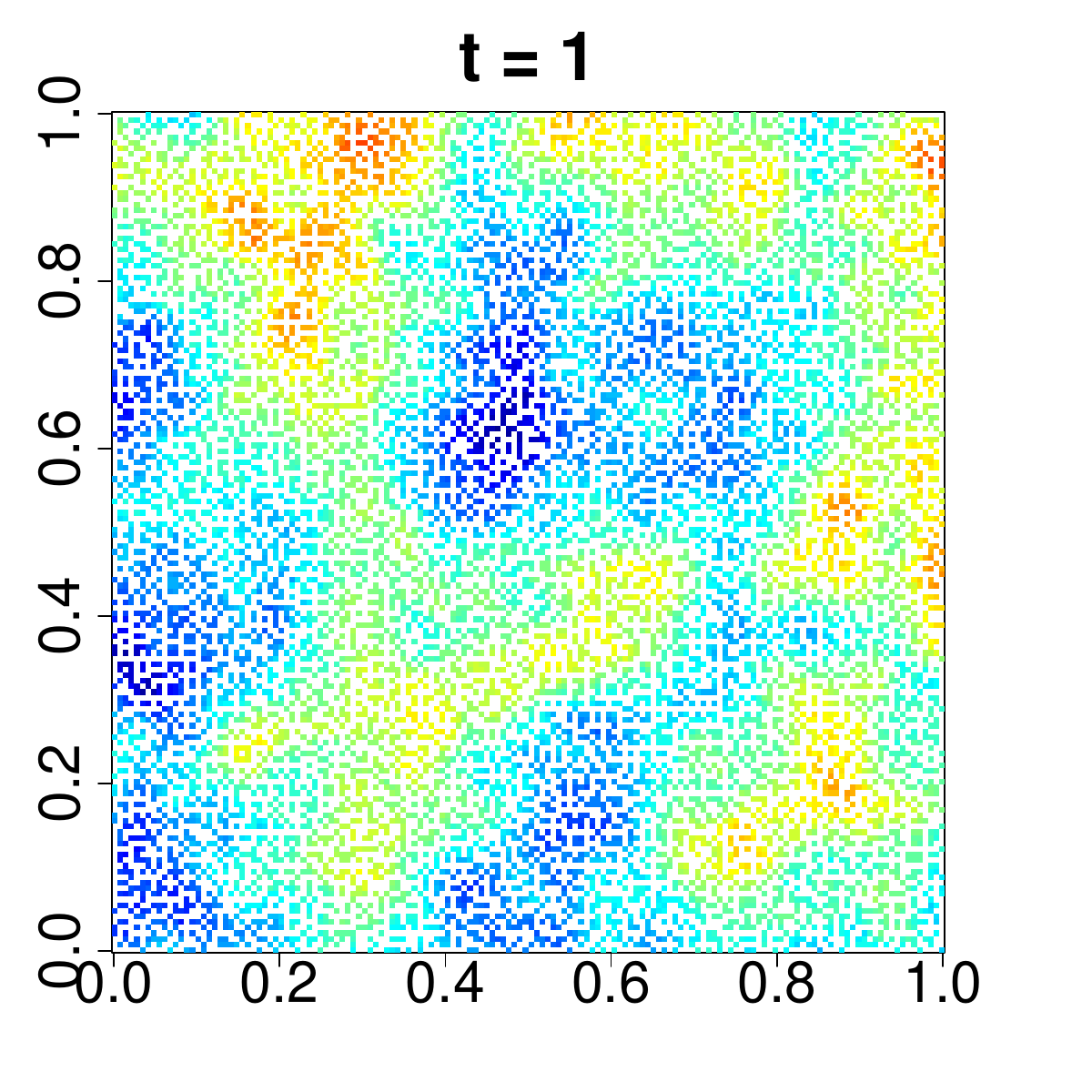}
     \end{subfigure}
     \hfill
     \begin{subfigure}[b]{0.32\linewidth}
         \centering
         \includegraphics[width=\linewidth]{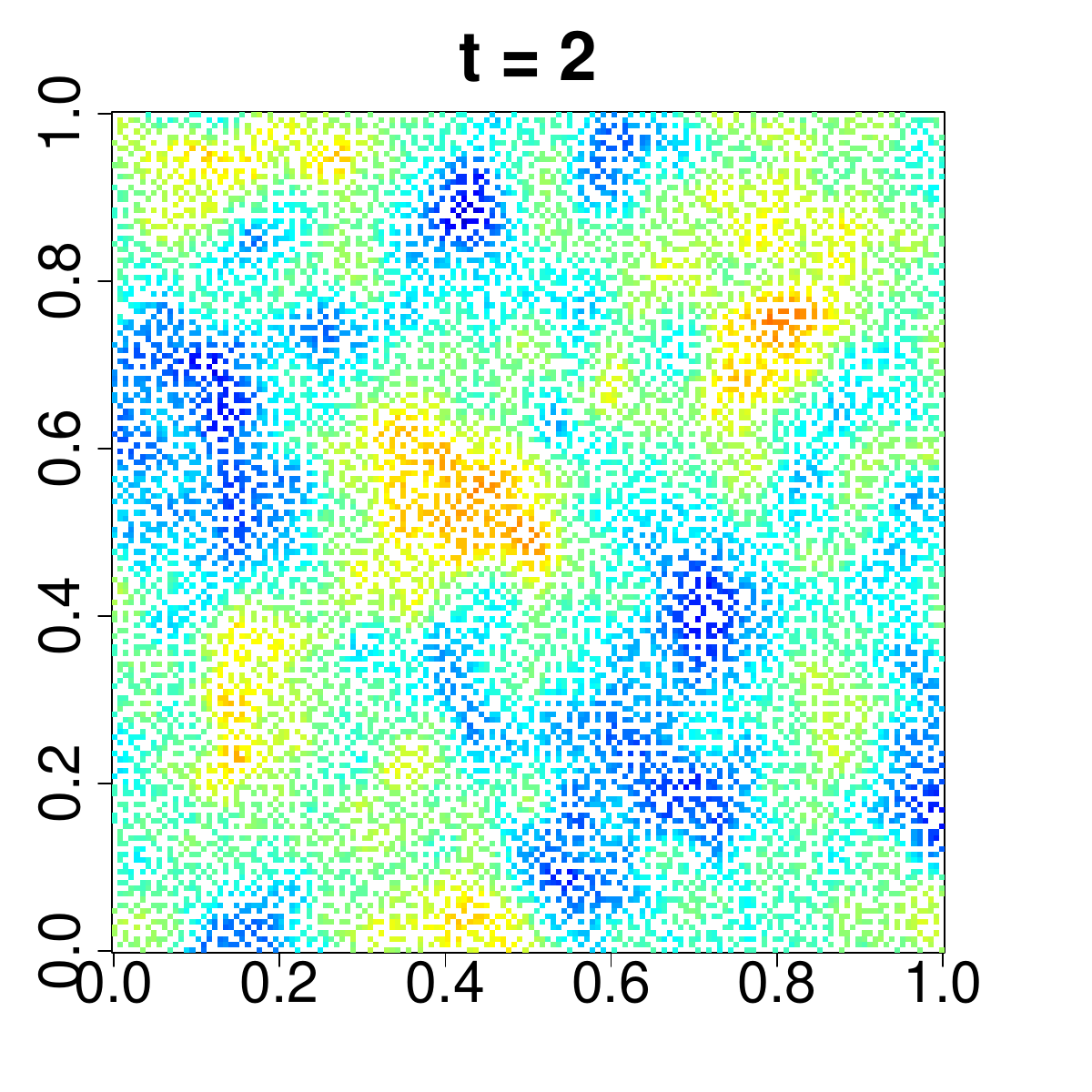}
     \end{subfigure}
     \hfill
     \begin{subfigure}[b]{0.32\linewidth}
         \centering
         \includegraphics[width=\linewidth]{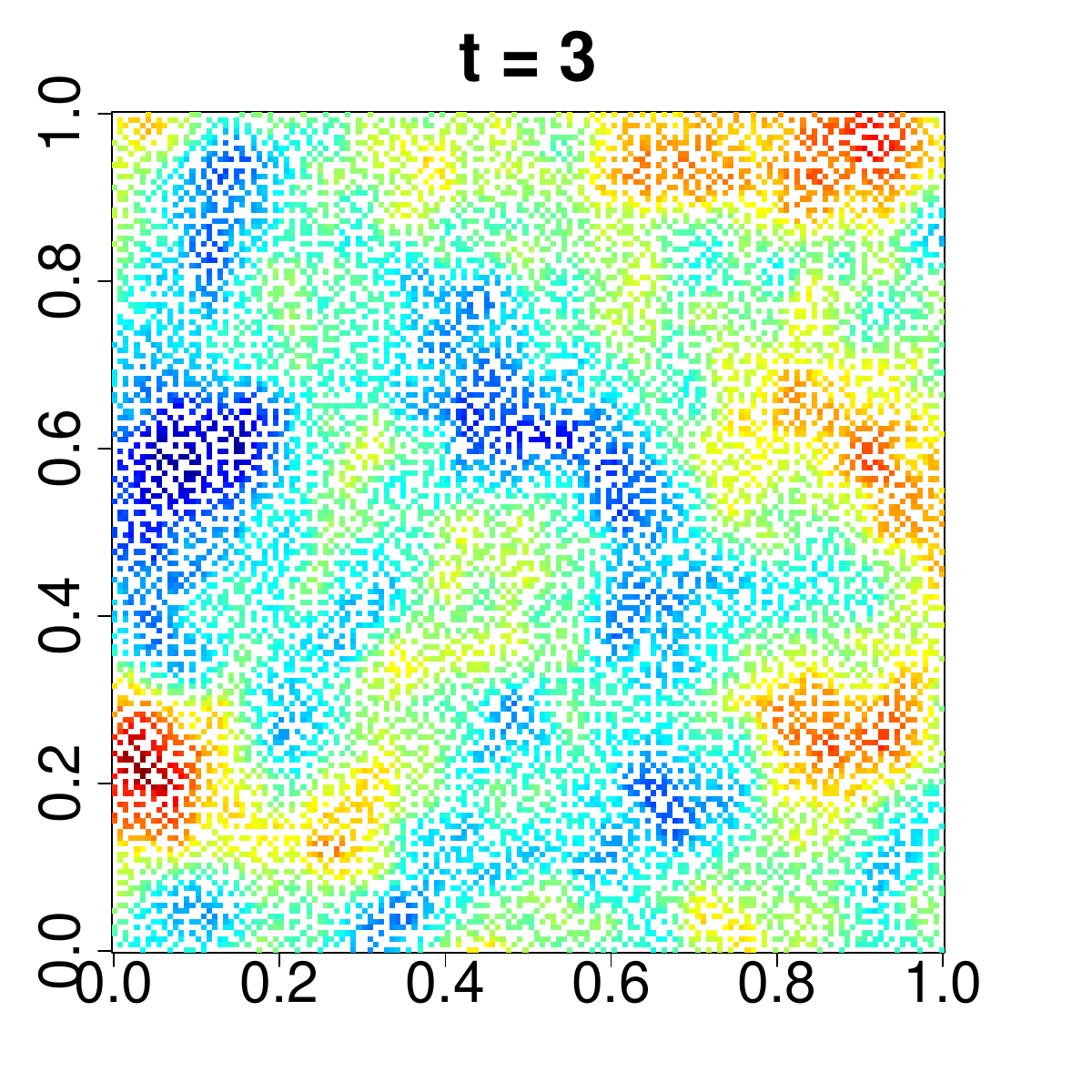}
     \end{subfigure}
     \hfill
     \vspace{-0.4cm}
     {\raggedright \Large \bfseries $\mathcal{D}_3:$ \par}
    \vspace{-0.2cm}
     \begin{subfigure}[b]{0.32\linewidth}
         \centering
         \includegraphics[width=\linewidth]{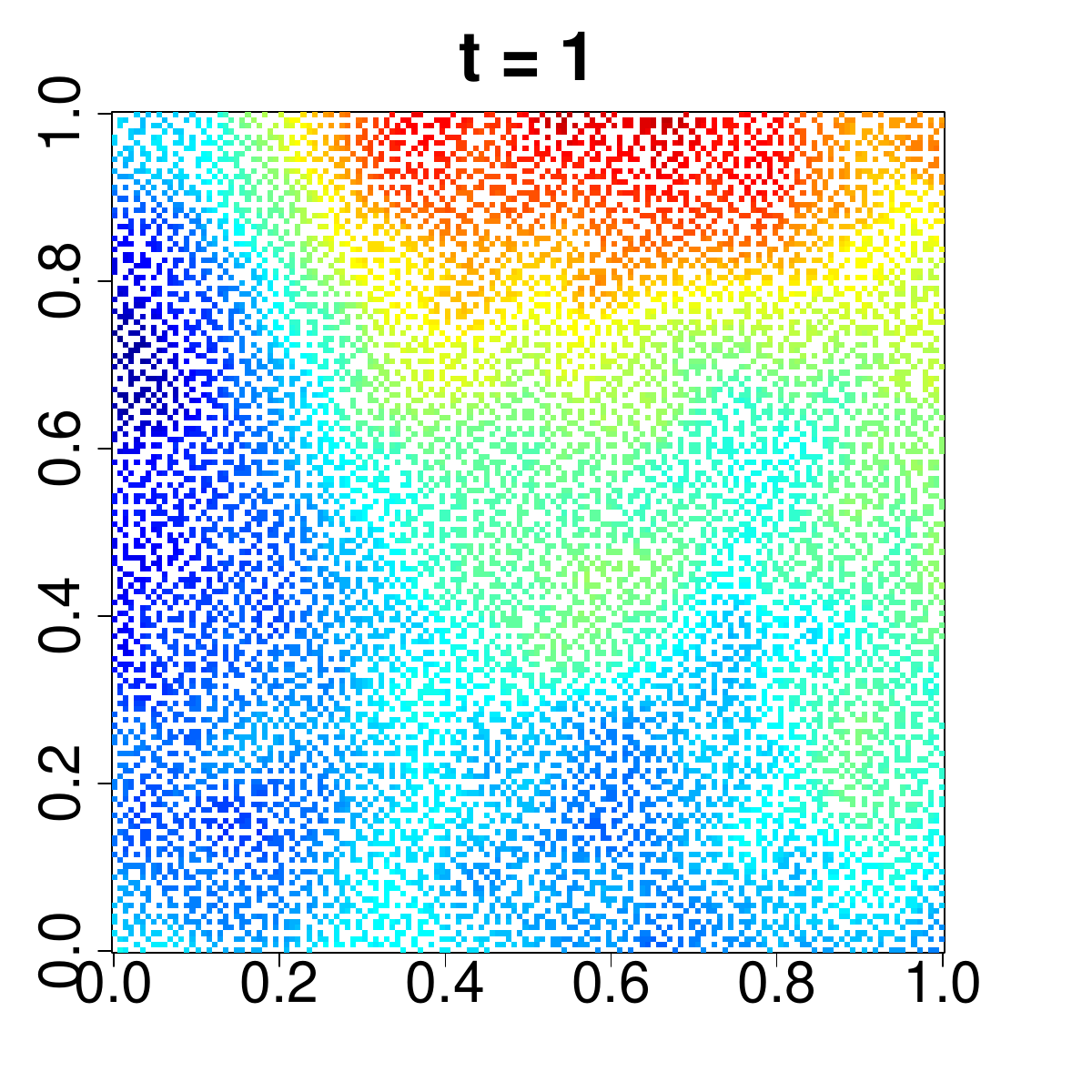}
     \end{subfigure}
     \hfill
     \begin{subfigure}[b]{0.32\linewidth}
         \centering
         \includegraphics[width=\linewidth]{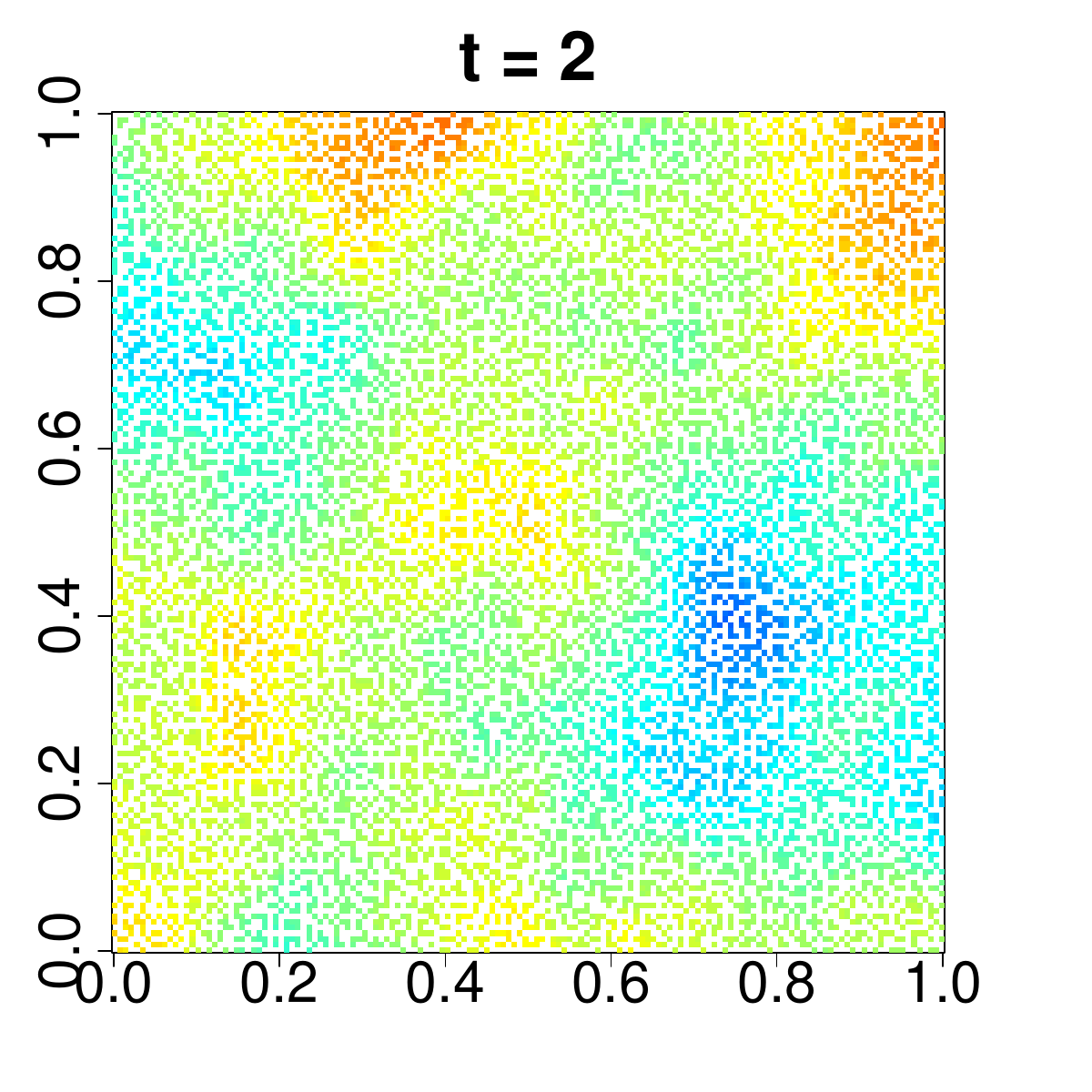}
     \end{subfigure}
     \hfill
     \begin{subfigure}[b]{0.32\linewidth}
         \centering
         \includegraphics[width=\linewidth]{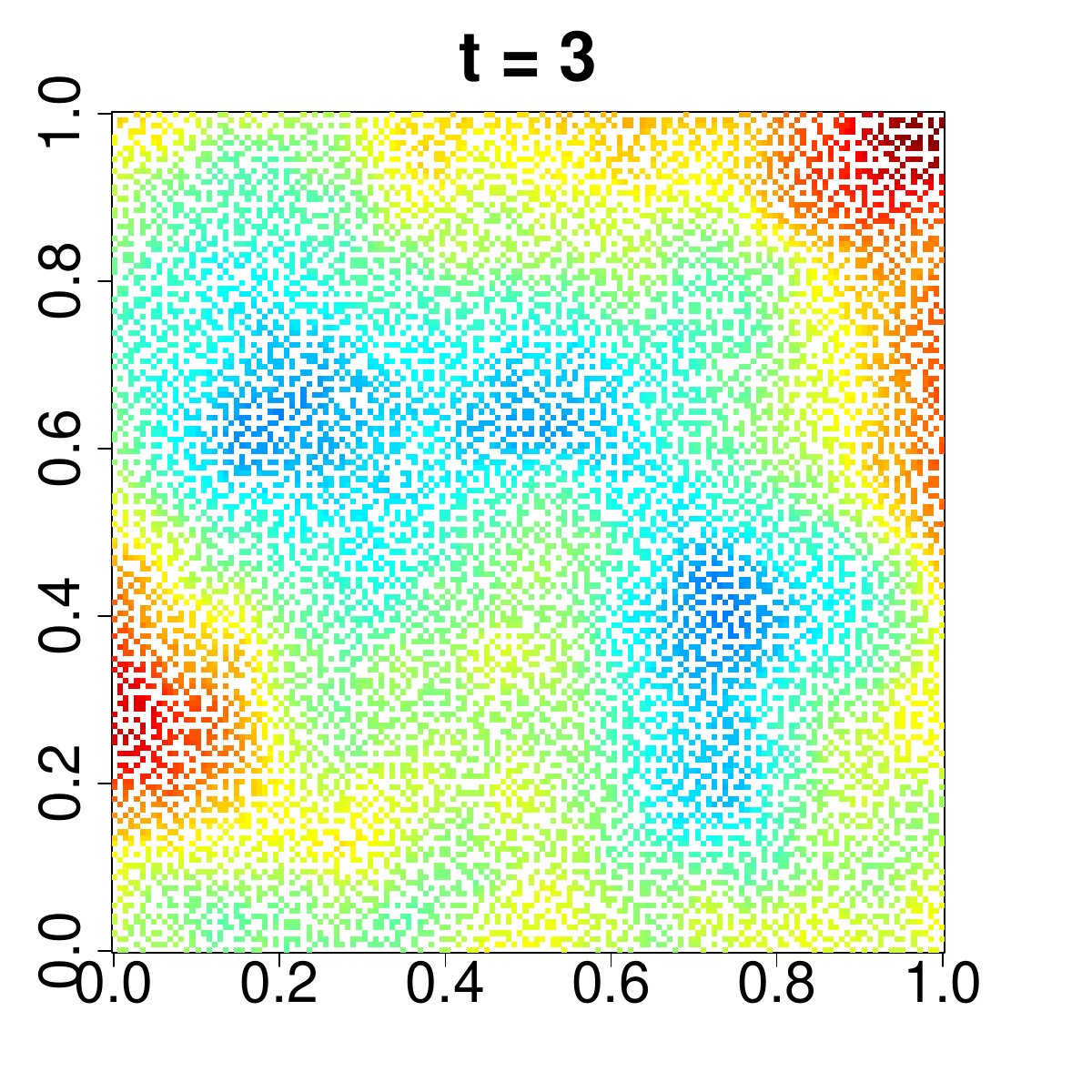}
     \end{subfigure}
     \caption{Synthetic space-time datasets $\mathcal{D}_1$ (top), $\mathcal{D}_2$ (middle), and $\mathcal{D}_3$ (bottom) for first three time points $t \in \{1,2,3\}$ with $10,000$ spatial locations per time point.}
        \label{fig:Bench_dat}
\end{figure}

%\section*{GPU acceleration}
This section reports runtimes for sub-competition~\textit{2b} of \citet{abdulah2022second}, comparing implementations with and without GPU acceleration (Table \ref{Relerrpara1}). It further includes additional experiments for more general Gaussian process settings, including models with non-Gaussian likelihoods fitted via Laplace approximation and high-dimensional inputs with automatic relevance determination (ARD) Matérn kernels. The results document runtime behavior across these settings and illustrate the impact of GPU acceleration under varying computational workloads, including scenarios in which neighbor search represents a substantial portion of the overall computational cost.

\begin{table}[ht!]
 \centering
 \begin{threeparttable}
 \caption{\label{Relerrpara1} Average end-to-end runtime (training + prediction, in seconds) for sub-competition \textit{2b} of \citet{abdulah2022second}, comparing CPU and GPU execution and the resulting speedups. }
 \begin{tabular}{ |p{.6cm}|p{0.4cm}||p{2cm}|p{2.cm}|p{2.cm}|p{2.4cm}|p{2.cm}|}
  \hline
   \multicolumn{2}{|c||}{\textbf{Time}}&\makecell[c]{\textbf{Vecchia} \\ Euclidean}& \makecell[c]{\textbf{Vecchia} \\ Correlation}& \makecell[c]{\textbf{FITC} \\ kMeans++ }& \makecell[c]{\textbf{FITC} \\ sts-kMeans++ }& \makecell[c]{\textbf{VIF} } \\
  \hline 
  \multicolumn{2}{|c||}{CPU}& \makecell[r]{1131 s}& \makecell[r]{18873 s} & \makecell[r]{2935 s}& \makecell[r]{2942 s} & \makecell[r]{69063 s}\\
  \hline 
  \multicolumn{2}{|c||}{GPU}& \makecell[r]{759 s}& \makecell[r]{1350 s} & \makecell[r]{743 s}& \makecell[r]{756 s} & \makecell[r]{5847 s}\\
  \hline 
  \multicolumn{2}{|c||}{Speedup}& \makecell[r]{1.49}& \makecell[r]{13.98} & \makecell[r]{3.95}& \makecell[r]{3.89} & \makecell[r]{11.81}\\
  \hline 
 \end{tabular}
     \end{threeparttable}
 \end{table}

Table~\ref{tab:gpu_speedup_operations} decomposes the GPU acceleration into the main computational operations for each approximation. 

\begin{table}[ht!]
\centering
\caption{GPU speedups (CPU runtime divided by GPU runtime) for the main computational operations across increasing sample sizes. Vecchia and VIF involve negative log-likelihood (NEGLL), gradient, and neighbor-search computations, whereas FITC does not require Vecchia neighbor searches.}
\label{tab:gpu_speedup_operations}
%\resizebox{\textwidth}{!}{
{\begin{tabular}{llccccc}
\toprule
Method & Operation 
& 10k & 100k & 200k & 500k & 1,000k \\
\midrule

\multirow{3}{*}{Vecchia (Euclidean)}
& NEGLL            & 1.0  & 1.1 & 1.1 &  1.1 &  1.1\\
& Gradient  &  1.0  & 1.1 & 1.1 &  1.1 &  1.2 \\
& Neighbor search  & 1.0  & 1.1 & 1.2 &  1.3 &  1.4\\
\midrule

\multirow{3}{*}{Vecchia (Correlation)}
& NEGLL            & 1.0  & 1.1 & 1.1 &  1.1 &  1.1\\
& Gradient  &  1.0  & 1.1 & 1.1 &  1.1 &  1.2 \\
& Neighbor search  & 1.7 &  3.1 & 6.9 & 10.8 & 17.5 \\
\midrule

\multirow{2}{*}{FITC}
& NEGLL            & 1.6 &  2.0 & 2.3 & 2.5 & 2.7 \\
& Gradient         & 2.4 &  3.0 & 3.3 & 3.4 & 3.5\\
\midrule

\multirow{3}{*}{VIF}
& NEGLL            & 1.7 &  2.1 & 2.4 & 2.6 & 3.2\\
& Gradient         & 2.8 &  3.5 & 3.9 & 4.3 & 4.5\\
& Neighbor search  & 2.1 &  4.1 & 5.3 & 7.2 & 14.1\\
\bottomrule
\end{tabular}
}
\end{table}

\begin{figure}[ht!]
    \centering
        
    \includegraphics[width=0.8\linewidth]{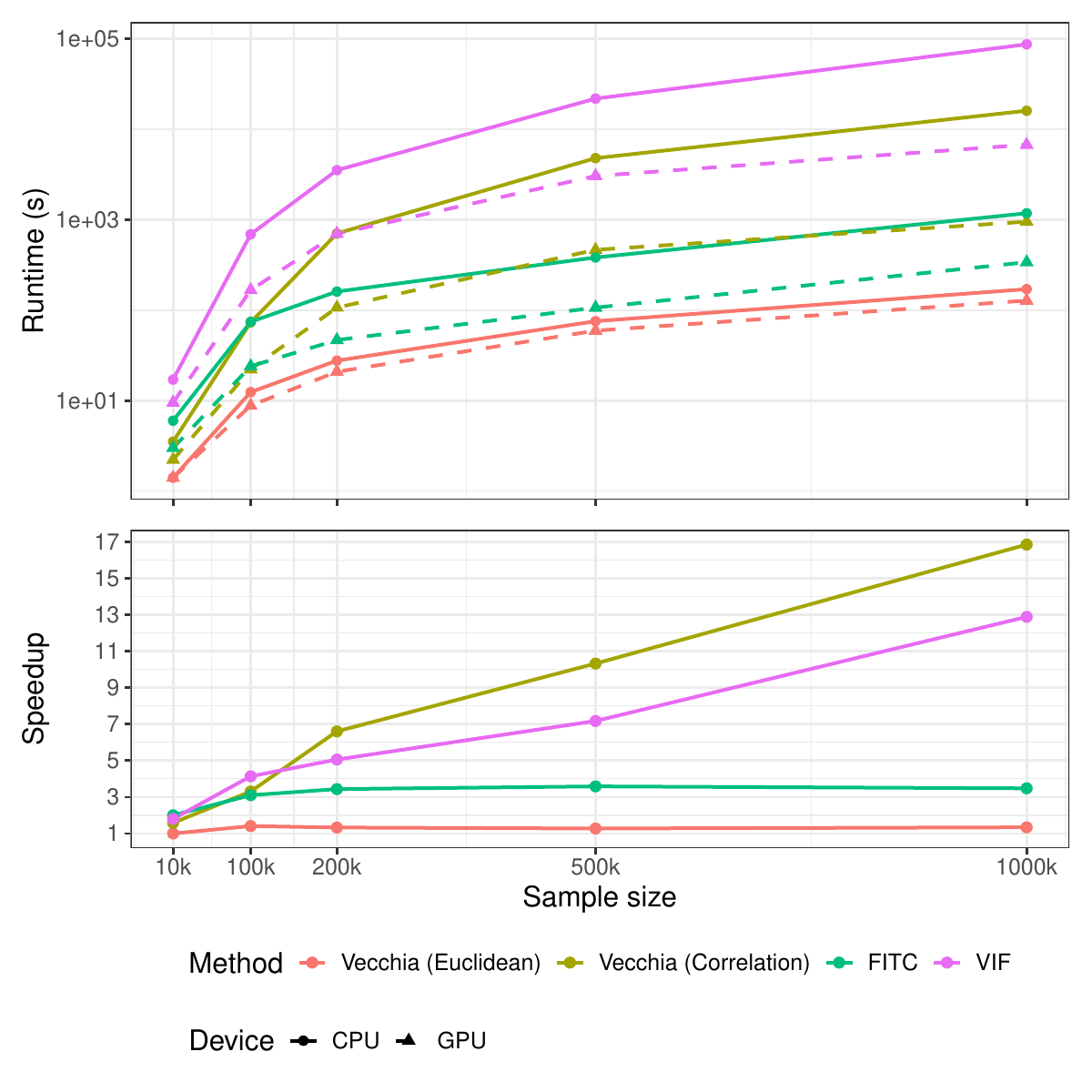}
    
    \caption{Runtimes in seconds (log-scale) and GPU speedups for a single parameter-optimization iteration of a latent GP model with a non-Gaussian likelihood, consisting of two Laplace-approximated evaluations of the negative log-marginal likelihood, gradient computations for all covariance parameters, and two Vecchia neighbor searches. Results are shown for varying sample sizes and for the Vecchia, FITC, and VIF approximations.}\label{fig:RWRGPU1}
\end{figure}

Figure~\ref{fig:RWRGPU1} reports runtimes and speedups for a single optimization iteration for GP models with a non-Gaussian likelihoods, comprising two marginal log-likelihood evaluations, gradient computations, and two Vecchia neighbor searches, across sample sizes and approximation strategies. 

\begin{figure}[!htbp]
    \centering
        
    \includegraphics[width=0.8\linewidth]{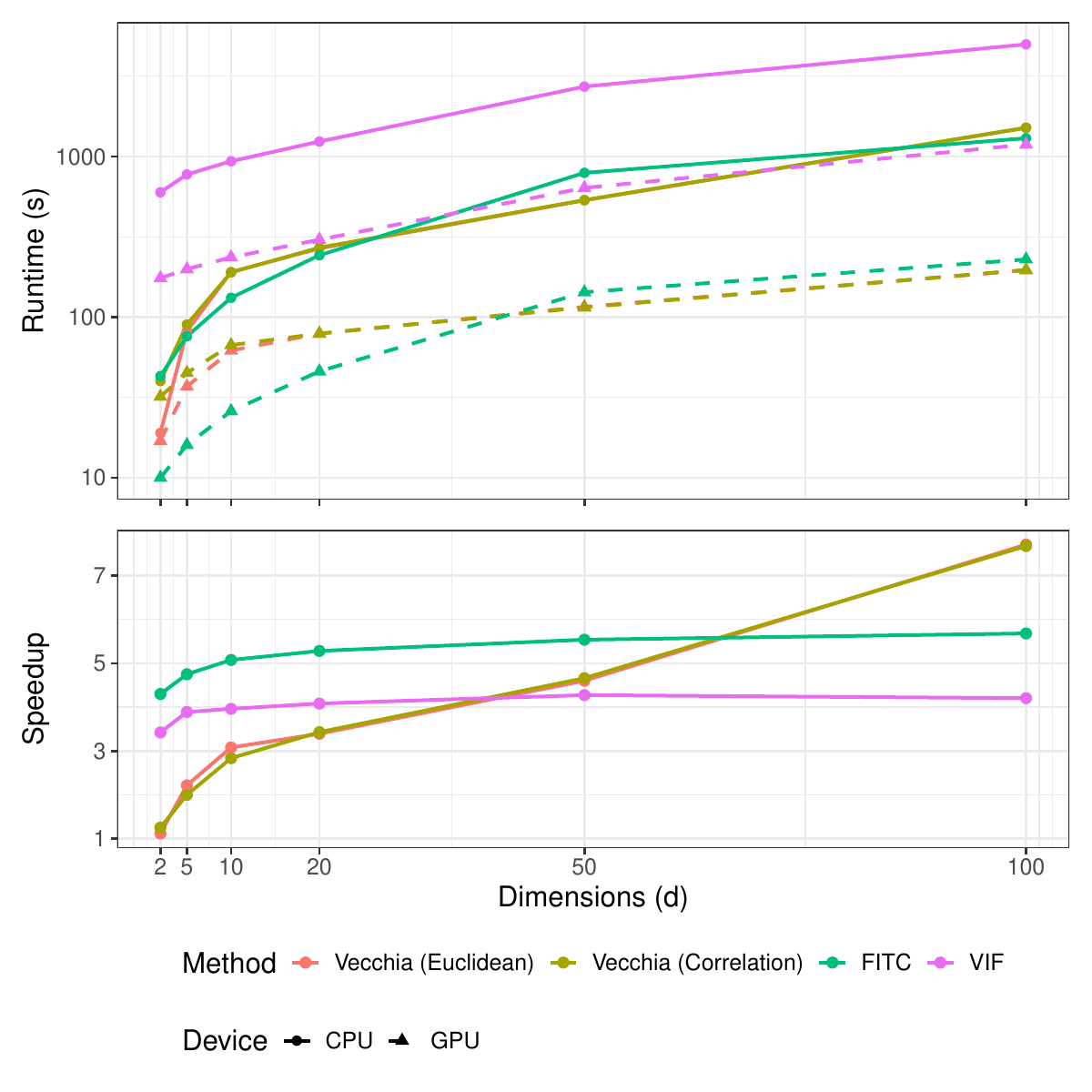}
    
    \caption{Runtimes in seconds (log-scale) and GPU speedups for a single parameter-optimization iteration of a GP model with $d$-dimensional inputs and an ARD Matérn covariance function, consisting of two negative log-likelihood evaluations, gradient computations for all covariance parameters, and two Vecchia neighbor searches. Results are shown for varying sample sizes and for the Vecchia, FITC, and VIF approximations with a fixed sample size of 200,000.}\label{fig:RWRGPUd}
\end{figure}

Figure~\ref{fig:RWRGPUd} reports runtimes and speedups for a single optimization iteration for high-dimensional
inputs modeled using ARD Matérn kernels, comprising two negative log-likelihood evaluations, gradient computations, and two Vecchia neighbor searches, across various input dimensions and approximation strategies.

\clearpage

\section{Real-world application: Additional plots and tables}\label{App:PandT}

\subsection{Short-term prediction of daily maximum temperature}
Additional plots, tables, and results for the short-term prediction of daily maximum temperature. 

\begin{figure}[!htbp]
    \centering
    \includegraphics[width=\linewidth]{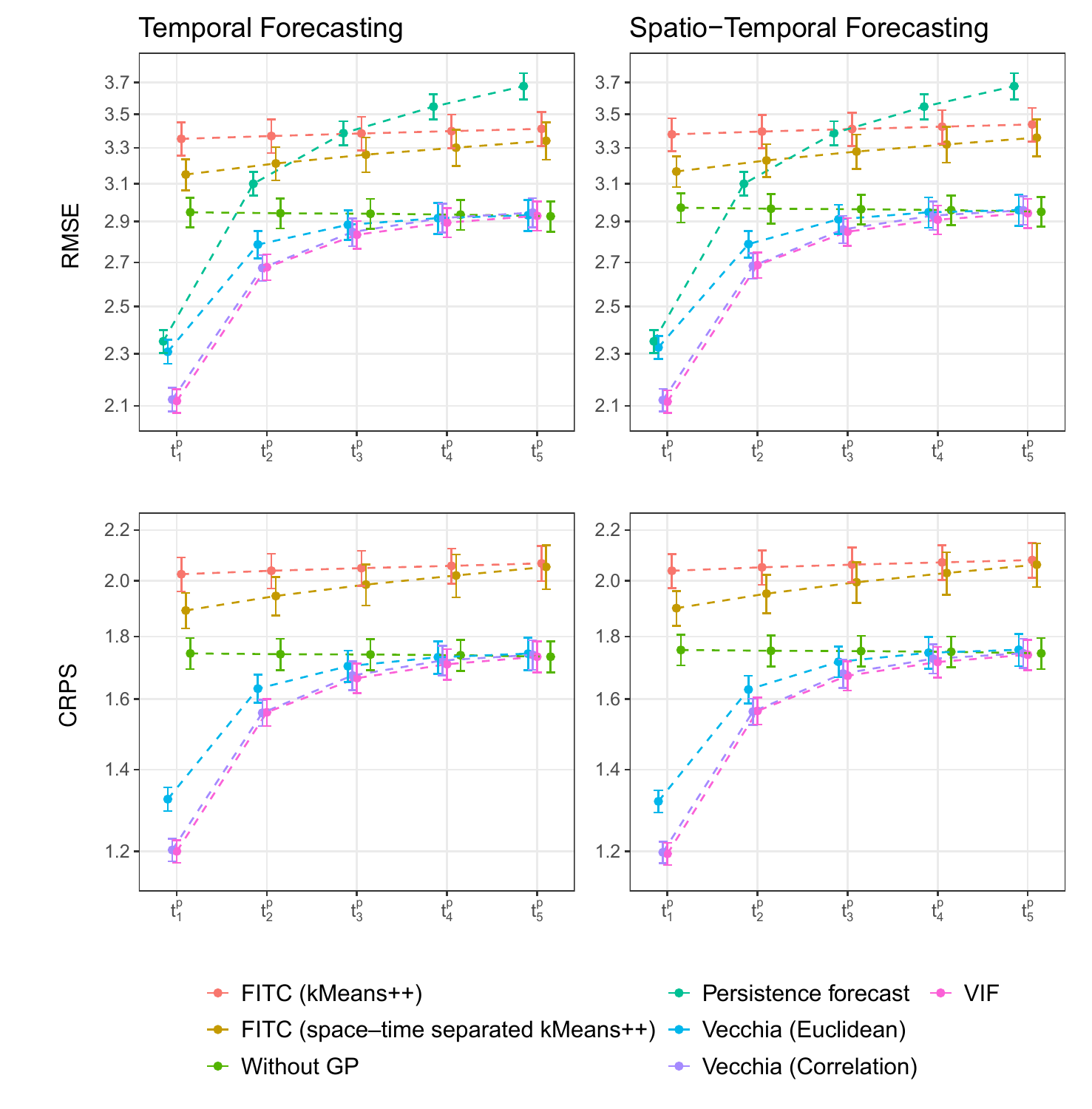}
    \caption{Day-wise RMSE and CRPS (log scale) with error bars (mean $\pm$ one standard error) for one- to five-day temperature forecasts under the temporal, and the spatio-temporal forecasting settings. Methods include the Vecchia approximation with Euclidean- and correlation-based neighbors ($m_v=30$), FITC with standard and space-time separated kMeans++ inducing points ($m=500$), and VIF ($m_v=30$, $m=500$). A persistence forecast and a fixed-effects-only regression model (without GP) are included for comparison.}
    \label{fig:RWR0}
\end{figure}

Figure \ref{fig:RWR0} illustrates additional results evaluating the models for temporal-only predictions at stations for which data has been observed.

%\newpage

\begin{figure}[ht!]
    \centering
    \includegraphics[width=\linewidth]{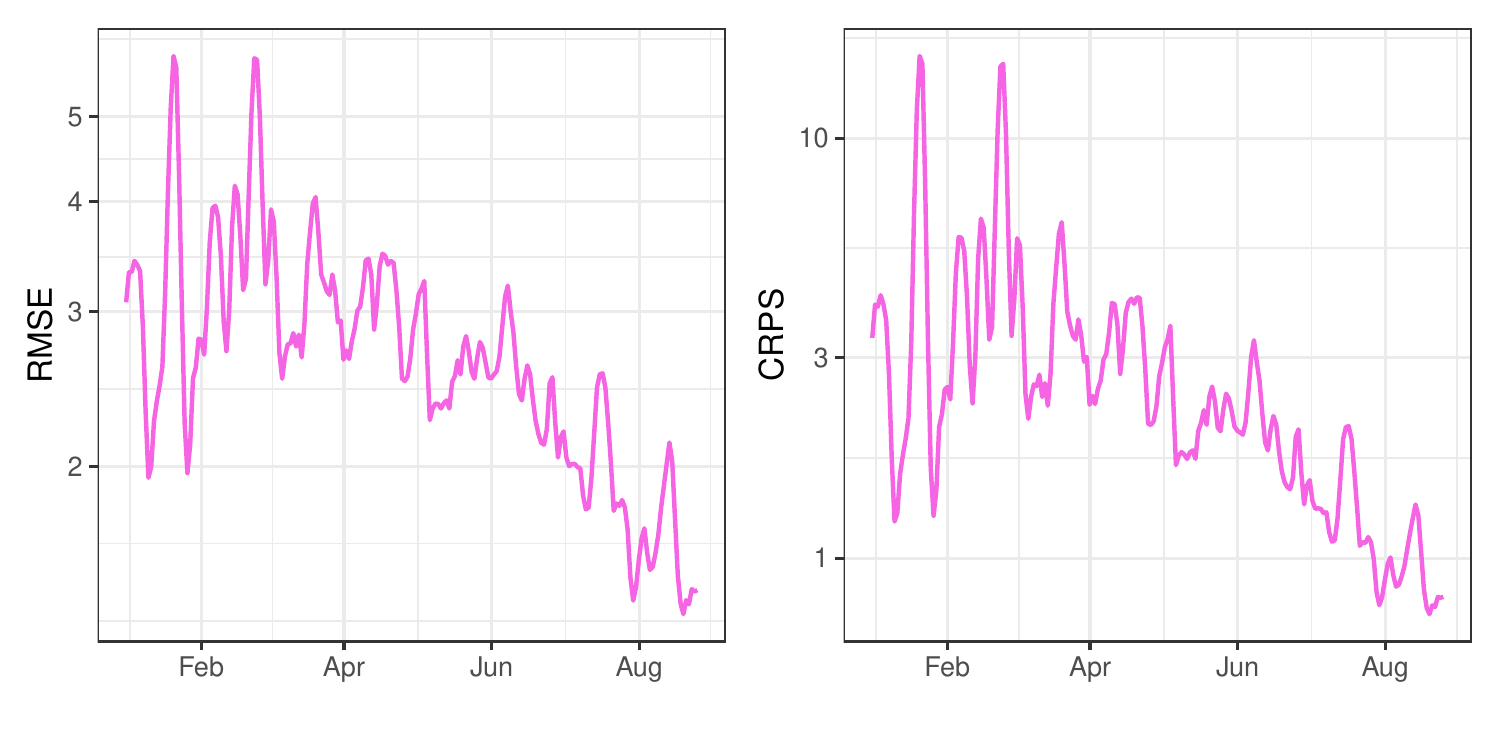}
    \caption{Evolution of overall five-day temperature forecast error (RMSE and CRPS, log scale) for the VIF approximation throughout 2025.}
    \label{fig:RWR12}
\end{figure}

Figure \ref{fig:RWR12} displays the forecast error for the VIF approximation throughout 2025. The accuracy improves over time as additional data accumulate, with notably strong performance in late summer, when temperature fields evolve more smoothly than in winter and spring.

\begin{figure}[ht!]
\centering
\includegraphics[width=\linewidth]{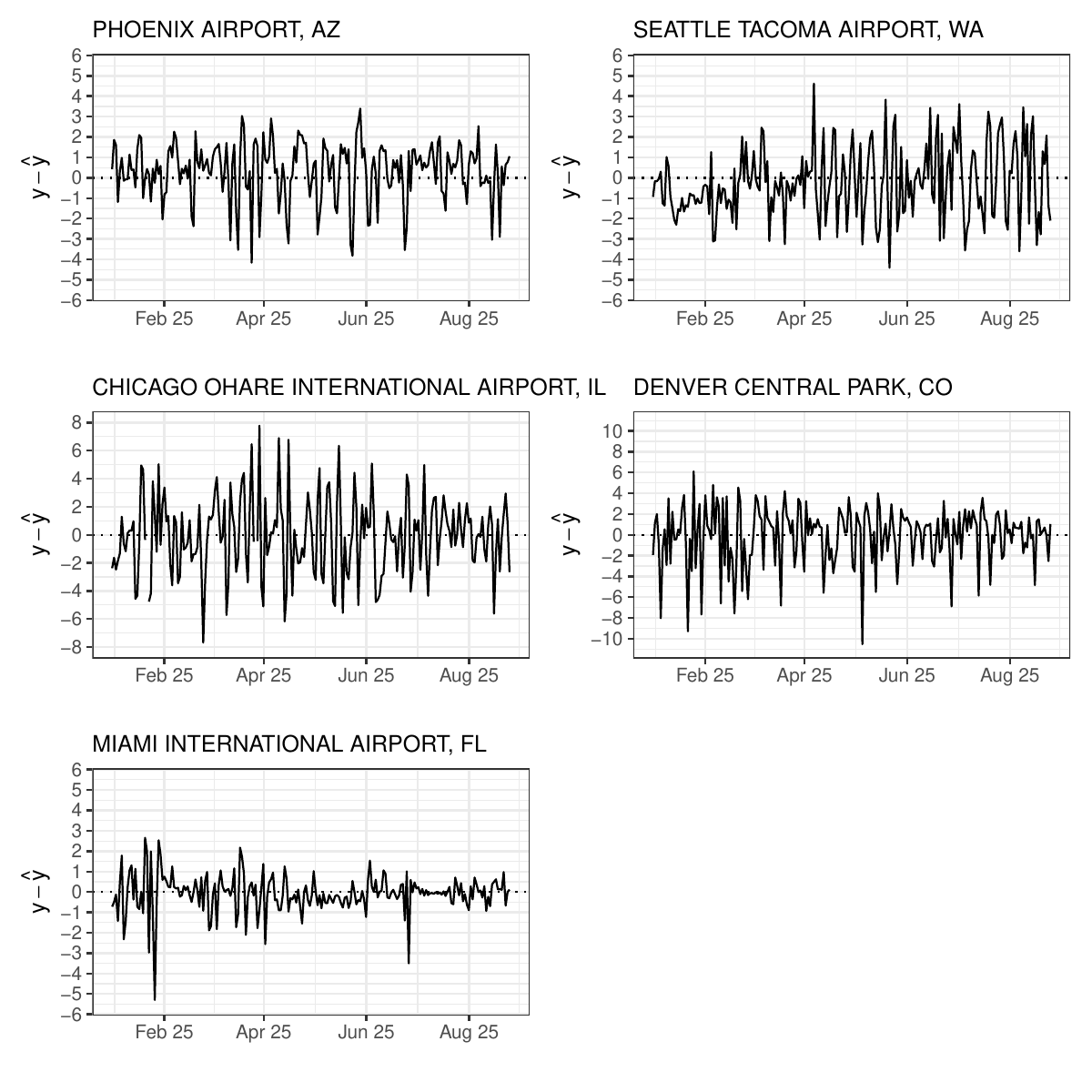}
\caption{One-day-ahead maximal temperature forecast residuals under the VIF approximation for five representative GHCNd stations across the full prediction period.}\label{fig:RWT12345}
\end{figure}

Figure \ref{fig:RWT12345} displays one-day-ahead temperature residuals under the VIF approximation for selected stations. The residuals show no systematic drift over the year, indicating that the VIF-approximated GP model successfully captures both the seasonal cycle and short-lead temperature dynamics. Residual magnitudes vary across stations, reflecting local variability in daily temperature fluctuations, but no persistent patterns of under- or over-prediction are evident.

%\begin{figure}[ht!]
 %   \centering
    %\includegraphics[width=\linewidth]%{figures/Real_World_Temp_aug_method.pdf}
    %\ifdefined\ARXIV
  %\includegraphics[width=\linewidth]{figures/Real_World_Temp_aug_method.png}
%\else
  %\includegraphics[width=\linewidth]{figures/Real_World_Temp_aug_method.pdf}
%\fi
    %\caption{Mean absolute error and mean CRPS of one-day-ahead forecasts for August 2025, computed per station for the Vecchia (Euclidean and correlation-based; $m_v=30$), FITC (standard and space-time separated kMeans++; $m=500$), and VIF ($m_v=30$, $m=500$) approximations.}
    %\label{fig:RWRA}
%\end{figure}

\begin{figure}[ht!]
    \centering
    \ifdefined\ARXIV
  \includegraphics[width=0.65\linewidth]{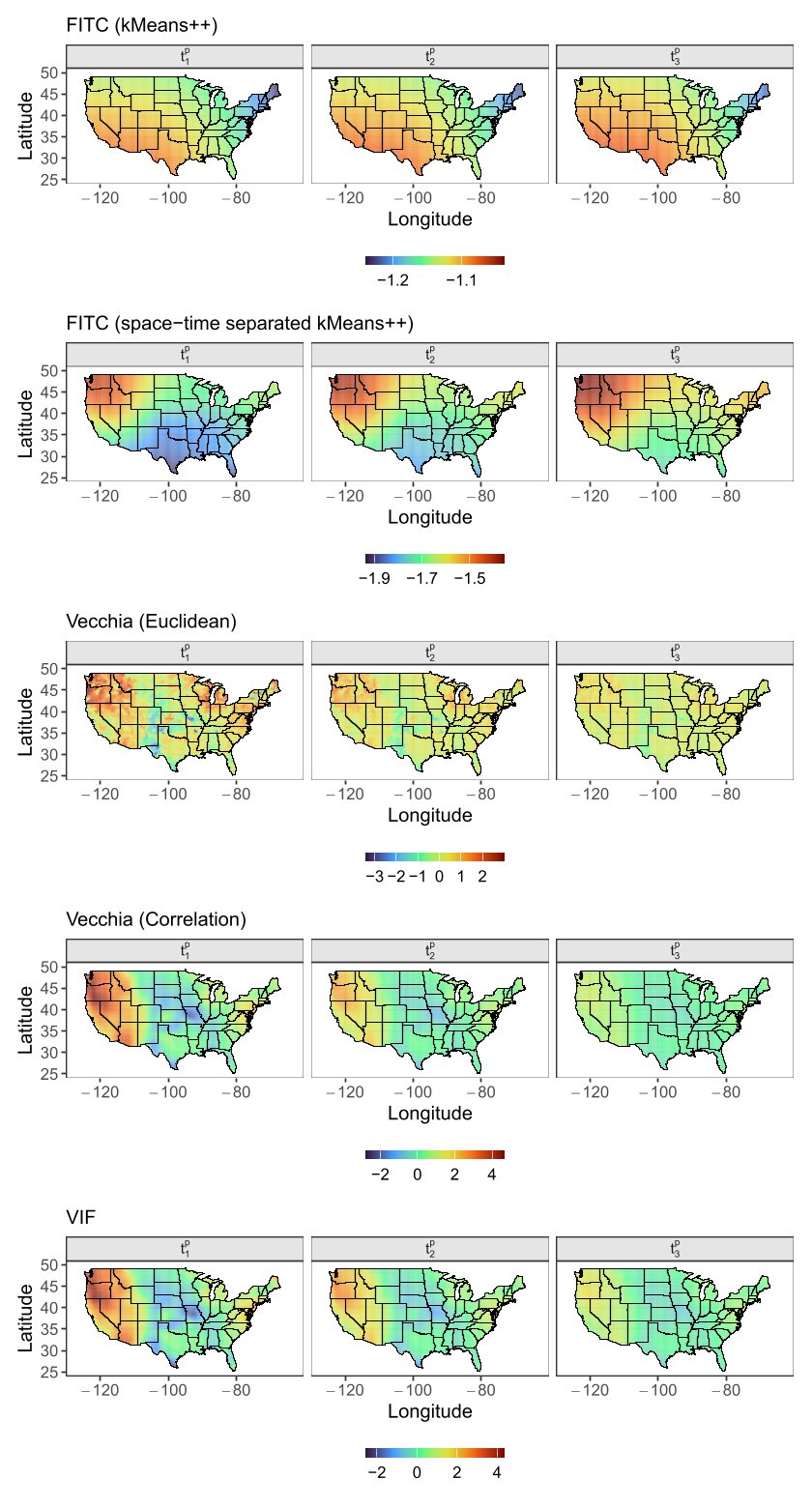}
\else
  \includegraphics[width=0.65\linewidth]{figures/Real_World_Temp_GP.pdf}
\fi
    \caption{Estimated GP random effects at the final re-estimation step for the temperature data across the conterminous United States for multiple temporal lags, shown for the Vecchia (Euclidean and correlation-based), FITC (standard and space-time separated kMeans++), and VIF approximations.}
    \label{fig:RWRA_GP}
\end{figure}

Moreover, Figure \ref{fig:RWRA_GP} displays the estimated GP random effects across the conterminous United States for several temporal lags, illustrating how each approximation captures the underlying spatial and temporal dependence structure.

\begin{figure}[ht!]
    \centering
    \includegraphics[width=0.4\linewidth]{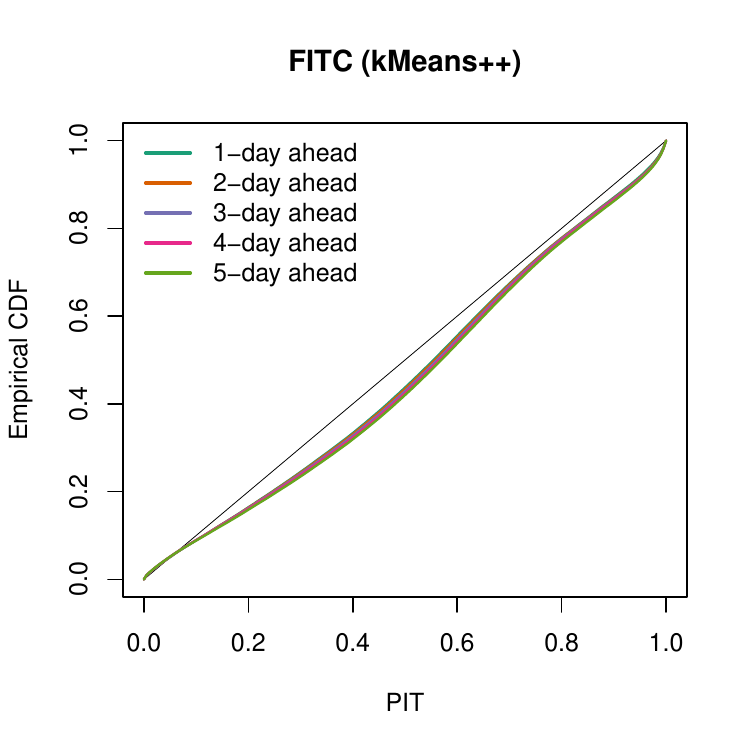}
    \includegraphics[width=0.4\linewidth]{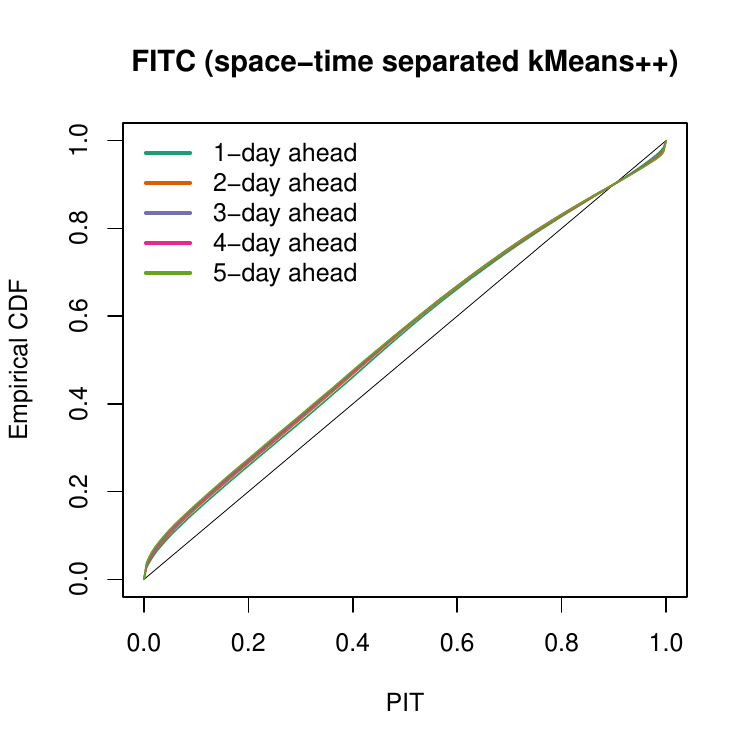}
\vspace{-2em}
    \includegraphics[width=0.4\linewidth]{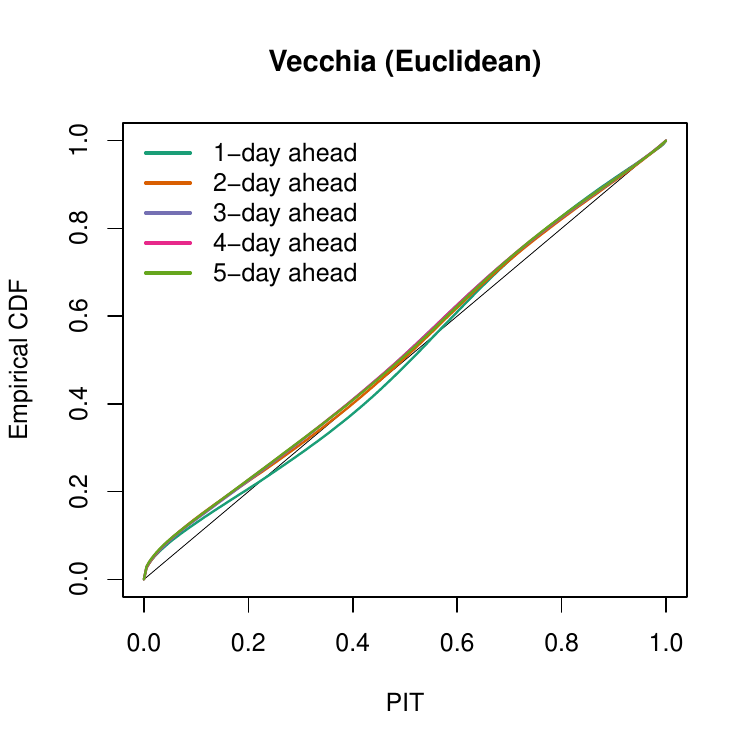}
    \includegraphics[width=0.4\linewidth]{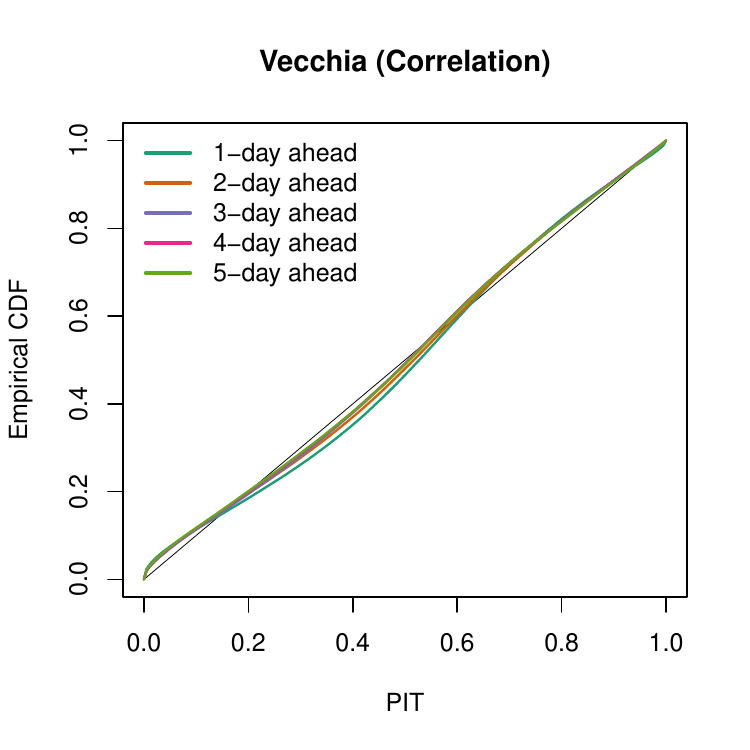}
\vspace{-2em}
    \includegraphics[width=0.4\linewidth]{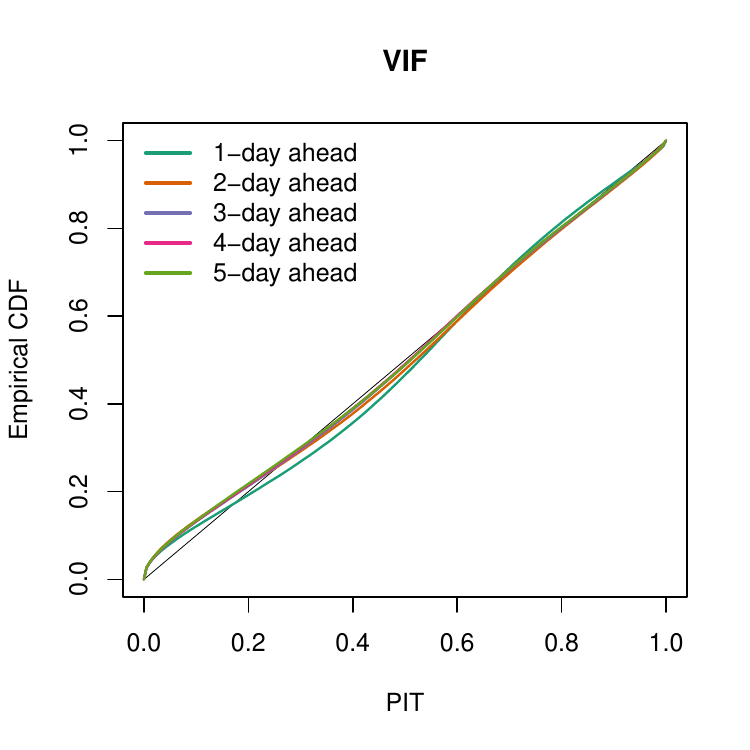}

    \caption{PIT reliability diagrams for temperature forecasts using VIF, FITC, and Vecchia approximations. Curves show uniform(0,1) QQ-plots for the empirical PIT CDFs for forecast instances pooled over all stations and shown separately for each lead time from one to five days. The black diagonal indicates perfect calibration.}
    \label{fig:PIT}
\end{figure}

In addition, we check the calibration of the forecasts using the probability integral transform (PIT). Uniform(0,1) QQ-plots of PIT transformed values. The PIT reliability diagrams in Figure \ref{fig:PIT} show generally good calibration for the VIF and Vecchia approximations across all lead times, with PIT CDFs closely tracking the diagonal uniform reference. Calibration remains relatively stable from one- to five-day lead times, with only minor dispersion differences across horizons. Differences between the Vecchia (Euclidean and correlation-based) and VIF approaches are small at the aggregate level. In contrast, both FITC variants exhibit noticeably poorer calibration across all lead times, with systematic deviations from the diagonal indicative of bias and misrepresentation of uncertainty. Residual departures from uniformity across all methods point to remaining lead-dependent dispersion errors and mild tail effects.

\clearpage

\subsection{Short-term prediction of precipitation}
Additional plots, tables, and results for the short-term prediction of precipitation.

\begin{figure}[ht!]
    \centering
    \includegraphics[width=\linewidth]{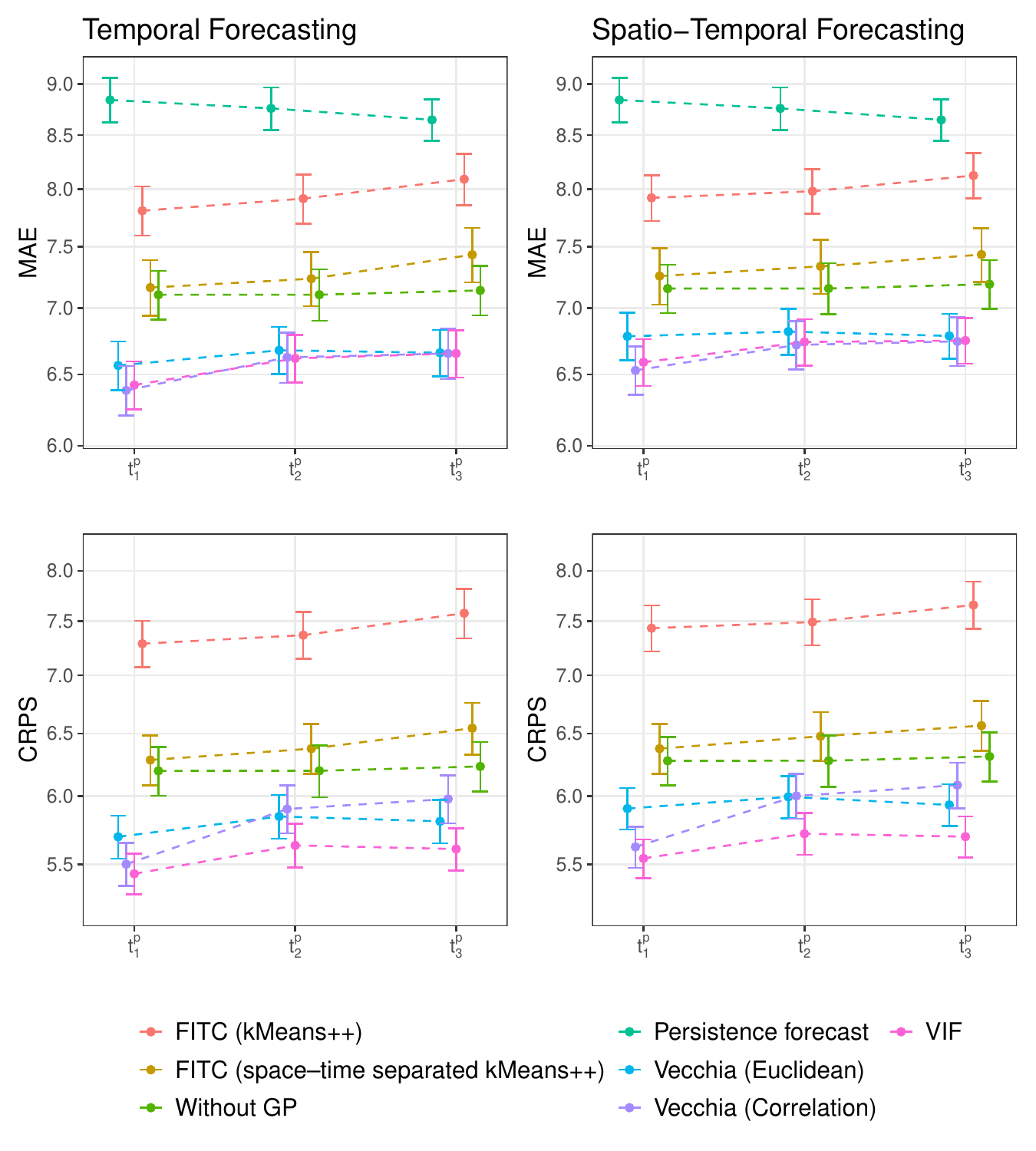}
    \caption{Day-wise MAE and CRPS (log scale) with error bars (mean $\pm$ one standard error) for one- to three-day precipitation amount forecast under the temporal, and the spatio-temporal forecasting
settings. Methods include the Vecchia approximation with Euclidean- and correlation-based neighbors ($m_v=30$), FITC with standard and space-time separated kMeans++ inducing points ($m=500$), and VIF ($m_v=30$, $m=500$). A persistence forecast and a fixed-effects-only regression model (without
GP) are included for comparison.}
    \label{fig:RWRnG}
\end{figure}

Figure \ref{fig:RWRnG} shows that the gap between temporal and spatio-temporal forecasting is slightly larger for precipitation than for temperature, yet still remains very modest. This is noteworthy given the inherently localized and rapidly evolving nature of rainfall. Convective storms, orographic lifting, and frontal passages should, in principle, make prediction at unobserved stations substantially more difficult. The relatively small deterioration in skill suggests that the GP models capture the relevant fine-scale spatial structure and, together with the covariates, effectively propagate information to new locations through their joint space-time covariance. 

\begin{figure}[ht!]
    \centering
    \includegraphics[width=\linewidth]{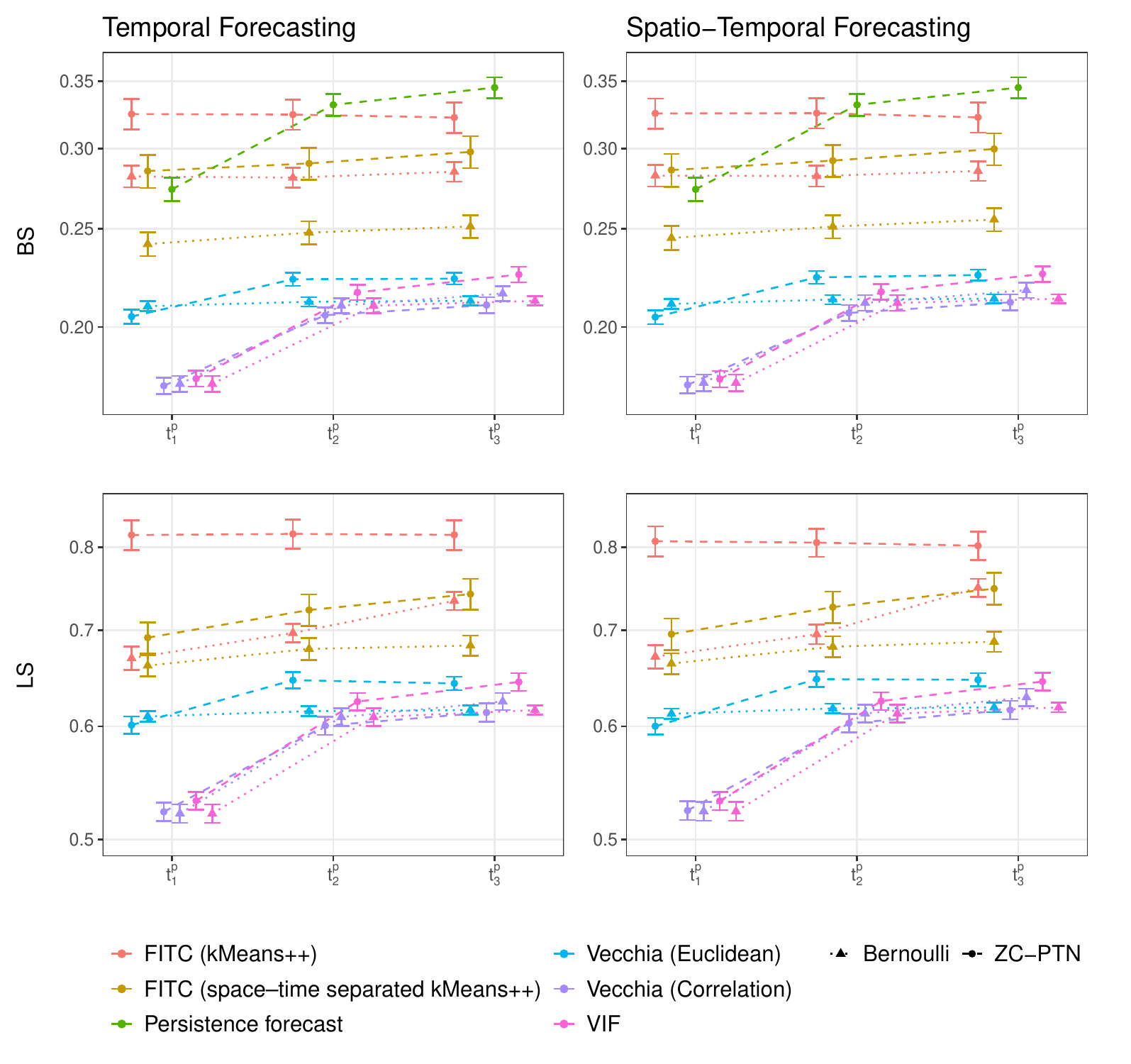}
    \caption{Day-wise BS and LS (log scale) with error bars (mean $\pm$ one standard error) for one- to three-day precipitation occurrence forecast under the temporal, and the spatio-temporal forecasting
settings. Methods include the Vecchia approximation with Euclidean- and correlation-based neighbors ($m_v=30$), FITC with standard and space-time separated kMeans++ inducing points ($m=500$), and VIF ($m_v=30$, $m=500$). A persistence forecast and results for the Bernoulli likelihood are included for comparison.}
    \label{fig:RWRnG1}
\end{figure}

Figure \ref{fig:RWRnG1} reports results for a latent GP model with a Bernoulli likelihood and a logit link function. Unlike the ZC-PTN model, this model only uses binary wet/dry data and ignores precipitation amounts. However, we observe only minor differences compared to the ZC-PTN likelihood for the most accurate GP models (a correlation-based Vecchia approximation and a VIF approximation).

\begin{figure}[ht!]
\centering
\includegraphics[width=\linewidth]{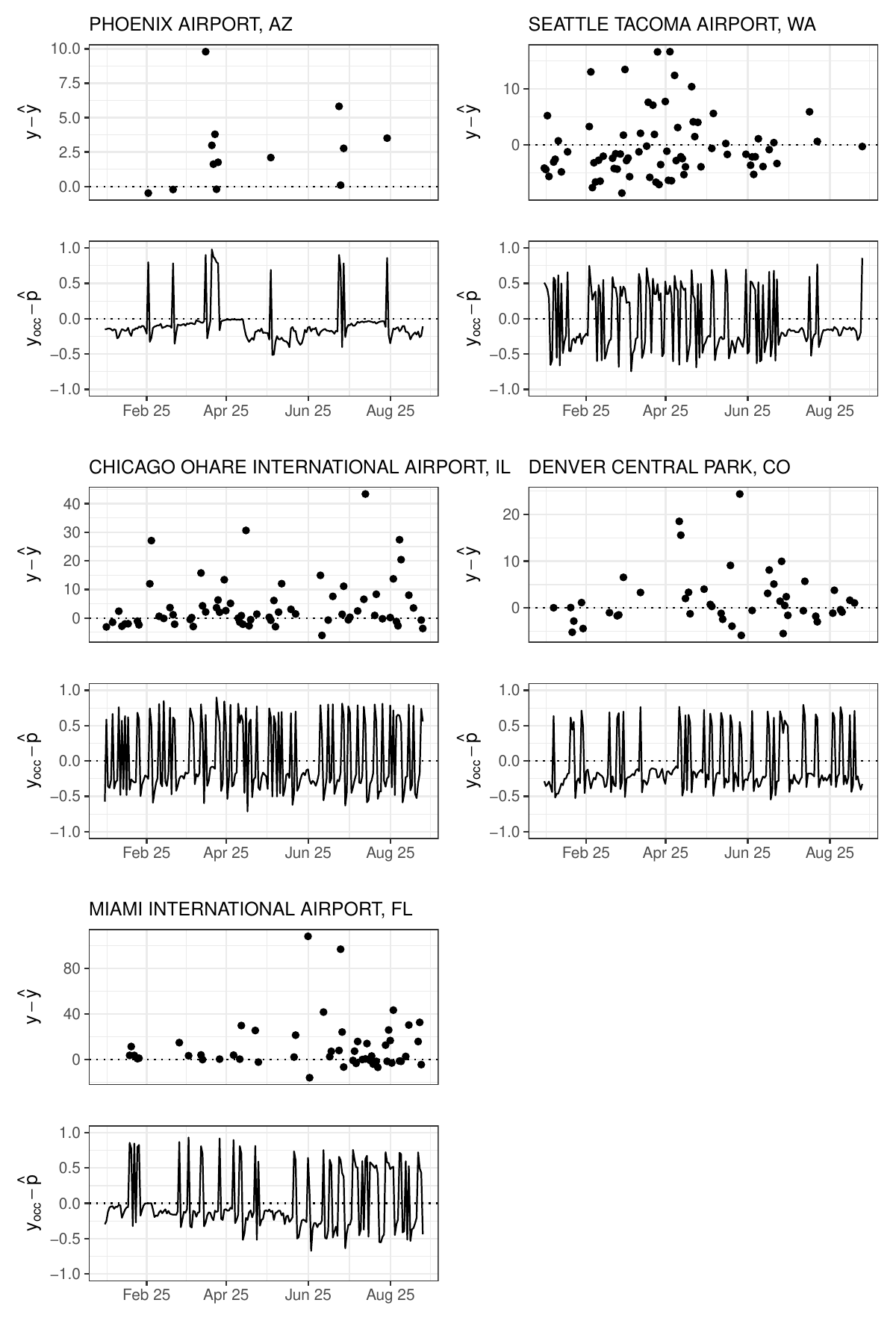}
\caption{One-day-ahead precipitation forecast residuals (conditional on rain) and rain-occurrence errors for five representative GHCNd stations across the full prediction period.}\label{fig:RWT123456}
\end{figure}

Station-level one-day-ahead precipitation residuals reported in Figure \ref{fig:RWT123456} show a tendency to underestimate rainfall amounts for larger events. This pattern is consistent with the challenges of predicting convective and storm-driven extremes at daily resolution. Regions dominated by frequent, moderate-intensity rainfall exhibit more balanced residuals, whereas areas with highly intermittent or topographically forced precipitation show larger conditional errors. Occurrence residuals remain centered near zero, with a mild tendency to underpredict rainy days. This reflects the difficulty of anticipating rare or localized precipitation events, while periods with more regular rainfall yield smaller occurrence errors. Overall, the lack of sustained bias in either component indicates stable performance of the GP model across diverse precipitation regimes.

\begin{figure}[ht!]
    \centering
    \ifdefined\ARXIV
  \includegraphics[width=0.65\linewidth]{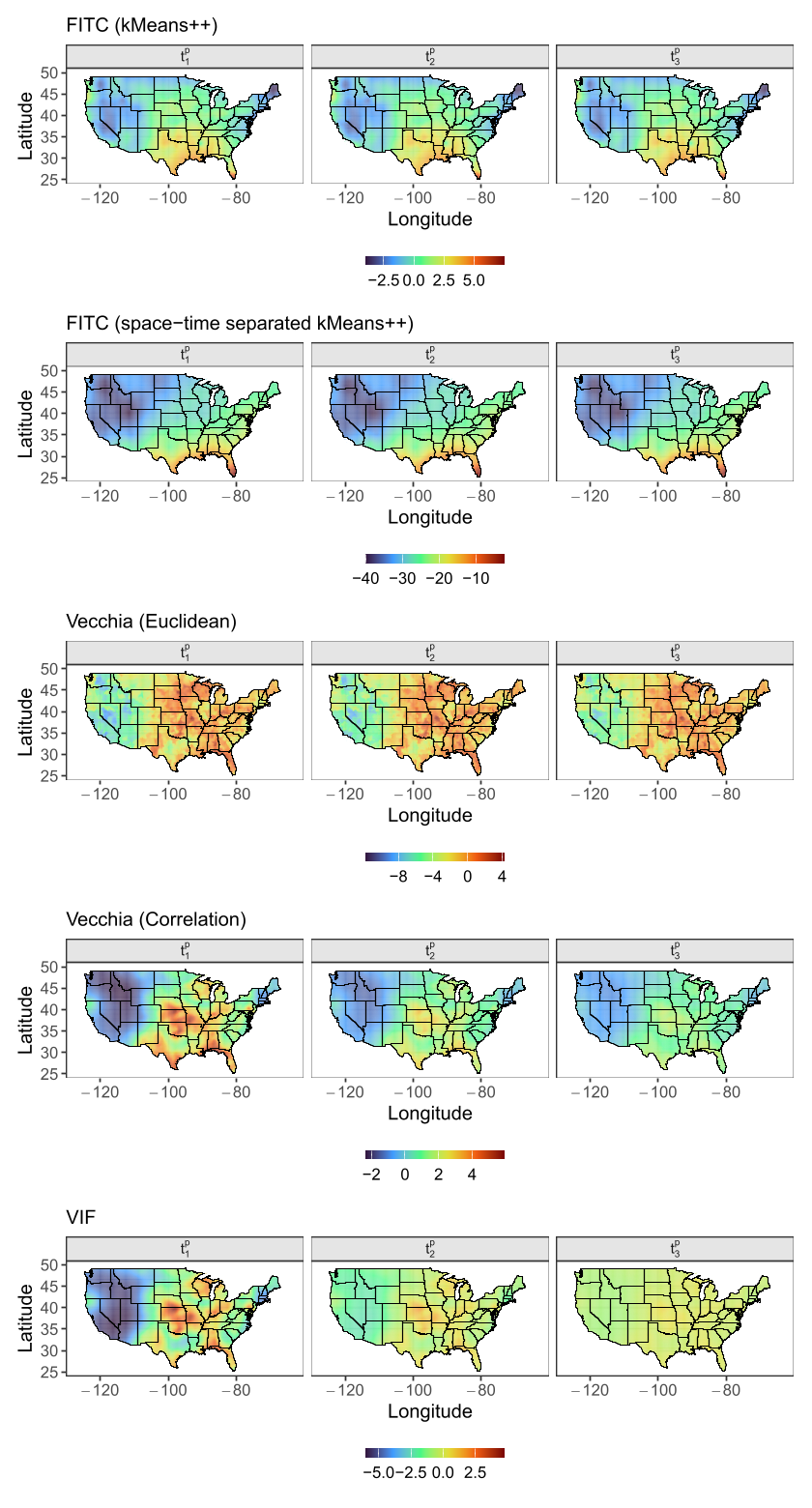}
\else
  \includegraphics[width=0.65\linewidth]{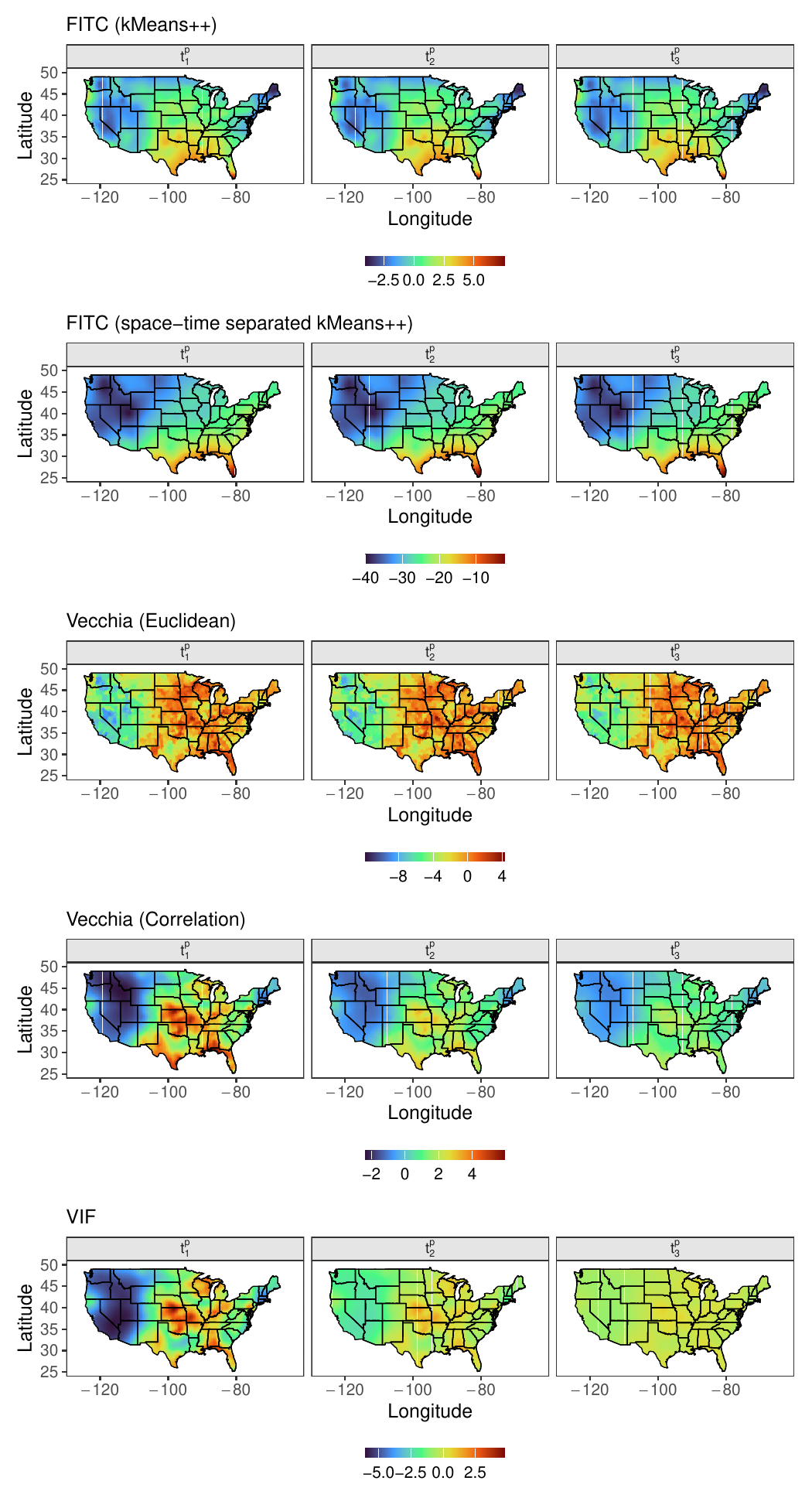}
\fi
    \caption{Estimated GP random effects at the final re-estimation step for the precipitation data across the conterminous United States for multiple temporal lags, shown for the Vecchia (Euclidean and correlation-based), FITC (standard and space-time separated kMeans++), and VIF approximations.}
    \label{fig:RWRA_GP_N}
\end{figure}

Figure \ref{fig:RWRA_GP_N} shows the estimated GP random effects over the conterminous United States at multiple temporal lags, providing a visual comparison of how the different approximations represent space-time dependence.

\begin{figure}[ht!]
    \centering
    \includegraphics[width=0.4\linewidth]{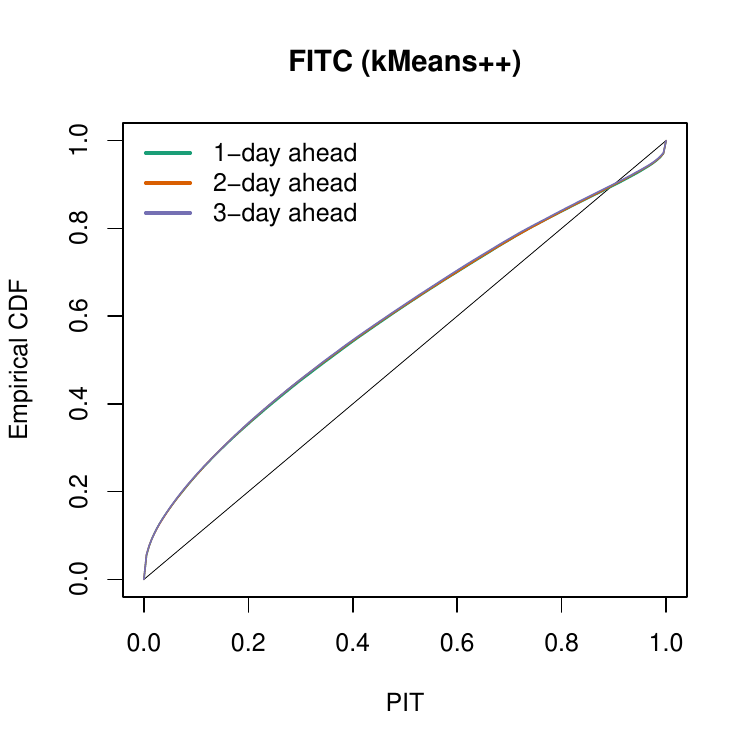}
    \includegraphics[width=0.4\linewidth]{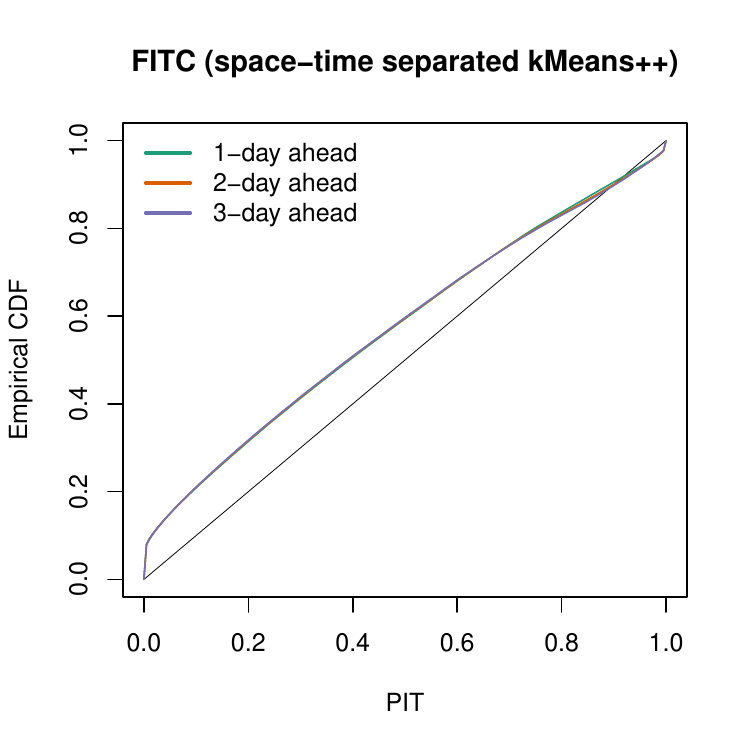}
\vspace{-2em}
    \includegraphics[width=0.4\linewidth]{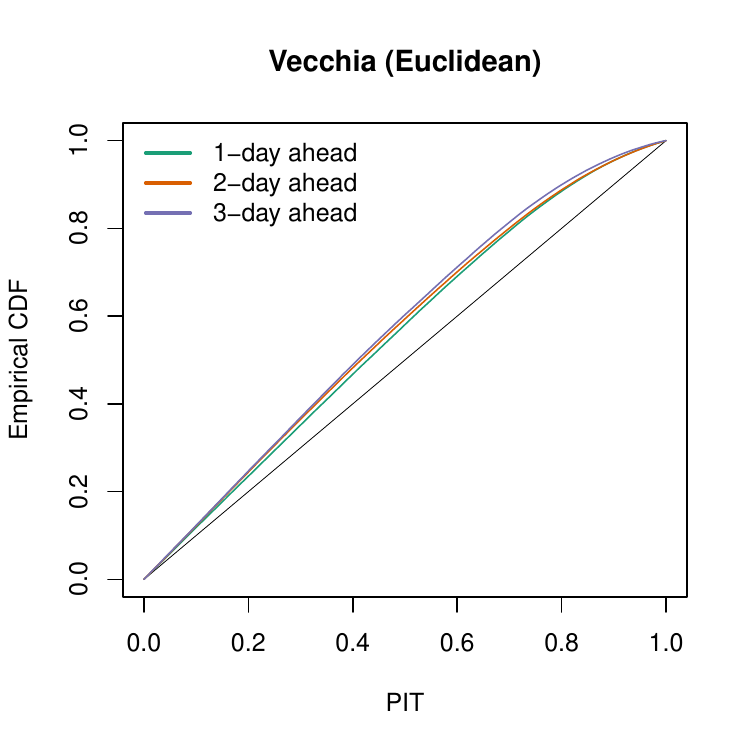}
    \includegraphics[width=0.4\linewidth]{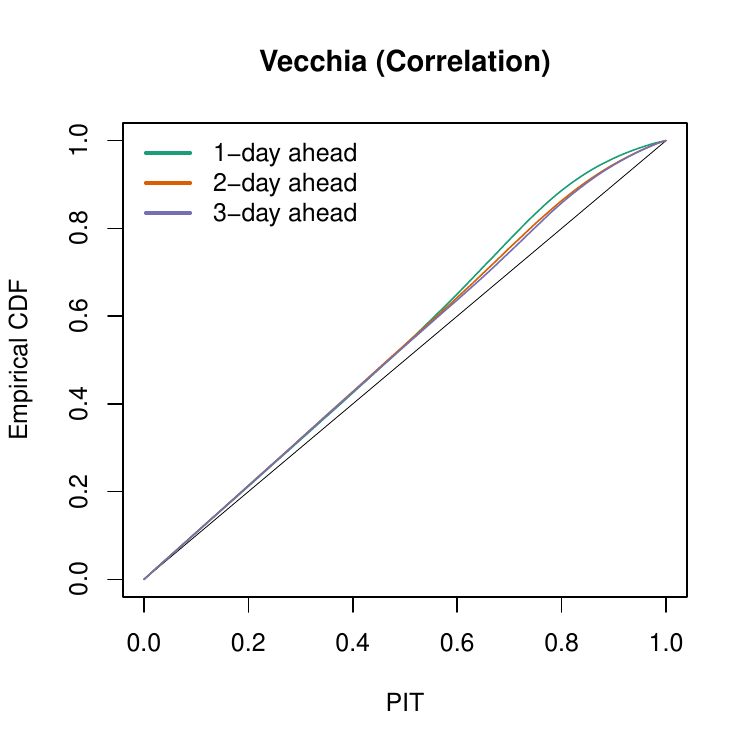}
\vspace{-2em}
    \includegraphics[width=0.4\linewidth]{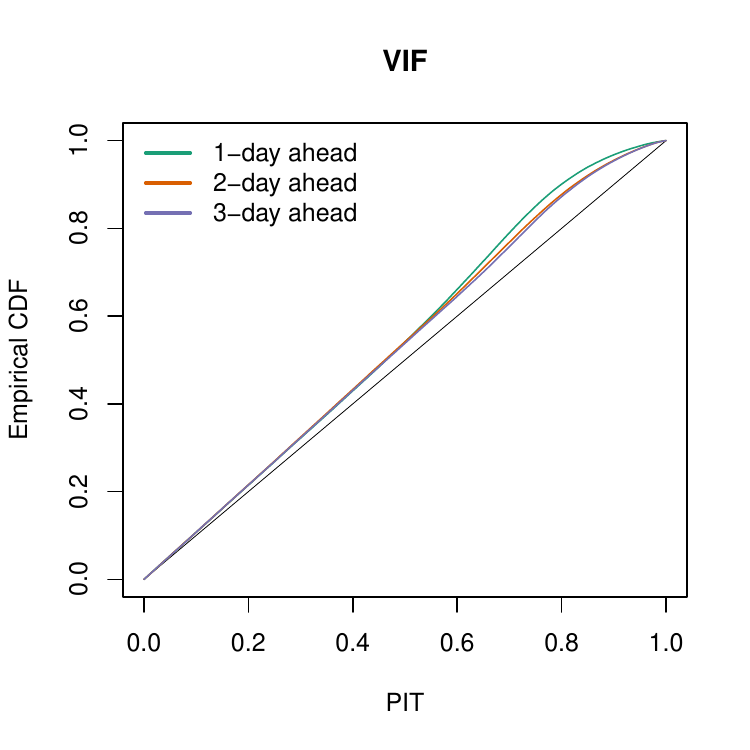}

    \caption{Randomized PIT reliability diagrams for precipitation forecasts using VIF, FITC, and Vecchia approximations under the ZC-PTN likelihood. Curves show uniform(0,1) QQ-plots for the empirical PIT CDFs for forecast instances pooled over all stations and shown separately for each lead time from one to three days. The black diagonal represents perfect calibration.}
    \label{fig:PIT_P}
\end{figure}

The randomized PIT reliability diagrams for precipitation in Figure \ref{fig:PIT_P} evaluate calibration of the full zero-censored power-transformed normal (ZC-PTN) predictive distribution, jointly assessing precipitation occurrence and positive amounts. %The FITC approximations exhibit substantial miscalibration across lead times, with PIT CDFs lying well above the diagonal, indicating systematic wet bias and overdispersion. In contrast, both Vecchia variants and VIF show markedly improved calibration, with PIT curves closer to the diagonal and relatively stable behavior across the one- to three-day prediction horizon, suggesting a more accurate representation of predictive uncertainty. Differences between the Vecchia (Euclidean and correlation-based) and VIF approximations are small at the aggregate level, indicating broadly comparable performance, with slight improvements for the correlation-based Vecchia and VIF approaches. Residual departures from uniformity across all methods point to remaining lead-dependent dispersion errors and mild tail effects.

\clearpage

\subsection{Additional results} Table~\ref{smootness_tab} shows that the forecasting performance of the VIF approximation is largely insensitive to the choice of the fixed smoothness parameter $\nu \in \{0.5,1.5,2.5\}$.

\begin{table}[H]
\begin{threeparttable}
\caption{Average three-day MAE, CRPS, BS, and LS for precipitation forecasts, and five-day RMSE and CRPS for temperature forecasts (mean $\pm$ one standard error), using the VIF approximation with fixed smoothness parameters $\nu \in \{0.5,1.5,2.5\}$ on a subset of the data spanning 2025-01-01 to 2025-01-31.}\label{smootness_tab}
\begin{tabular}{ |p{3.cm}|p{2cm}||p{3cm}|p{3cm}|p{3cm}|}
  \hline
   \multicolumn{2}{|c||}{}&\makecell[c]{$\nu = 0.5$}& \makecell[c]{$\nu = 1.5$}& \makecell[c]{$\nu = 2.5$}\\
  \hline 
  \makecell[l]{Short-term \\ prediction of \\precipitation}& \makecell[l]{MAE \\ CRPS \\ BS \\ LS} & \makecell[l]{$4.405 \pm 0.145$ \\
  $3.588 \pm 0.133$ \\ 
  $0.211 \pm 0.008$ \\ 
  $0.601 \pm 0.032$}& \makecell[l]{$4.393 \pm 0.145$ \\
  $3.585 \pm 0.133$ \\ 
  $0.211 \pm 0.008$\\ 
  $0.600 \pm 0.032$}& \makecell[l]{$4.388 \pm 0.144$ \\
  $3.579 \pm 0.132$ \\
  $0.210 \pm 0.008$\\ 
  $0.600 \pm 0.031$}\\
  \hline 
  \makecell[l]{Short-term \\ prediction of \\daily maximum \\temperature}& \makecell[l]{RMSE \\ CRPS}& \makecell[l]{$3.417 \pm 0.145$ \\
  $2.019 \pm 0.105$}& \makecell[l]{$3.417 \pm 0.145$ \\
  $2.019 \pm 0.105$}& \makecell[l]{$3.417 \pm 0.145$ \\
  $2.019 \pm 0.105$}\\
  \hline 
 \end{tabular}
     \end{threeparttable}
 \end{table}

\end{appendices}

\clearpage
\renewcommand{\refname}{Appendix References}
\putbib[bib_IterartiveFSA] % prints only citations from this bibunit
\end{bibunit}